%% file: mhongoose_mnras.r2.tex
\documentclass[a4paper,fleqn,usenatbib]{mnras}

\usepackage{newtxtext,newtxmath}

\usepackage[T1]{fontenc}
\usepackage{ae,aecompl}

\usepackage{graphicx}	
\usepackage{amsmath}	
\usepackage{amssymb}	

\usepackage{amssymb}
\usepackage{lipsum}
\usepackage{afterpage}
\usepackage{threeparttable}
\usepackage{placeins}
\usepackage{multirow,makecell,booktabs}
\usepackage{subfigure}
\usepackage{cleveref}
\Crefname{figure}{Fig.}{Figs.}
\usepackage{enumitem}

\newcommand{\hi}{\mbox{H{\sc i}}}

\newcommand{\kms}{km s$^{-1}$}

\newcommand{\ml}{$M_\odot/L_\odot$}
\newcommand{\Mpcc}{\rm M_\odot\,pc^{-3}}

\newcommand{\Mo}{\rm M_{\odot}}
\newcommand{\Ms}{{\rm M_\star}}
\newcommand{\Mhi}{{\rm M_{\textsc{Hi}}}}
\newcommand{\dg}{^{\circ}}
\newcommand{\mjybeam}{\rm mJy\,beam^{-1}}

\newcommand{\cm}{\rm cm^{-2}}

\newcommand{\e}[1]{\times 10^{#1}}
\newcommand{\HI}{\textsc{Hi}}
\newcommand{\nhi}{N_\HI}

\newcommand{\miriad}{\textsc{miriad}}
\newcommand{\gipsy}{\textsc{gipsy}}

\newcommand{\rotmas}{\textsc{rotmas}}

\newcolumntype{H}{>{\setbox0=\hbox\bgroup}c<{\egroup}@{}} 

\title[Early Observations of the MHONGOOSE Galaxies]{Early Observations of the MHONGOOSE Galaxies: Getting Ready for MeerKAT}

\author[Sorgho et al.]{A. Sorgho$^{1}$\thanks{sorgho@ast.uct.ac.za}, 
C. Carignan$^{1,2}$,
D. J. Pisano$^{3,4}$\thanks{Adjunct Astronomer, Green Bank Observatory},
T. Oosterloo$^{5,6}$,
W. J. G. de Blok$^{5,1,6}$,
\newauthor
M. Korsaga$^{1,7}$,
N. M. Pingel$^{3,4}$,
Amy Sardone$^{3,4}$,
S. Goedhart$^{8}$,
S. Passmoor$^{8}$,
\newauthor
A. Dikgale$^{8}$,
S. K. Sirothia$^{8,9}$
\\
$^{1}$ Department of Astronomy, University of Cape Town, Private Bag X3, Rondebosch 7701, South Africa\\
$^{2}$ Laboratoire de Physique et Chimie de l'Environnement (LPCE), Observatoire d'Astrophysique de l'Universit\'{e} de Ouagadougou (ODAUO),\\ BP 7021, Ouagadougou 03, Burkina Faso\\
$^{3}$ Department of Physics \& Astronomy, West Virginia University, P.O. Box 6315, Morgantown, WV, 26506, USA\\
$^{4}$ Center for Gravitational Waves and Cosmology, West Virginia University, Chestnut Ridge Research Building, Morgantown, WV 26505, USA\\
$^{5}$ Netherlands Institute for Radio Astronomy (ASTRON), Postbus 2,
7990 AA Dwingeloo, The Netherlands\\
$^{6}$ Kapteyn Astronomical Institute, University of Groningen, P.O. Box
800, 9700 AV Groningen, The Netherlands\\
$^{7}$ Aix Marseille Univ, CNRS, CNES, LAM, Marseille, France\\
$^{8}$ South African Radio Astronomy Observatory (SARAO), Pinelands 7405, South Africa\\
$^{9}$ Department of Physics and Electronics, Rhodes University, PO Box 94, Grahamstown, 6140, South Africa\\
}

\date{Accepted October 12th}

\pubyear{2018}

\begin{document}
\label{firstpage}
\pagerange{\pageref{firstpage}--\pageref{lastpage}}
\maketitle

\begin{abstract}
 We present early observations of 20 galaxies in the MHONGOOSE survey galaxies using KAT-7, the GBT, and MeerKAT. We present the best calibrators for five of the MHONGOOSE galaxies observed with the KAT-7, and search for signs of gas accretion in the GBT data, down to $3\sigma$ column density levels of $2.2\e{18}\,\cm$ over a 20 \kms\, line width, but identify none. Using the KAT-7 and MeerKAT data, we have derived rotation curves and mass models for NGC 3621 and NGC 7424 out to an unprecedented extent. As a precursor to the SKA, the MeerKAT telescope combines both a high spatial resolution and a large field of view, necessary to map the extended neutral hydrogen in local galaxies. The mass models of the two galaxies were constructed for both the Dark Matter (DM) models (the pseudo-isothermal model and the Navarro-Frenk-White model) and MOND. Overall, we find that the DM models provide a better fit than MOND to the galaxies' rotation curves. Furthermore, the pseudo-isothermal model is found to be the most consistent with the observations.
\end{abstract}



\section{Introduction}

How galaxies accrete their gas and then convert that gas into stars remains one of the hot topics in the area of galaxy formation and evolution. Current star formation rates (SFRs) of nearby galaxies are, as derived from interferometer observations, an order of magnitude higher than what can be sustained by minor mergers and accretion \citep{Sancisi2008,Kauffmann2010}.
This implies that local galaxies must somehow be accreting gas from the intergalactic medium (IGM).

Gas accretion from the IGM comes in two modes \citep{Keres2005}: the ``cold'' mode, which dominates in low mass galaxies residing in low density and high redshift environments, and the ``hot'' mode predominant in high mass galaxies evolving in high density environments. Recent surveys such as The \hi\, Nearby Galaxies Survey \citep[THINGS;][]{Walter2008} and the WSRT Hydrogen Accretion in Local Galaxies Survey \citep[HALOGAS;][]{Heald2011} have begun to solve the puzzle of gas accretion by not only studying the \hi\ content and distribution in and around galaxies, but also by investigating the relation between the \hi\, content and the star formation of galaxies. HALOGAS has mapped 22 disk galaxies down to a column density level of $\sim10^{19}\cm$ at spatial resolutions of a few kpc and spectral resolution of 16 \kms, an order of magnitude lower than typically found in the main \hi\, disks. However, the survey focused on obtaining a high column density sensitivity but lacks spatial resolution. On the other hand, the THINGS survey was conducted with the VLA high spatial resolution but was limited in column density sensitivity. The challenge of the MHONGOOSE \citep[MeerKAT \hi\, Observations of Nearby Galaxies; Observing Southern Emitters,][]{DeBlok2018} survey will be to combine high column density sensitivity and high spatial resolution over a large field of view, making it the first survey to benefit from this triple advantage, and that will therefore provide information on the processes driving the transformation and evolution of galaxies in the nearby universe at high resolution and also to low column densities. MHONGOOSE is a deep \hi\, survey that will make use of the MeerKAT telescope to observe 30 nearby galaxies down to a targeted $3\sigma$ column density sensitivity of $7.5\e{18}\cm$, at a resolution of $30''$ integrated over 16 \kms. At $90''$ resolution, the corresponding $3\sigma$ limit is $5.5\e{17}\cm$. Several science topics will be addressed by MHONGOOSE, some of which are related to gas accretion in galaxies and the relation between the \hi\, content and star formation in galaxies. In effect, one of the goals of the survey consists of showing how the low column density gas is connected with the cosmic web, and where gas accretion occurs.

Before the MeerKAT telescope is ready to conduct the full survey, it is essential to use existing telescopes to get a first characterisation of the candidate galaxies. In the present work, we use the Karoo Array Telescope \citep[KAT-7;][]{Carignan2013} and the Green Bank Telescope\footnote{GBT is a facility of the U.S.A National Science Foundation operated under cooperative agreement by Associated Universities, Inc.} \citep[GBT,][]{Prestage2009} to conduct the early observations of the MHONGOOSE sample. As an engineering testbed for the MeerKAT telescope, KAT-7 is an extremely compact array of seven 12m-diameter antennas, with baselines ranging from 26 m to 185 m. These short baselines of the telescope and its large primary beam of $\sim1\dg$ make it suitable to map the low column density structures. As for the GBT, the 100m diameter and excellent surface brightness sensitivity of the single-dish telescope give it the ability to survey the diffuse \hi\, environments typically undetected by most interferometers; the telescope has been used regularly to detect low column density structures around galaxies \citep[e.g,][]{Chynoweth2008,Chynoweth2009,Lockman2012,Wolfe2013,Pisano2014}.

The MeerKAT telescope is designed such that, at full capacity, it should reach a dynamic range of 1:$10^4$ in spectral line mode; the telescope will be able to detect structures at $\sim10^{17}\,\cm$ in the presence of $\sim10^{21}\,\cm$ emission. Because of this high dynamic range, it is essential to select suitable calibrators -- calibrators that will cause the least errors in the flux and/or the phase of the targets -- to optimise the quality of the observations with the telescope.
Also, with the initial 16-antenna release of MeerKAT, one can test the capability of the upcoming full array by conducting science observations of nearby \hi-rich galaxies.

The paper is organised as follows: we begin with a description of the MHONGOOSE sample in \Cref{sec:sample}. The observations with the KAT-7, GBT and MeerKAT telescopes as well as the data reduction are discussed in \Cref{sec:obs}, and the derived \hi\, properties of the observed galaxies are given in \Cref{sec:hi-properties}. The rotation curves of NGC7424 (observed with KAT-7) and NGC 3621 (observed with MeerKAT) are derived in \Cref{sec:kinematics}, and \Cref{sec:massmodel} provides a study of the mass (luminous and dark) distribution.
Finally in \Cref{sec:discussion}, we summarise the main results of the work and discuss them.


\section{Sample}\label{sec:sample}
The MHONGOOSE galaxies were selected based on their membership to both the HIPASS \citep[\hi\, Parkes All Sky Survey,][]{Meyer2004} catalogue and the SINGG \citep[Survey for Ionization in Neutral Gas Galaxies,][]{Meurer2006} survey. These are galaxies with HIPASS flux higher than 50 mJy, having a galactic latitude $|b|>30\dg$ and a Galactic standard of rest velocity $>200$ \kms.
Further constraint was put on the declination ($\delta < -10\dg$) and the distance ($<30$ Mpc) to ensure that the physical resolution is comparable to that of the THINGS survey. This resulted in a sample of 88 galaxies that were later divided in 6 bins of $\log{(\Mhi)}$, and 5 galaxies of various inclinations were selected in each bin: edge-on, face-on galaxies and galaxies of intermediate inclination ($50\dg-60\dg$) were chosen. The latter inclination is optimal for kinematical modelling. A more complete description of the sample selection is given in  \citet{DeBlok2018}.

Five of the 30 MHONGOOSE galaxies were observed with KAT-7. We observed fifteen of the remaining galaxies (plus one already observed with KAT-7) with the GBT, which brings our total sample size to 20 galaxies.

On top of the MHONGOOSE sample, the galaxy NGC 3621 was also observed with the MeerKAT telescope as part of commissioning observations. The total sample is listed in \Cref{tb:sample}.

\begin{table*}
	\centering
	\begin{tabular}{ l  c c c c c c}
      \hline
      \hline
      \multicolumn{1}{c}{Object} & R.A & Dec.   &  Type & B mag. & D (Mpc) & Tel.\\
			 
					
      \multicolumn{1}{c}{(1)} & \multicolumn{2}{c}{(2)} & (3) & (4) & (5) & (6) \\ 
      \hline
      ESO 300-G014\dotfill & 03 09 37.8 & -41 01 50 & SABm & 12.76 & 12.9 & K/G\\
      ESO 300-G016$^a$\dotfill & 03 10 10.5 & -40 00 11 & IAB & 15.57 & 9.3 & G\\
      ESO 302-G014\dotfill & 03 51 40.9 & -38 27 08 & IBm & 14.59 & 11.7 & G\\
      ESO 357-G007$^b$\dotfill & 03 10 24.3 & -33 09 22 & SBm & 14.61 & 17.8 & G\\
      IC 4951\dotfill      & 20 09 31.8 & -61 51 02 & SBdm & 13.61 & 11.2 & K\\
      KK98-195$^a$\dotfill     & 13 21 08.2 & -31 31 45 & I & 17.17 & 5.2 & G\\
      KKS2000-23$^a$\dotfill   & 11 06 12.0 & -14 24 26 & I & 15.80 & 12.7 & G\\
      NGC 0625\dotfill     & 01 35 04.6 & -41 26 10 & SBm & 11.59 & 4.1 & K\\
      NGC 1371\dotfill     & 03 35 01.3 & -24 56 00 & SABa & 11.41 & 20.4 & G\\
      NGC 1592\dotfill     & 04 29 40.1 & -27 24 31 & Pec & 14.38 & 13.0 & G\\
      NGC 3511\dotfill     & 11 03 23.8 & -23 05 12 & SAc & 11.36 & 14.2 & G\\
      NGC 3621$^c$\dotfill     & 11 18 16.5 & -32 48 51 & SAd & 10.16 & 6.6 & M\\
      NGC 5068\dotfill     & 13 18 54.8 & -21 02 21 & SABcd & 9.93 & 6.9 & G\\
      NGC 5170$^b$\dotfill & 13 29 48.8 & -17 57 59 & SAc & 11.54 & 28.0 & G\\
      NGC 7424\dotfill     & 22 57 18.4 & -41 04 14 & SABcd & 10.69 & 13.6 & K\\
      NGC 7793\dotfill     & 23 57 49.8 & -32 35 28 & SAd & 9.61 & 3.9 & K\\
      UGCA 015\dotfill     & 00 49 49.2 & -21 00 54 & IBm & 14.99 & 3.3 & G\\
      UGCA 250\dotfill     & 11 53 24.0 & -28 33 11 & Sd & 12.75 & 24.4 & G\\
      UGCA 307\dotfill     & 12 53 57.3 & -12 06 21 & IBm & 13.99 & 8.6 & G\\
      UGCA 320\dotfill     & 13 03 16.7 & -17 25 23 & IBm & 13.07 & 7.7 & G\\
      \hline
	\end{tabular}
	\caption{Total sample of galaxies observed. This includes the MHONGOOSE galaxies observed with KAT-7 and GBT, plus NGC 3621 observed with MeerKAT. Column (1): Object name; Column (2): J2000 optical position from NED; Column (3): Morphological type from the RC3 catalogue; Column (4): The B$_j$-band magnitude from \citet{Doyle2005}; Column (5): Distance of the galaxy in Mpc from \citet{Meurer2006}; Column (6): Telescope used for observation: K=KAT-7, G=GBT, M=MeerKAT. Notes: $^a$ Morphological type and total B-magnitude from HyperLEDA, $^b$ distance from \citet{Sorce2014}, $^c$ distance from \citep{Freedman2001}.}
	\label{tb:sample}
\end{table*}


\section{Observations}\label{sec:obs}


\subsection{KAT-7 observations and reduction}\label{sec:katobs}

The KAT-7 observations were conducted between late 2014 and early 2015, and three different phase calibrators were observed for each galaxy. The observing times are summarised in \Cref{tb:obs}, and the correlator mode used provided a total bandwidth of $12.5$ MHz over 4096 channels, which corresponds to a velocity resolution of about 0.65 \kms. The observation technique consists in alternating between the target and a phase calibrator about every 20 min to calibrate the phase stability, and observe the flux calibrator two or three times per session. Given the large primary beam of the telescope ($\sim1\dg$), a single pointing was sufficient to individually cover the galaxy candidates.

The complete data reduction was performed with the \miriad\, package \citep{Sault1995}, and the procedure followed is similar to the standard WSRT data reduction. However, since the KAT-7 telescope specifications are not implemented in \miriad, special care was taken to manually input, upon importing the data, some parameters such as the FWHM ($58'$), the system temperature (26 K), the telescope gain ($37.5\,\rm Jy\,K^{-1}$) and the \hi\, rest frequency (1420.4 MHz). Furthermore, we performed a spectral and time averaging over 4 channels and 20 min respectively, to improve the signal-to-noise ratio of the data. This brings the total number of channels down to 1024 and the spectral resolution to 2.6 \kms.
Because KAT-7 is an engineering testbed telescope, not all the RFI (Radio Frequency Interference) sources are known since the monitoring was still ongoing during the observations. So the RFI inspection had to be manually done to remove eventual artifacts from the data, including the effects of Galactic \hi\, on low velocity sources. The observing sessions were individually reduced and for each session a second order polynomial fit over line-free channels was used to perform continuum subtraction in the {\it uv} plane using the \miriad's task {\sc uvlin}. The emission from the galaxies typically spans less than 100 channels, providing a large number of line-free channels for the baseline fits. We then proceeded to correct the velocity coordinates of the continuum-subtracted {\it uv} data with the {\sc cvel} task in {\sc aips} \citep{Greisen2003}. This is because KAT-7 does not use Doppler tracking and, as mentioned before, the \miriad\, package does not contain the telescope's parameters. The resulting datasets for a source's individual sessions were then combined and a datacube was produced with a {\it natural} weighting.

\begin{figure*}
\includegraphics[width=\textwidth]{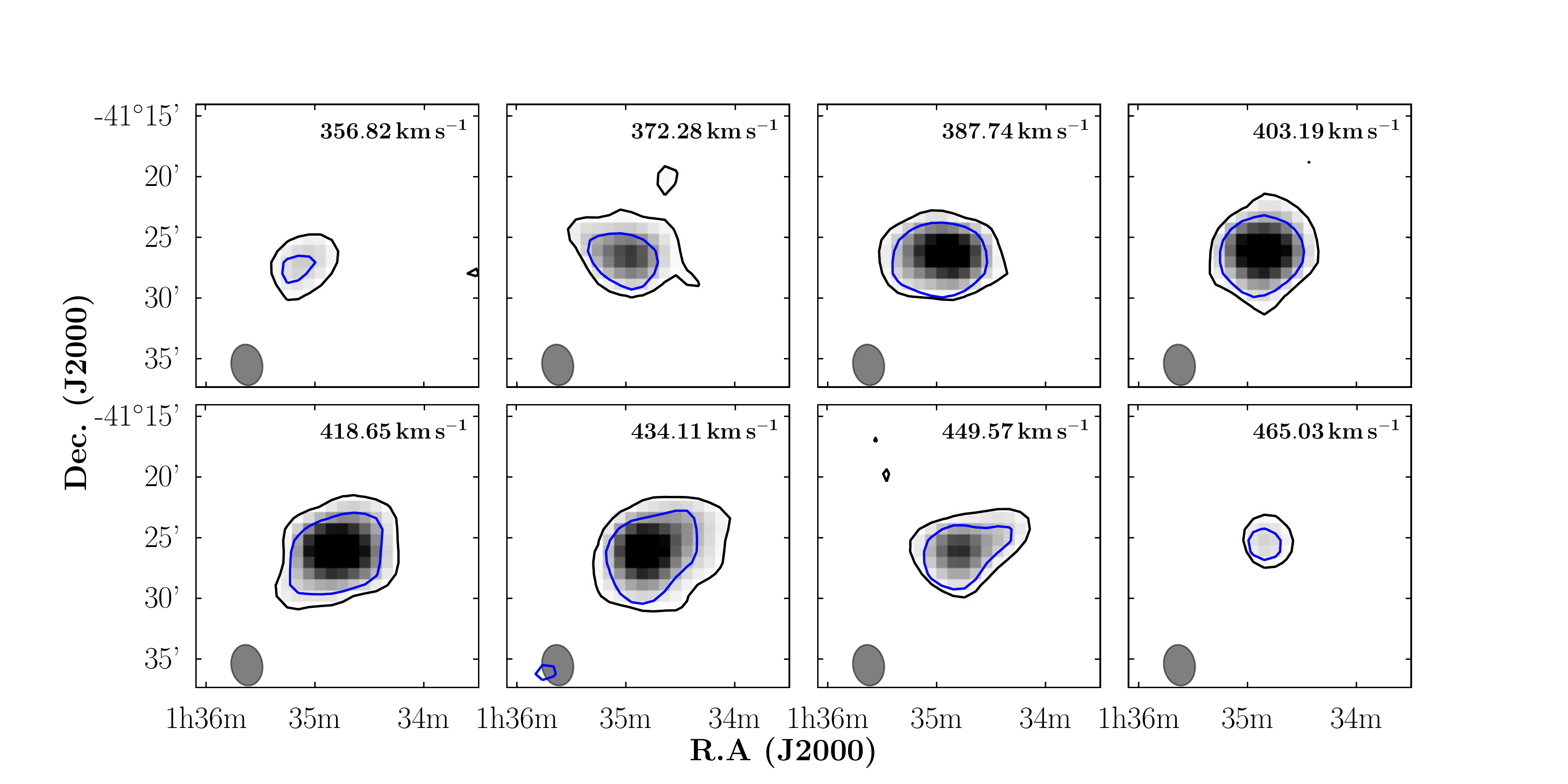}
\caption{Channel maps of  NGC 625 from two different cubes, obtained with two different phase calibrators: 0201-440 ({\it grayscale and black contours}) and 0039-445 ({\it blue contours}). The channels are sampled at 15.46 \kms, and both the black and blue contours are taken at the same flux level,  $0.02 \rm\,Jy\,beam^{-1}$. The synthesised beam of the data is represented by an ellipse in the lower left of each panel.}\label{fig:n625-cals}
\end{figure*}

To determine the best calibrators for the different sample galaxies, the process described above was repeated for each and every calibrator source. An ideal calibrator source can be approximated as a point source to the telescope. The selection of the best calibrators is based on their brightness and structure: the sources have to be bright and not confused, and also should not be at large angular distance from the target. In the data, these criteria result in the following:
\begin{itemize}[label={--}]
  \item the source must provide the least rms noise in the calibrated data cube of the target (the reference for the rms value being the theoretical noise),
  \item the scatter in the calibrator's phase vs amplitude graph for the best source must be the least.
\end{itemize}
The second criterion is to ensure that the source has the least variation both in phase and amplitude. To conduct a meaningful comparison between the flux calibrators, data with the same amount of observing time were considered. In \Cref{tb:obs} we list the best flux and phase calibrators, as well as the total observing time for each of the five targets considered. To illustrate the effects of the choice of the phase calibrators on the target source, we present in \Cref{fig:n625-cals} the channel maps of the NGC 625 data cubes calibrated with two different phase calibrators: 0039-445 and 0201-440. This is the most extreme case amongst the galaxies in the sample of five, where we notice a large difference in the end products with different calibrators although the same calibration scheme was used. \Cref{fig:n625-cals} also clearly demonstrates that we obtain a much higher signal-to-noise ratio (SNR) with the calibrator 0201-440.

\subsubsection*{Self-calibration}
\begin{figure}
\hspace{-25pt}
\includegraphics[width=1.2\columnwidth]{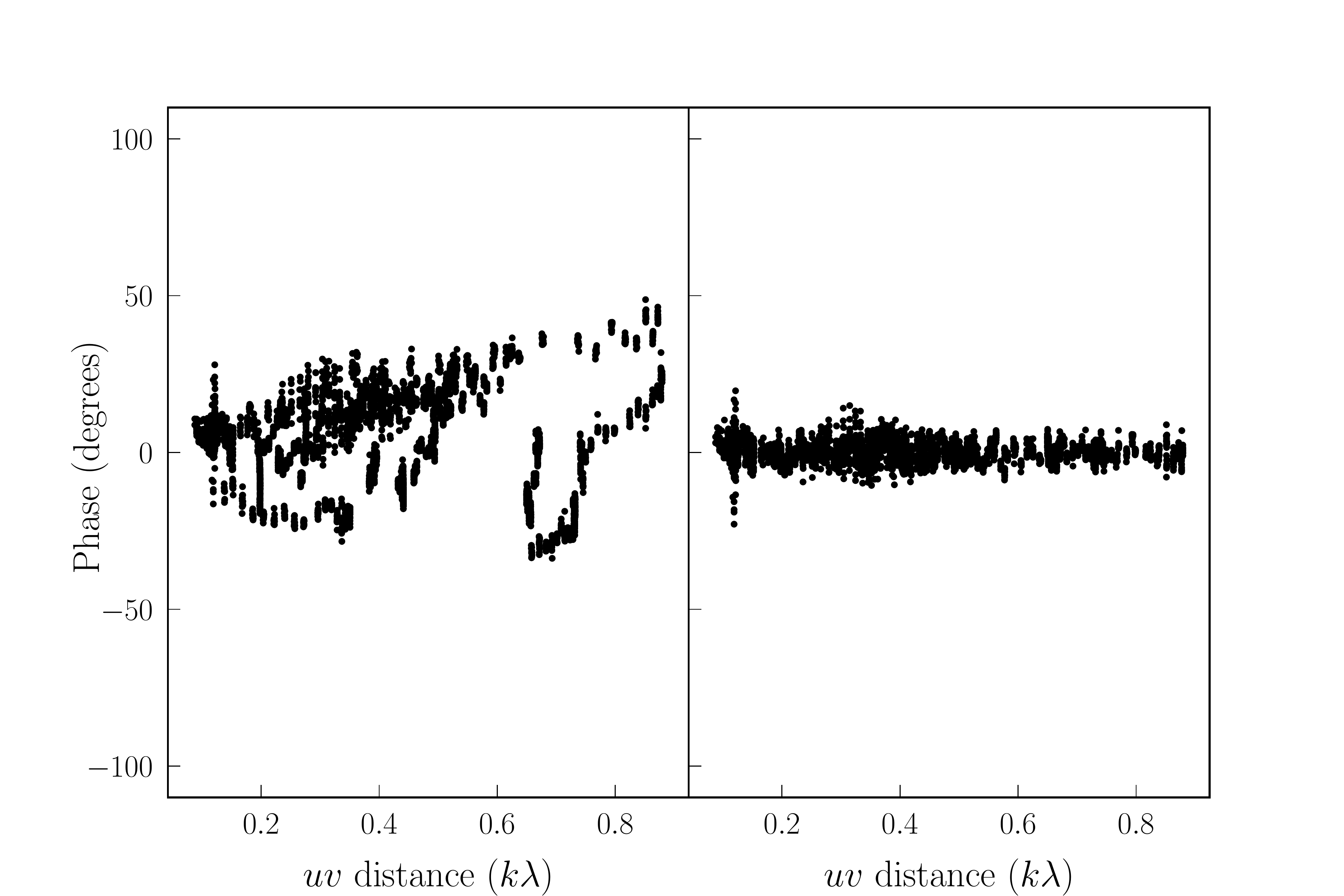}
\caption{Phase variation of the calibrator source 0039-445 observed with the KAT-7, before ({\it left}) and after ({\it right}) applying self-calibration.}\label{fig:phase-0039}
\end{figure}

A drawback of the KAT-7's large field of view is mostly noticed when observing calibrators. Ideally one would want to have only the compact calibrator source in the field. However, this is not easily achieved when the FWHM is as wide as it is for KAT-7. In effect, unwanted secondary sources in the calibrators' field are likely to affect the phase variation of the data. Because of this, the observed spectra of the visibilities are not flat. As a result, assuming a point source for deriving the bandpass introduces ripples in the bandpass calibration.
To solve the phase stability problem, we applied a self-calibration on the calibrator source. It consists of {\sc clean}ing a dirty map of the source previously produced with the {\sc invert} task in \miriad; the output cleaned component is in turn used as a model with the {\sc selfcal} task. To illustrate the effect of the self-calibration, we show in \Cref{fig:phase-0039} the variation of phase of the calibrator source 0039-445 as a function of the baseline, after standard calibration (left panel) and after self-calibration (right panel). Although the standard calibration significantly reduces the scatter in the source's phase -- the standard deviation in the phase goes from $99.3\dg$ to $15.5\dg$ -- one does not quite achieve a stable phase throughout the probed baseline. However, in the right panel of the figure, all the visibility points lie around the $0\dg$ phase.

\begin{table}
\begin{center}
\begin{tabular}{l c c c c}
\hline
\hline
Galaxy & Flux cal. & \multicolumn{2}{c}{Observing time ($h$)} & Best phase cal.\\
 & (best cal. *) & Total & On source & \\
\hline
\multirow{2}{*}{NGC 625}      & 0407-658* & 11.2 & 8.3 & \multirow{2}{*}{0201-440}\\
 & PKS 1934-638 & 13.2 & 11.1 \\
 \hline
\multirow{2}{*}{ESO 300-G014} & 0407-658* & 13.5 & 6.4 & \multirow{2}{*}{0220-349}\\
 & 3C138 & 4.5 & 3.0\\
 \hline
IC 4951      & PKS 1934-638* & 22.7 & 22.4 & PKS 1934-638\\
\hline
\multirow{2}{*}{NGC 7424}     & 0407-658 & 12.9 & 6.7 & \multirow{2}{*}{2259-375}\\
 & PKS 1934-638* & 14.0 & 9.2\\
 \hline
\multirow{2}{*}{NGC 7793}     & 0407-658 & 13.0 & 7.4 & \multirow{2}{*}{0008-421}\\
 & PKS 1934-638* & 15.1 & 8.7 \\
\hline
\end{tabular}
\end{center}
\caption{Observing time spent on each of the KAT-7 sample galaxies. These are estimated after flagging} \label{tb:obs}
\end{table}

\subsection{GBT observations and reduction}\label{sec:gbtobs}
The GBT telescope was used as an alternate instrument to observe the rest of the sample galaxies after KAT-7 became unavailable. Out of the 25 remaining galaxies, a total of 16 observable from the GBT were selected for mapping. The sample of 16 was observed in 2016 during 55 observing sessions\footnote{As part of the project GBT16B-212} and for each session, an area of $2\dg\times2\dg$ around each galaxy was observed for 10 hours, requiring a total observing time of 160 hours for the whole sample. The observational technique used is similar to that of \citet{Pisano2014}. The observations were done with the L-band receiver and the VEGAS spectrometer, using a 23.4 MHz bandwidth at 0.7 kHz resolution and a frequency switching at $\pm 2.5$ MHz. The primary flux calibrator source used during the individual sessions was either 3C48, 3C147 or 3C295 and the $T_{\rm cal}$ values determined for the noise diode were constant throughout individual sessions varied between 1.5 K and 1.7 K (with an average of $1.66\pm0.03$ K) from one session to another, for both the XX and the YY polarisations. This dispersion in $T_{\rm cal}$ corresponds to a flux uncertainty of 2\%. 

The frequency-switched spectra obtained after each observing session were individually reduced using \textsc{gbtidl}\footnote{http://gbtidl.nrao.edu/} where a third-order polynomial fit was applied to the line-free regions of the spectra to remove residual baseline structures and eventual continuum sources. The reduction was done with the same script as in \citet{Pingel2018}. The data were then flagged to remove recurrent RFIs at $\sim1416$ MHz. The flagged data were then smoothed to a 6.4 \kms\, resolution using a boxcar kernel, and exported into \textsc{aips} via \textsc{IdlToSdfits}\footnote{https://safe.nrao.edu/wiki/bin/view/GB/Data/IdlToSdfits} where they were combined by source, before being imaged into data cubes. For most sources, the baseline subtraction did not succeed to completely remove the strong continuum sources in the field. For these, an additional continuum subtraction was performed using a fourth-order polynomial with the \textsc{aips} task \textsc{xbasl}.

The typical noise achieved per galaxy candidate is $\sigma \sim 20\,\rm mK$ over $6.4\,\rm km\,s^{-1}$, equivalent to $2.3\e{17}\,\rm cm^{-2}$ assuming optically thin emission. This corresponds to a flux sensitivity of $\sigma \sim 10\,\rm mJy/beam$, or a $5\sigma$ \hi\, mass detection limit of $M_\HI = 4.7\e{5}\,d^2\,\Mo$ for a 20 \kms\, line width source at a distance $d$ Mpc.

\begin{table}
	\begin{center}
  \begin{tabular}{ p{3cm} | c | c | c}
      \hline\hline
      Parameter & KAT-7 & GBT & MeerKAT\\
      \hline
			Weighting function & natural & -- & robust = 0 \\
			Channel width (\kms)&  2.6 & 6.4 & 11.2\\
			FWHM of beam & $\sim3.5'\times3.2'$ & $9.1'$ & $1.9'\times1.2'$\\
			Pixel size & $55''\times55''$ & $105''\times105''$ & $15''\times15''$\\
			Map size & $3.9\dg\times3.9\dg$ & $2\dg\times2\dg$ & $\sim2\dg\times2\dg$\\
			RMS noise ($\mjybeam$)& 7.0 & $10$ & $1.7$\\
			$\nhi$ ($1\sigma\e{18}\,\rm cm^{-2}$, over 16.5 \kms)$^a$ &	3.2 & 0.59 & 3.8\\
			\hline 
  \end{tabular}
  \caption{Summary of the KAT-7, GBT and MeerKAT image cubes. The channel width listed for the MeerKAT data cube is the width obtained after Hanning-smoothing. Note: $^a$ the $\nhi$ limit is assuming that any emission completely fills the beam.}\label{tb:obs-cubes}
  \end{center}
\end{table}


\subsection{MeerKAT commissioning observations of NGC 3621}\label{sec:meerkatobs}
As of April 8th 2017, date of NGC 3621's observations, only half of the 32 then completed MeerKAT's antennas were available since the ROACH2 correlator then used was limited to a total of 32 inputs \citep[see][for a description of MeerKAT]{Camilo2018}. The observations were made as part of MeerKAT commissioning activities, and were not taken as part of MHONGOOSE. Two polarisations were used from each of the 16 antennas, which were chosen to give the best collection of shortest baselines available. Besides the high column density sensitivity and resolution of the telescope ($1.9'\times1.2'$ as used), what makes it suitable for detecting extended \hi\, features is its large field of view of $\sim1\dg$. NGC 3621 has an optical size of $12.3'\times7.1'$, which required no more than one MeerKAT pointing to map its \hi\, distribution. The observations were conducted on the evening and at night for 8.2 hours in total, and made use of 1117-248 as complex gain calibrator and the sources PKS 0408-65 and PKS 1934-638 as flux and bandpass calibrators. The observation process is similar to KAT-7's, with the bandpass calibrator observed for 10min every hour during the observation, and the complex gain calibrator observed for 2min every 5min with a dump rate of 4 seconds. The observations were performed with the 32k correlator mode, which has 32,768 channels over a total bandwidth of 856 MHz covering the frequency range of 856 MHz to 1712 MHz. This gives a channel width of 26.1 kHz (5.6 \kms).

Overall, after flagging, the total observing time of the target is 253 mins, that of the flux calibrators is 73 mins, while the phase calibrator received 110 mins. The data were reduced with \miriad\, following the same procedure as the KAT-7 data. In general, the MeerKAT data cannot be reduced with \miriad\, because of the {\it w}-projection algorithm not being available in the package.
However, the {\it w}-projection is only necessary when the observations are affected by the {\it non-coplanar baselines} effect, which happens when the Fresnel number, defined as $N_F = D^2/B\lambda$ (where $D$ is the antenna diameter, $B$ the maximum baseline and $\lambda$ the observing wavelength), is $N_F < 1$ \citep{Cornwell2008}. In the case of the MeerKAT observations, the baseline length at which $N_F < 1$ is 868m, which is larger than the maximum baseline of 715m of the array as used. This means that the non-coplanar effects are negligible, and {\it w}-projection can be ignored.




\section{\hi\, properties}\label{sec:hi-properties}

\subsection{\hi\, Distribution}\label{sec:hidist}
\begin{figure*}
\subfigure{\includegraphics[width=\columnwidth]{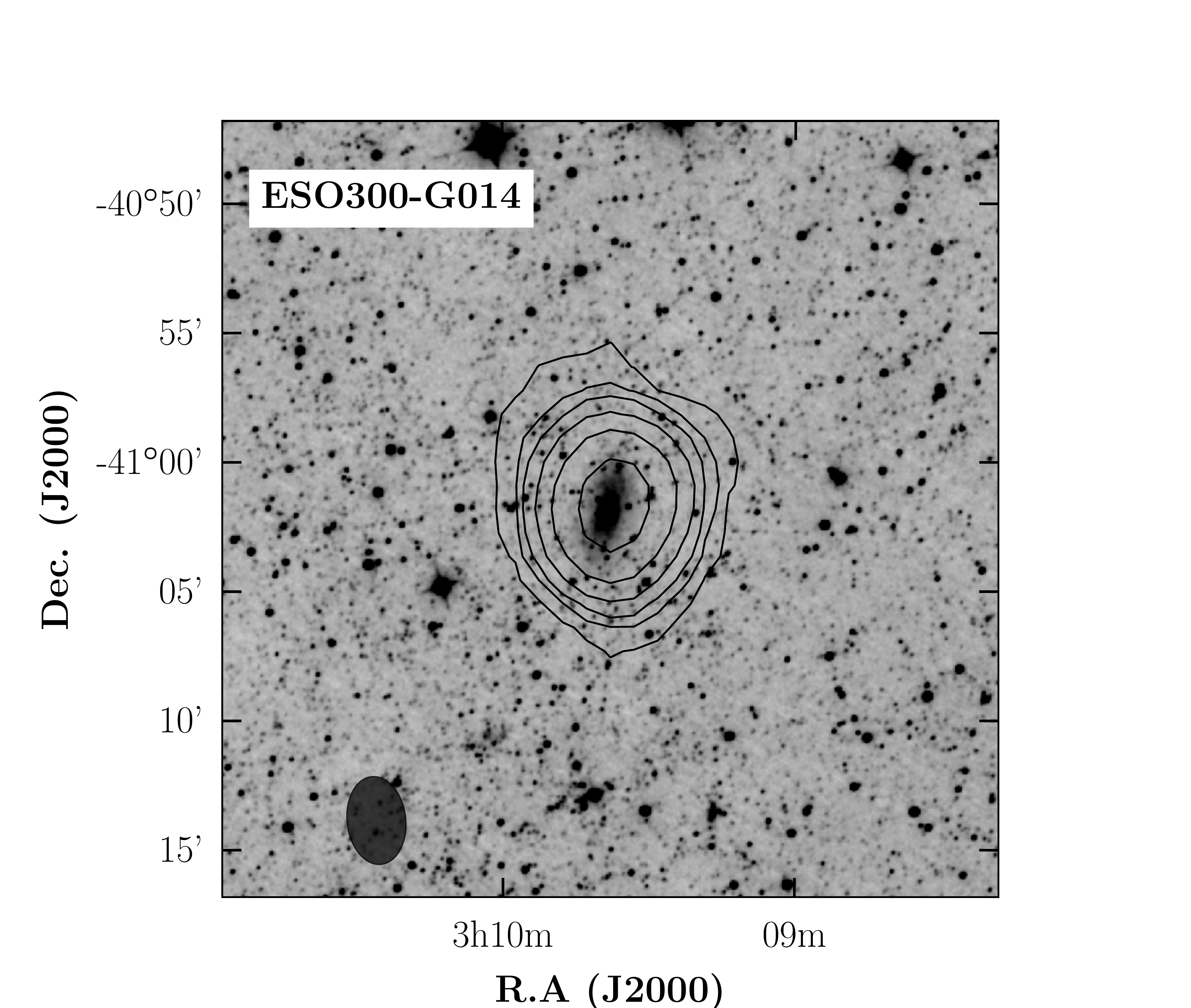}}\quad
\subfigure{\includegraphics[width=\columnwidth]{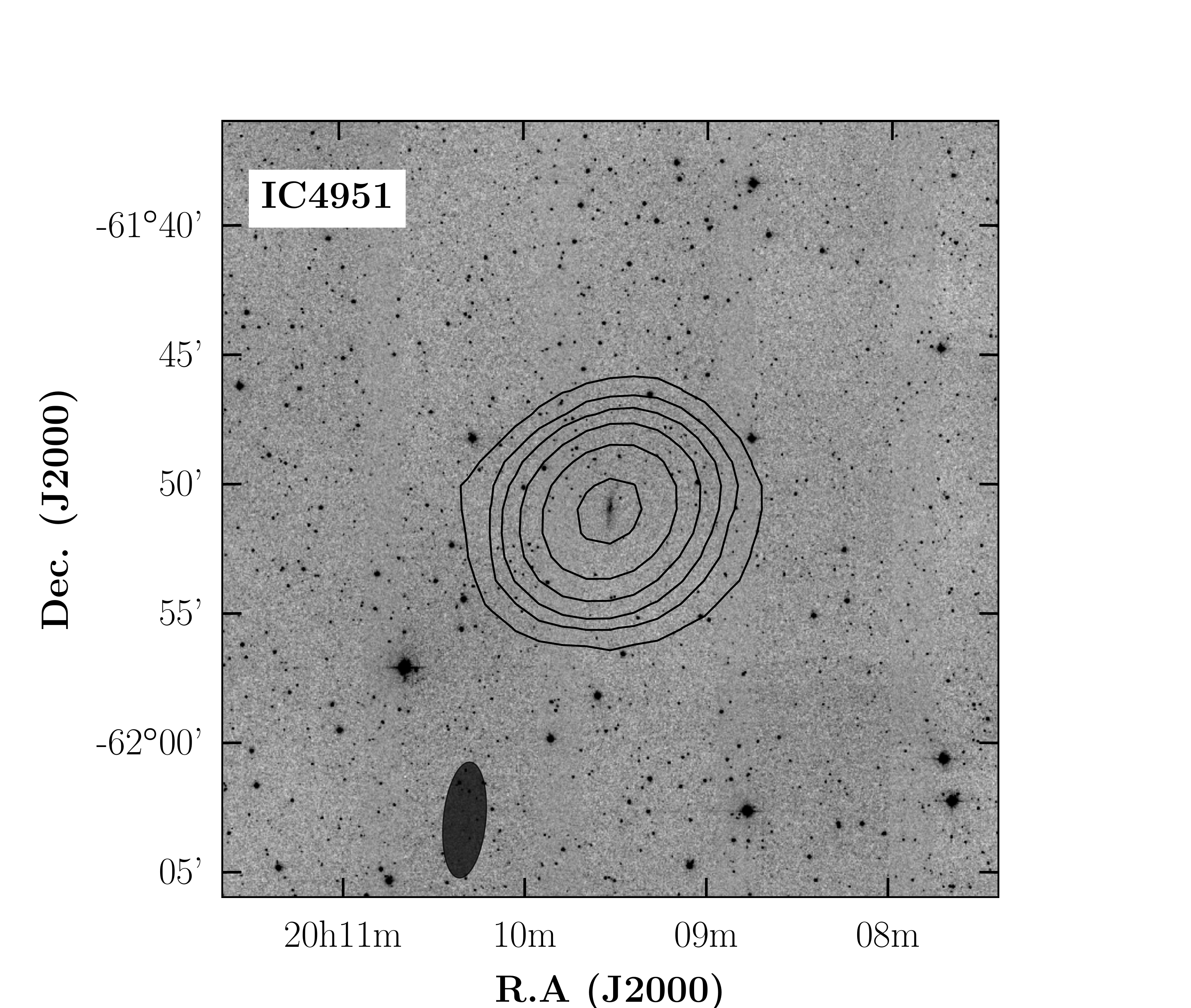}}\quad
\subfigure{\includegraphics[width=\columnwidth]{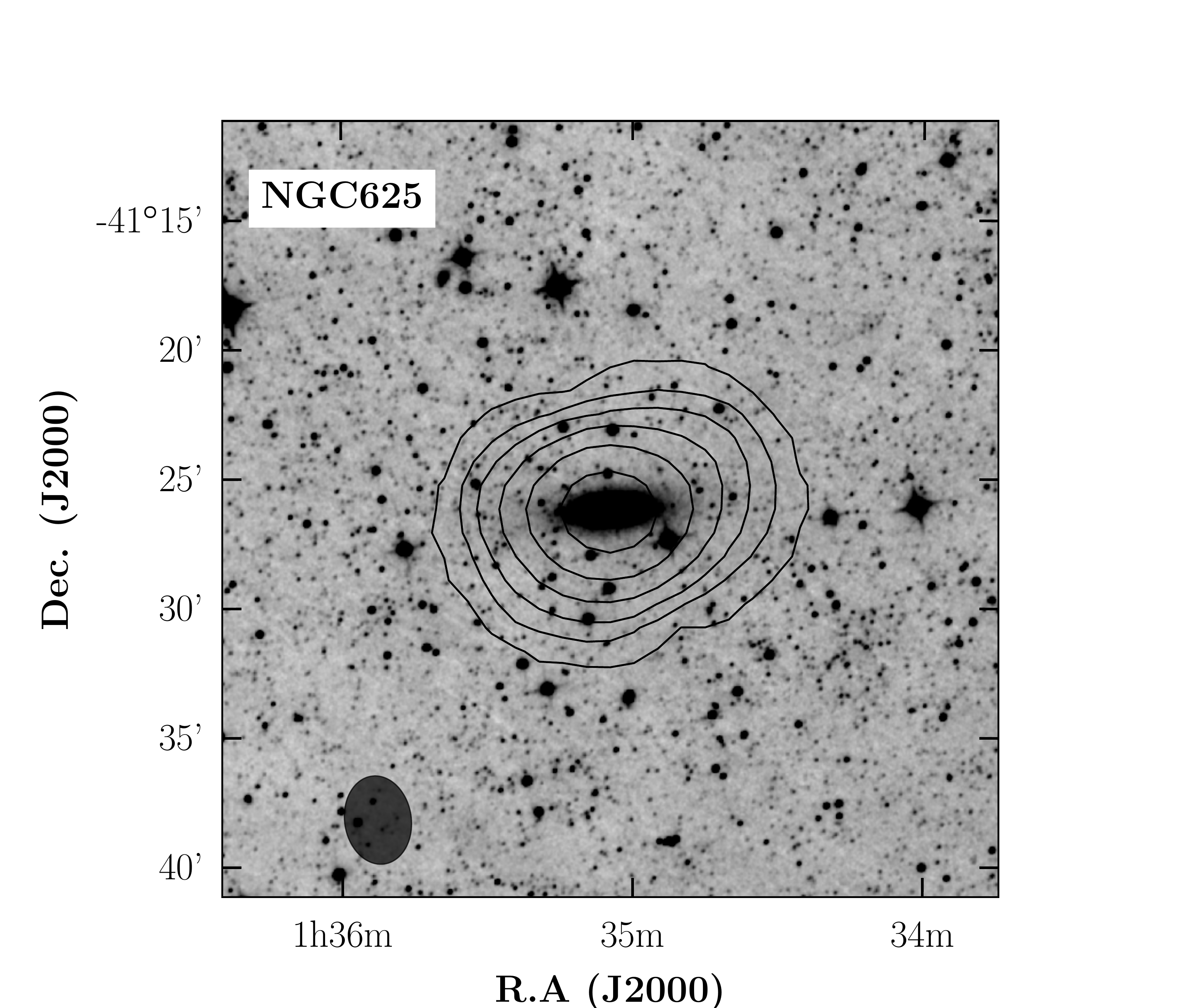}}\quad
\subfigure{\includegraphics[width=\columnwidth]{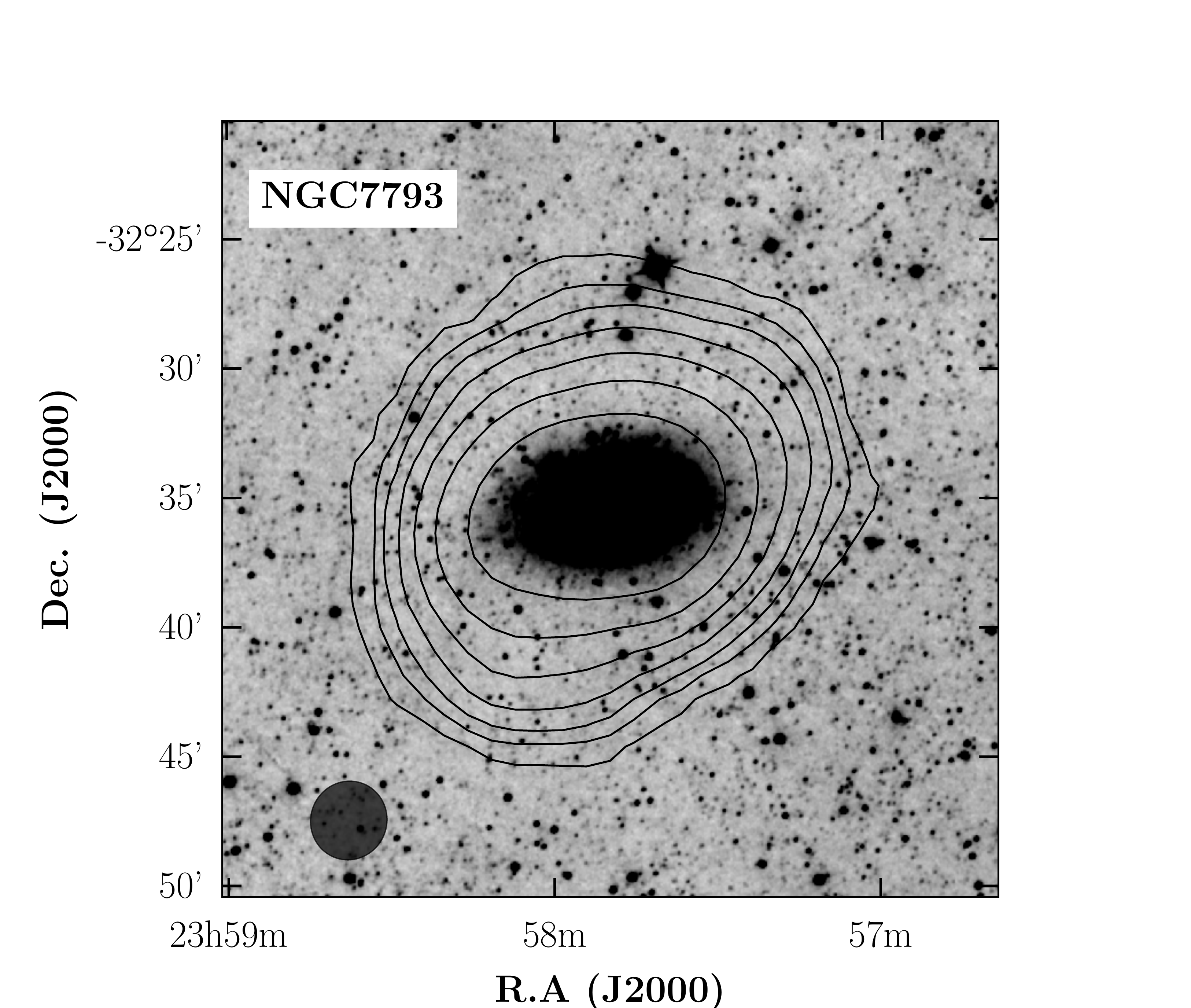}}
\caption{Column density maps of four of the KAT-7 galaxies overlaid on their {\it WISE} W1 images: ESO 300-G014 ({\it top left}), IC 4951 ({\it top right}), NGC 625 ({\it bottom left}) and NGC 7793 ({\it bottom right}). The contour levels are 0.5, 1.0, 2.0, 4.0, \dots $\times10^{19}\rm\,cm^{-2}$. The KAT-7 synthesised beam is shown as an ellipse at the bottom left of each panel.}\label{fig:coldens}
\end{figure*}

\begin{figure*}
\centering
\includegraphics[width=\columnwidth]{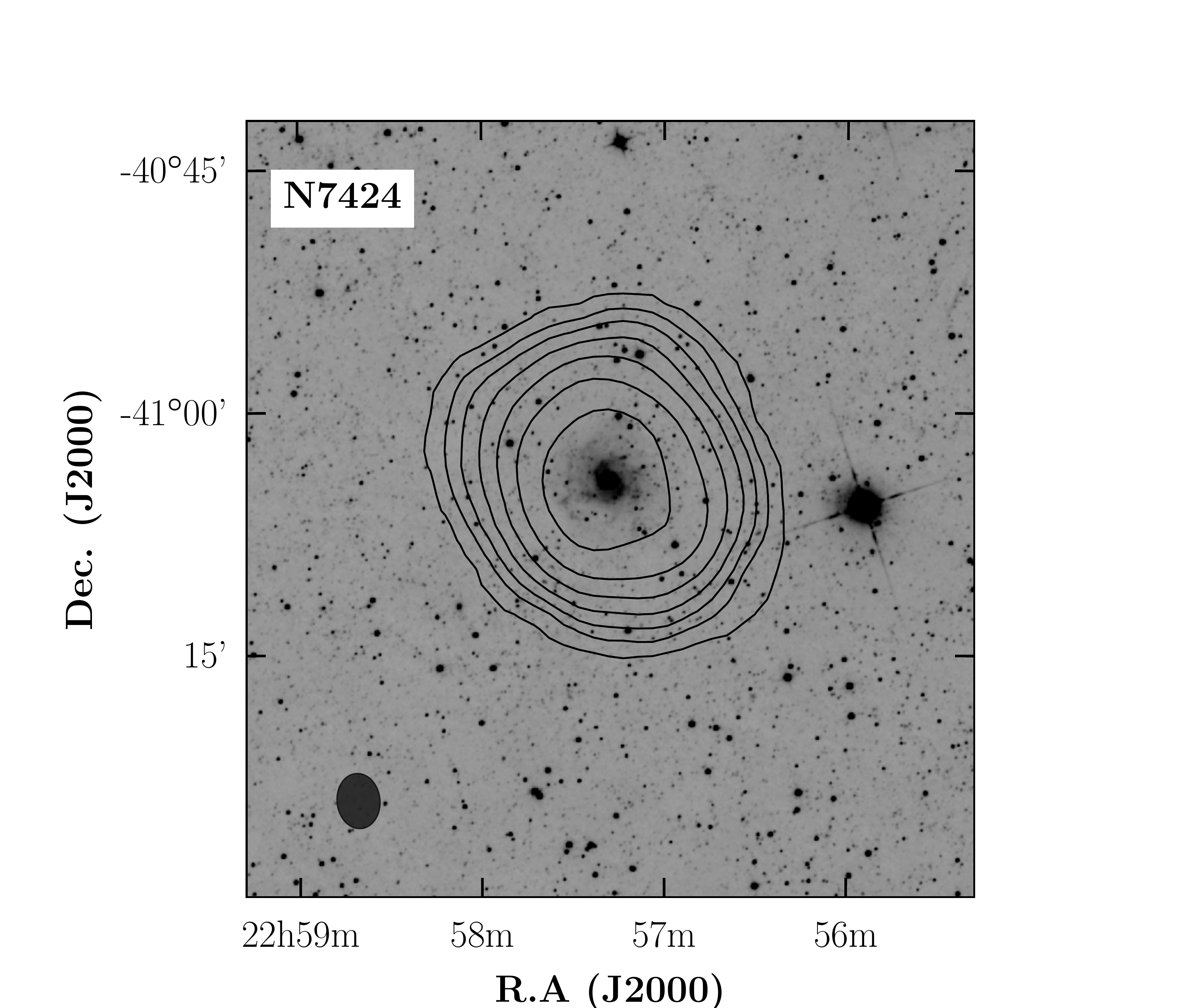}
\includegraphics[width=\columnwidth]{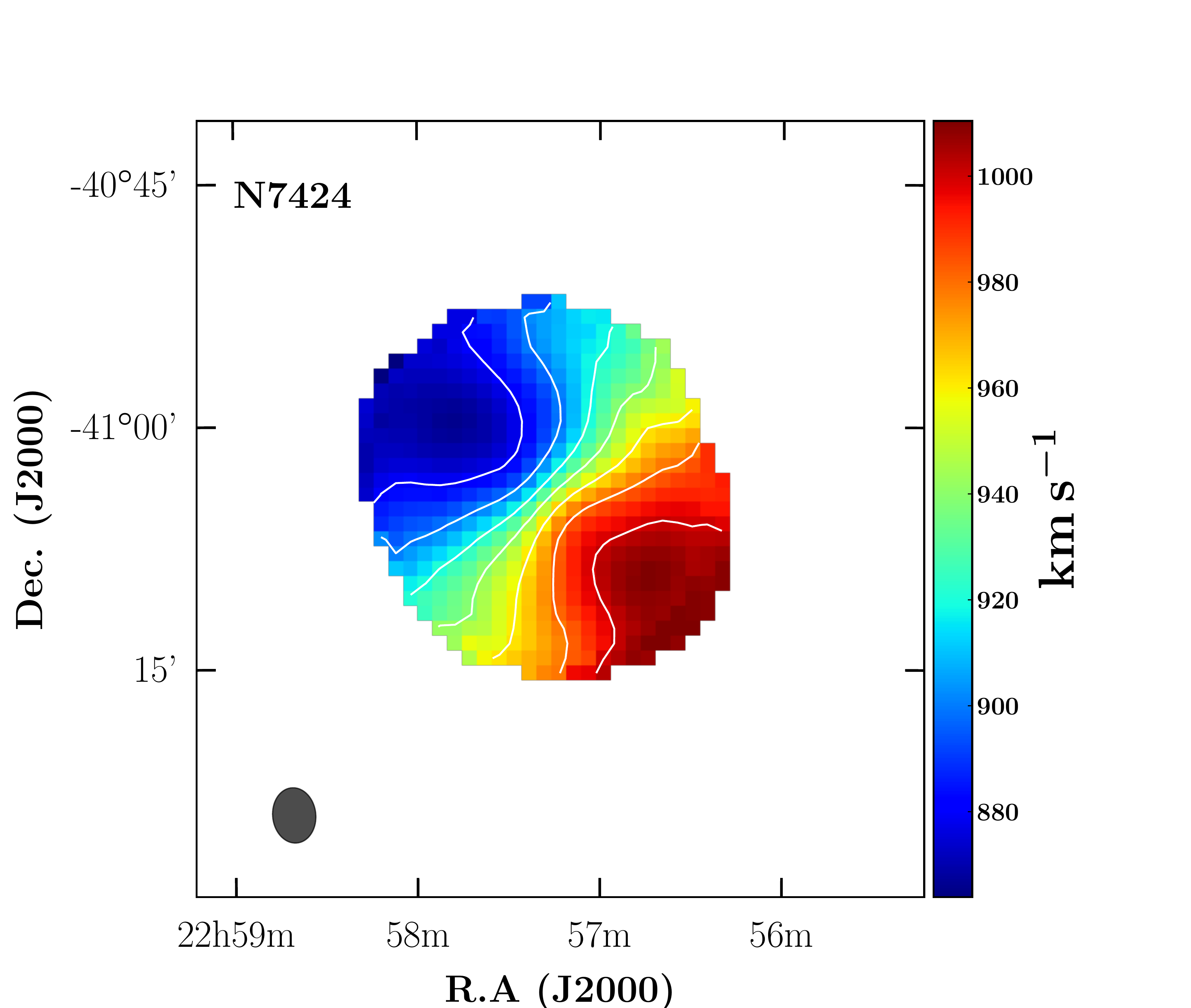}
\caption{KAT-7 \hi\, column density map ({\it left}), and velocity field ({\it right}) of NGC 7424. The contours of column density are 0.5, 1.0, 2.0, 4.0, \dots $\times10^{19}\rm\,cm^{-2}$, those of the velocity are 880, 900, 920, \dots, 1000 \kms.}\label{fig:n7424-maps}
\end{figure*}

\begin{figure*}
\includegraphics[width=\columnwidth]{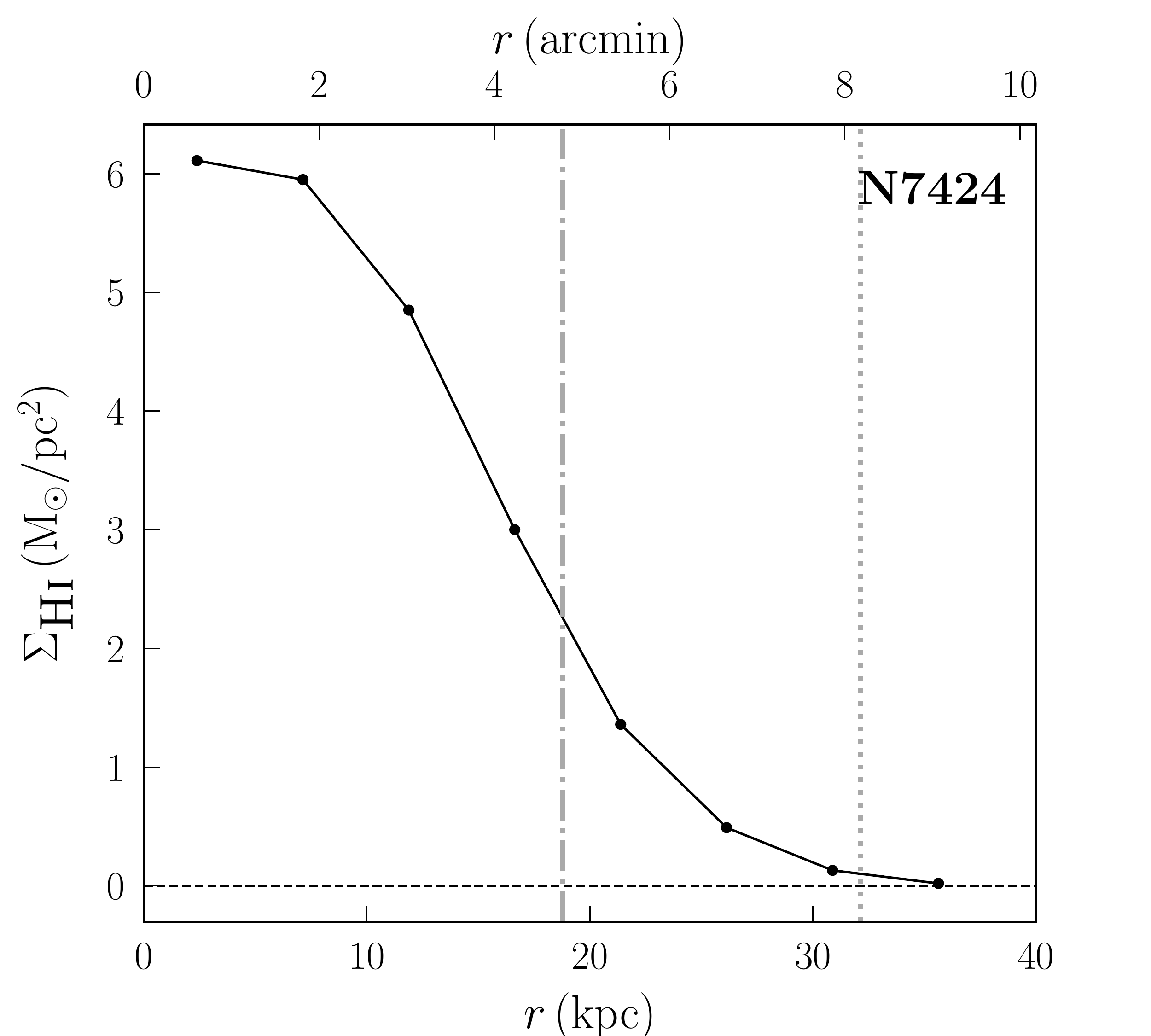}
\includegraphics[width=\columnwidth]{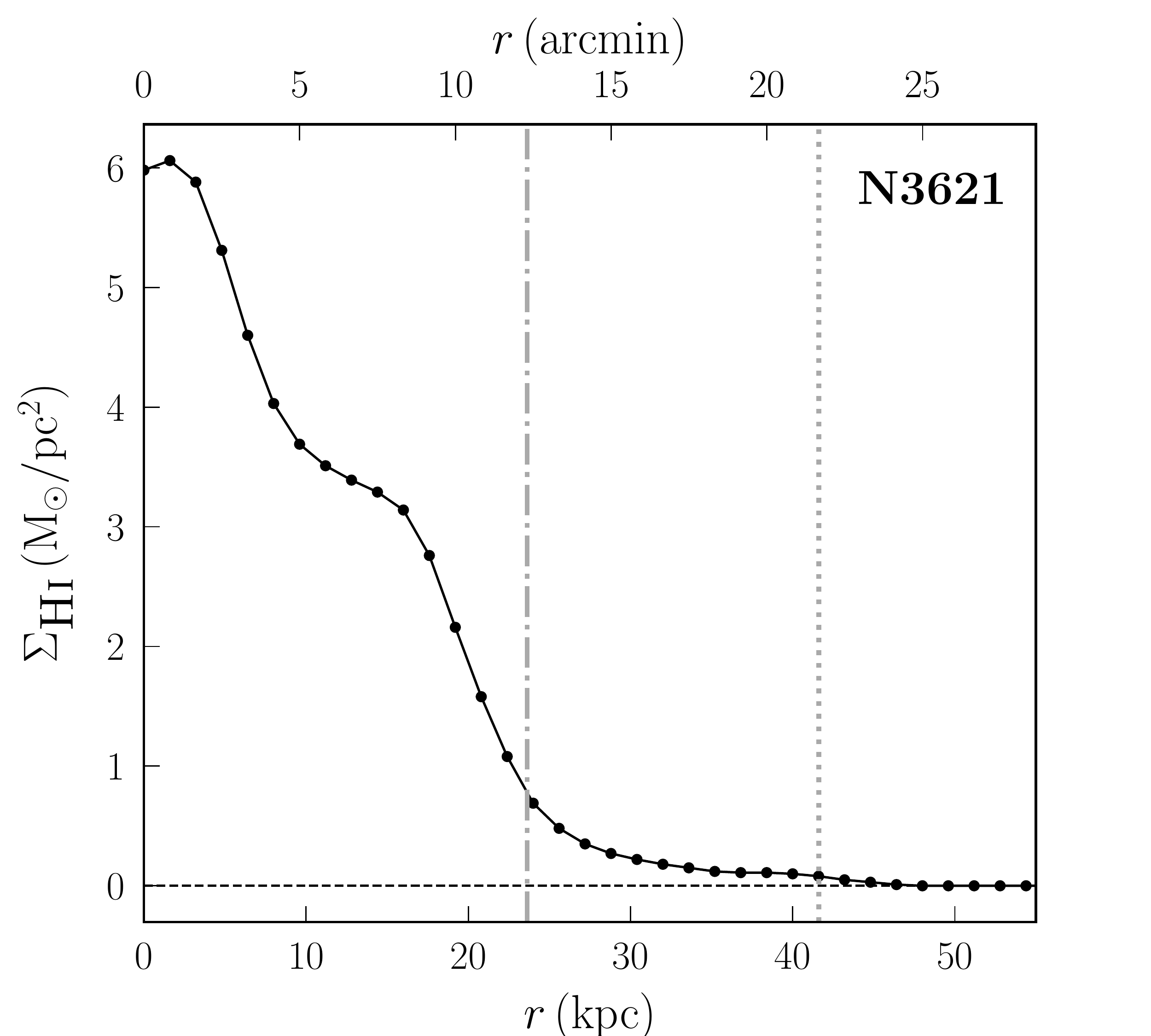}
\caption{Radial distribution of the \hi\, surface density of NGC 7424 ({\it left}) and NGC 3621 ({\it right}.). The {\it gray dashdot} and {\it dotted} lines show respectively the RC3 optical 25th magnitude and \hi\, ($10^{19}\,\cm$ column density level) radii. The optical and \hi\, diameters of NGC 7424 are listed in \Cref{tb:hi-props} and those of NGC 3621 are $d_{25}=24.6'$ and $d_{\HI}=43.3'$ respectively.}\label{fig:n3621-hidensity}
\end{figure*}

We present in \Cref{fig:coldens} the \hi\, column density maps of four of the five galaxies observed with the KAT-7, overlaid on optical DSS images. For the galaxy NGC 7424, for which we present a kinematical study in \Cref{sec:kinematics}, we show the column density map and velocity field in \Cref{fig:n7424-maps}. \Cref{fig:n3621-hidensity} (left panel) contains its radial \hi\, surface density profile, obtained with the \gipsy's task {\sc ellint}. The axial ratio of the ellipses used in {\sc ellint} were determined by the optical inclination of the galaxy (see \Cref{tb:hi-props}), and their orientation is determined from the galaxy's velocity field. In \Cref{tb:hi-props} we list the \hi\, fluxes and masses of all observed galaxies, as derived from the observations. We also list, for comparison, the HIPASS \hi\, fluxes of the galaxies as reported in \citet{Koribalski2004}, \citet{Doyle2005} and \citet{Kobulnicky2008}. The \hi\, mass, \hi\, diameters measured at the $10^{19}\rm\,cm^{-2}$ column density level and optical diameters measured at the 25th magnitude \citep[from the RC3 catalogue;][]{DeVaucouleurs1991a} are also presented. Several authors \citep[e.g.][]{Broeils1997,Staveley-Smith2003,Noordermeer2005,Wang2014,Koribalski2018} estimate the \hi\ diameter at a surface density level of $1\,\Mo\,pc^{-2}$ (corresponding to $1.25\e{20}\cm$), which is smaller than the diameter at $10^{19}\,\cm$. However, as is noted in \citet[][for early type galaxies]{Serra2012} and \citet{Wang2016}, a significant fraction of \hi\ mass is generally found outside the diameter defined at $1\,\Mo\,pc^{-2}$. This makes the diameter at $10^{19}\,\cm$ more representative of the full extent of the \hi\ disk, as is further seen in \Cref{fig:n3621-hidensity}.
For each of the galaxies, except UGCA 319, the recovered KAT-7 flux is comparable to that of the HIPASS; also, the \hi\, envelope is, as expected, more extended than their optical disk, as shown by their \hi\, to optical diameter ratio listed in column (10) \citep[see][]{Broeils1997}.
For UGCA 319, the measured flux is $\sim3$ times lower than the reported HIPASS flux, likely due to the close proximity of the galaxy to UGCA 320. Indeed, the angular separation between the two galaxies is $\sim0.31\dg$, barely larger (about 1.3 times) than the Parkes resolution and only $\sim2.1$ times the GBT resolution. One therefore expects the two sources to be confused in the HIPASS (and even GBT) observations, making it difficult to accurately measure their individual fluxes. This is further seen when comparing the total line flux of the two galaxies in both observations: we obtain $126.4\pm0.7$ Jy \kms\, in the GBT data, comparable to $121.6$ Jy \kms\, from the HIPASS data. Moreover, the Parkes flux of UGCA 320, as reported in \citet{Pisano2011}, is $107.3\pm0.3$ Jy \kms, about $\sim12\%$ lower than the value derived in the present work. The total \hi\ intensity maps of the GBT data are presented in \Cref{sec:gbtmaps}.

\subsection{NGC 3621}\label{sec:n3621props}
\begin{figure}
\includegraphics[width=\columnwidth]{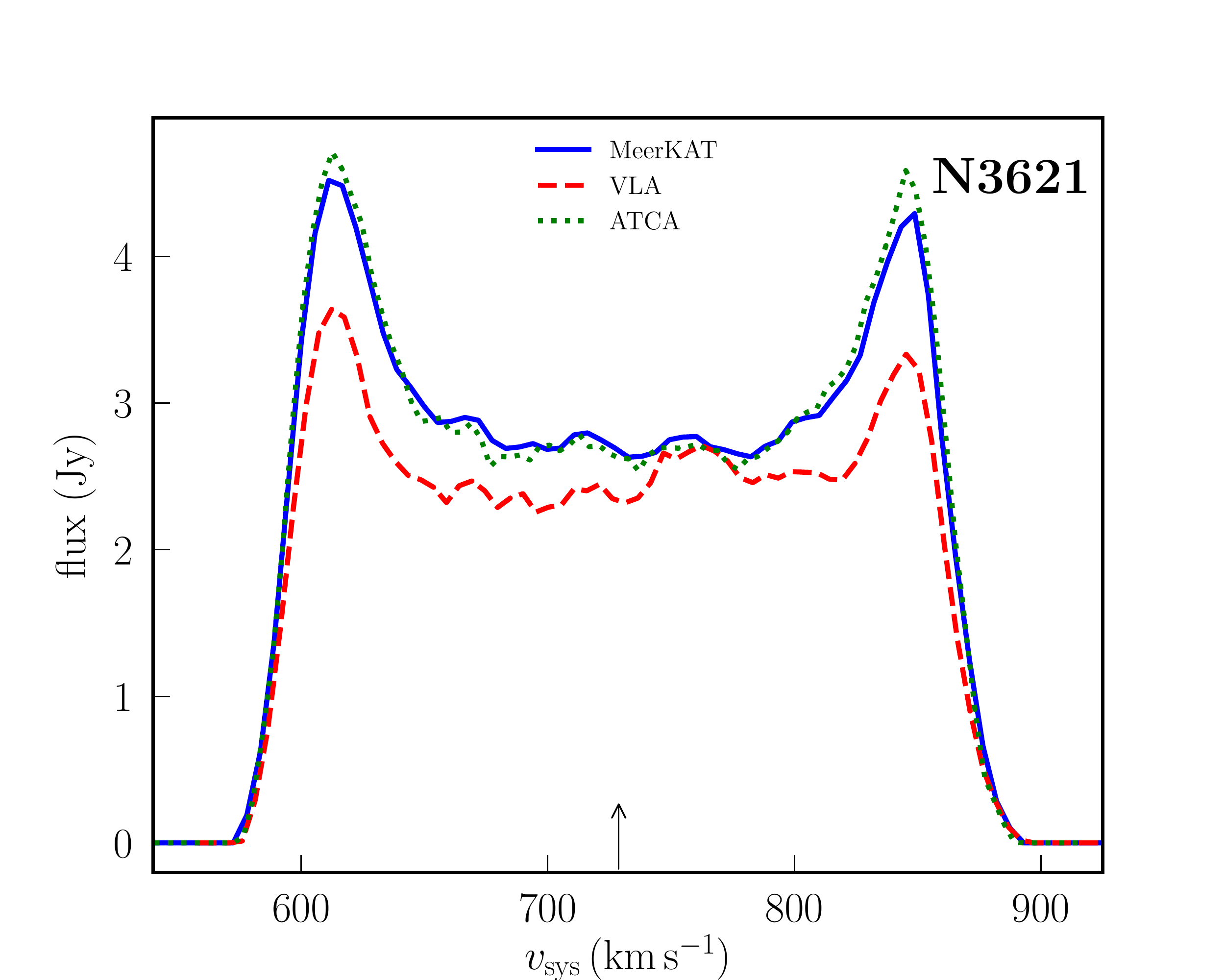}
\caption{Comparison of the global profiles of NGC 3621 as derived from the MeerKAT ({\it blue line}), VLA ({\it red dashed line}) and ATCA ({\it green dotted line}) data cubes. The systemic velocity of the galaxy is marked by an upward arrow.}\label{fig:specs-meerkat-things}
\end{figure}

\begin{figure*}
\centering
\includegraphics[width=0.35\textwidth]{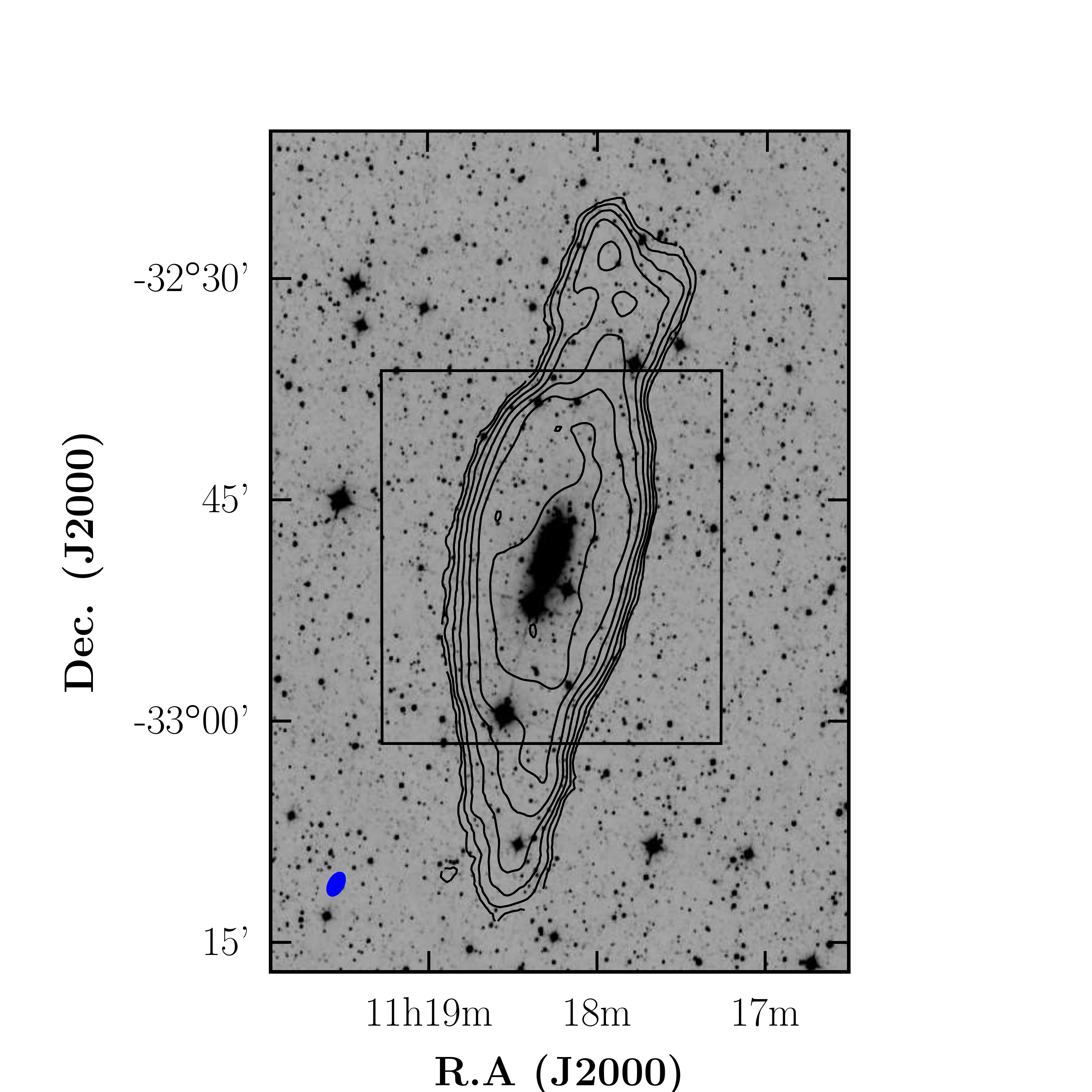}
\hspace{-25pt}
\includegraphics[width=0.35\textwidth]{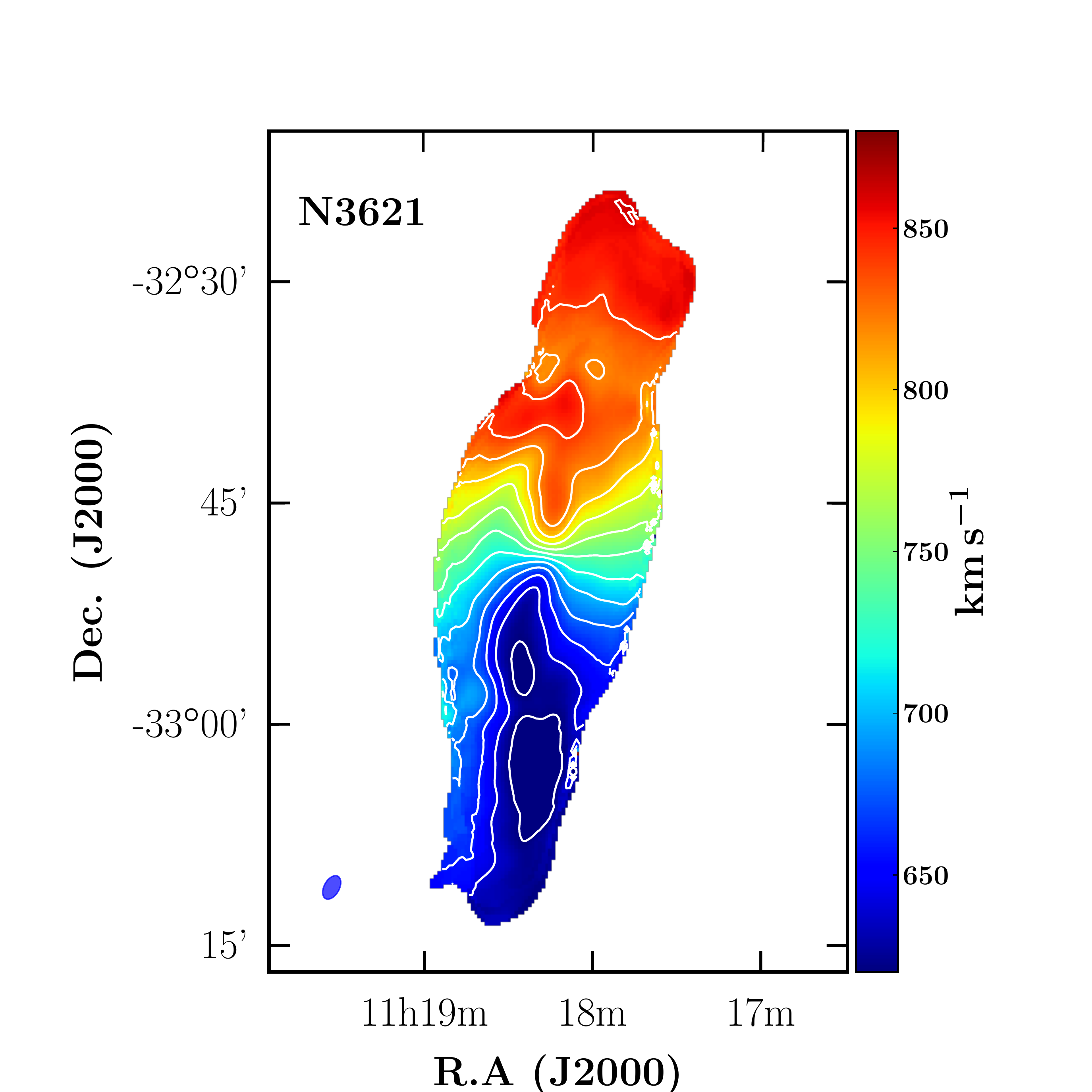}
\hspace{-15pt}
\includegraphics[width=0.35\textwidth]{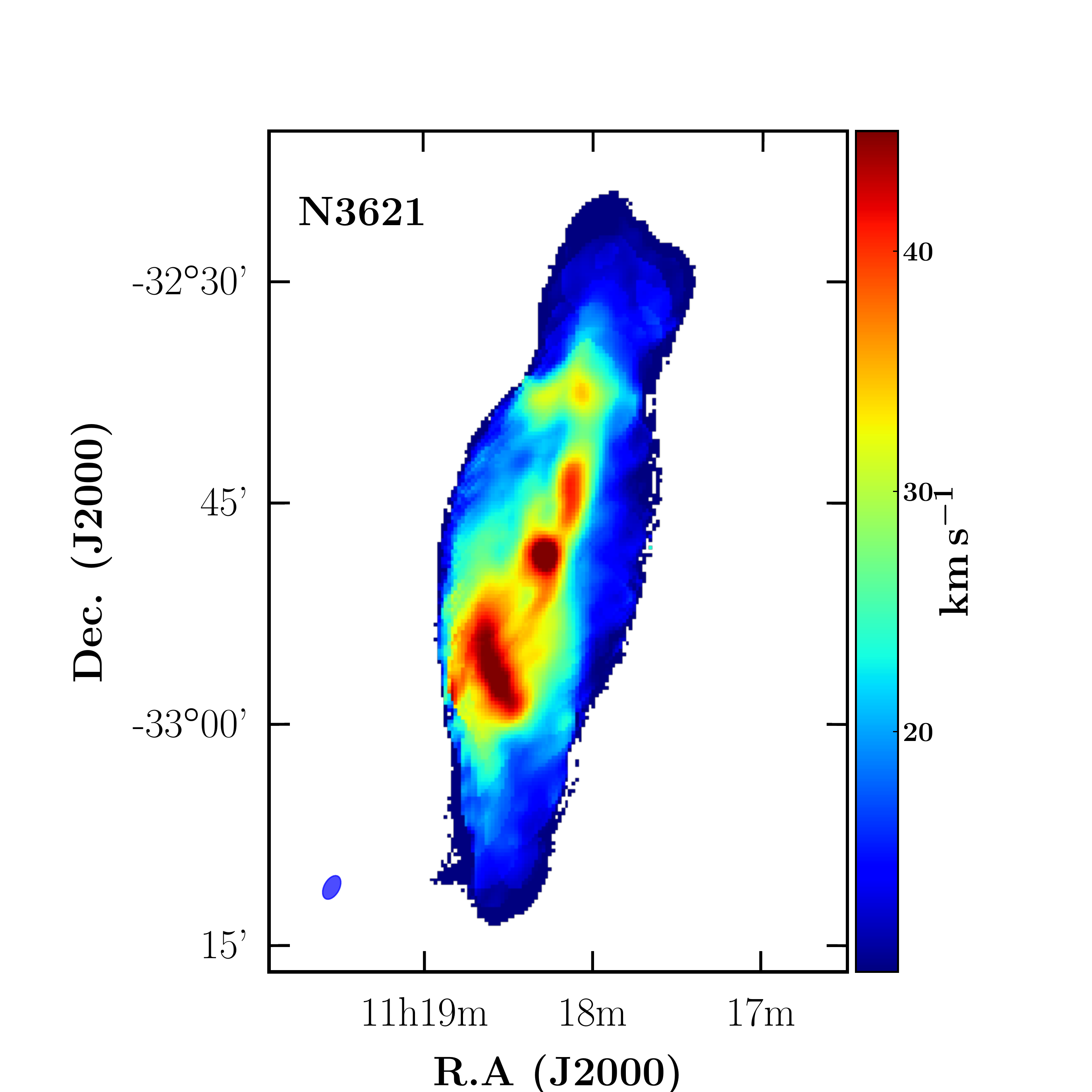}
\caption{MeerKAT maps of NGC 3621. {\it Left:} column density contours overlaid on the W1 grayscale map. The contours are $0.1, 0.2, 0.4,..., 12.8\e{20}\,\cm$, and the rectangle represents the approximate spatial size of the THINGS data. {\it Middle:} velocity field of the galaxy. Contours are $600, 620, 640,...,900$ \kms. {\it Right:} velocity dispersion map of the galaxy.}\label{fig:n3621-maps}
\end{figure*}

The \hi\, line flux of NGC 3621, as measured from the MeerKAT data, is $877\pm14$ Jy \kms. This is in agreement with the LVHIS \citep[Local Volume \hi\, Survey;][]{Koribalski2018} flux of 856.8 Jy \kms\, measured with the ATCA and $29\%$ higher than the THINGS measurement of 679 Jy \kms\, with the VLA. The difference with the VLA measurement is not surprising given the limited field of view and column density sensitivity of the THINGS observations. In effect, the total \hi\ intensity map presented in \citet{DeBlok2008} does not cover the full \hi\, extent of the galaxy, but only the inner $\sim25'$. On the other hand, although the field of view of the ATCA telescope is similar to that of the VLA, the LVHIS observations  of NGC 3621 consist of a mosaic of several fields to ensure that the full extent of the galaxy's \hi\ disk was covered.
In \Cref{fig:specs-meerkat-things} we present a comparison of the global profiles of the galaxy as obtained from the MeerKAT, THINGS and LVHIS observations, respectively. While the profiles obtained with MeerKAT and ATCA are comparable, it is obvious that VLA misses much of the galaxy's flux. This is consistent with the flux values derived for the three observations. The total \hi\ intensity map, velocity field and velocity dispersion maps are presented in \Cref{fig:n3621-maps}: a region of high velocity dispersion is seen South-East of the nucleus. The different morphologies of the northern and southern sides of the disk imply that the galaxy does not have a nicely rotating disk, but is warped in the line-of-sight \citep[this was also seen in the THINGS data,][]{DeBlok2008}, and there could be a cloud-like structure located on the receding side of the galaxy. We also present, in \Cref{fig:n3621-hidensity}, the radial distribution of its \hi\, surface density. As discussed in \Cref{sec:hidist}, the profile of the galaxy shows that the \hi\ diameter captures almost all the detected \hi, unlike the $1\,\Mo\,pc^{-2}$ isophote diameter which is approximately equal to the optical diameter of the galaxy.

\begin{figure*}
\centering
\includegraphics[width=\columnwidth]{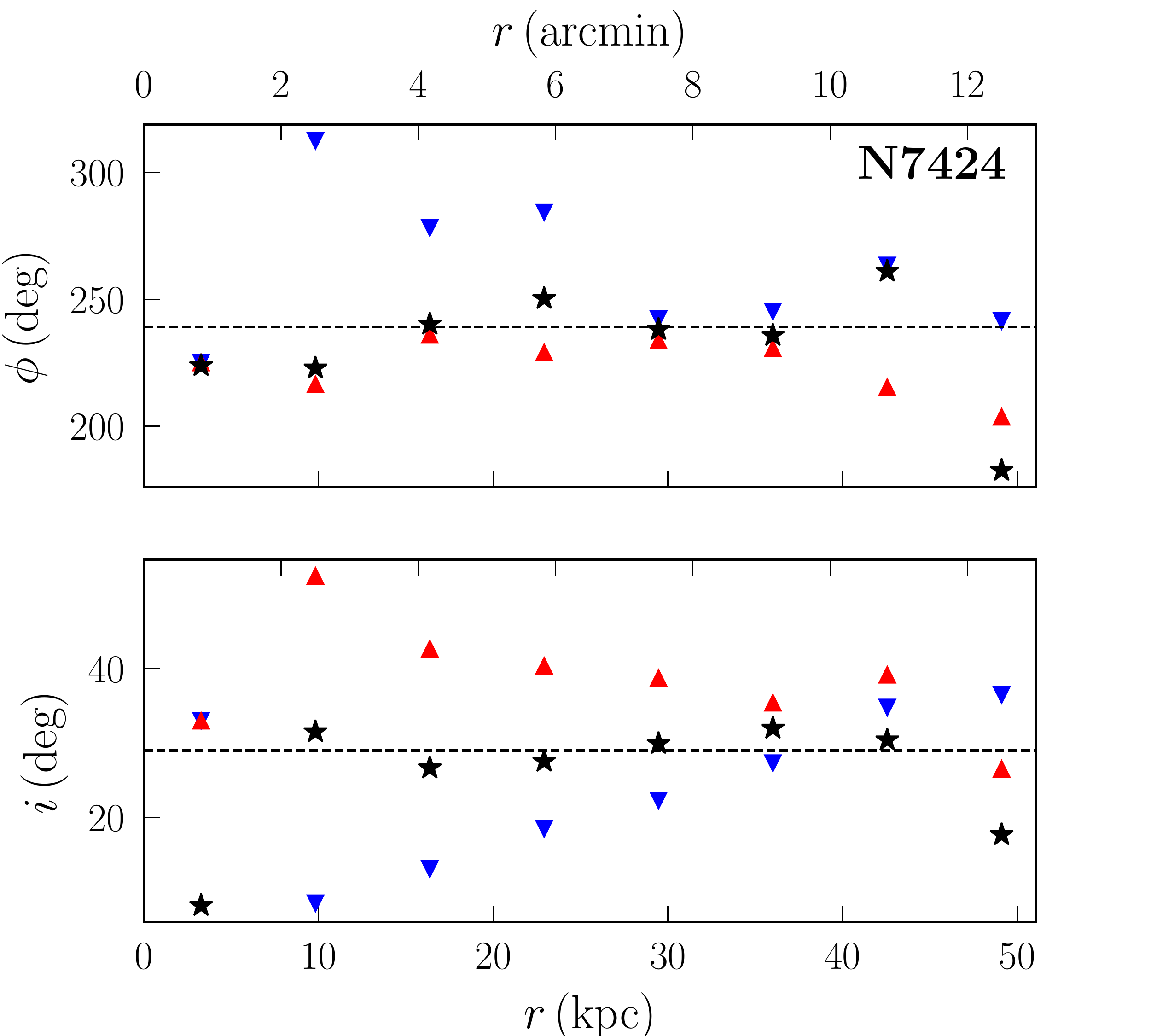}
\includegraphics[width=\columnwidth]{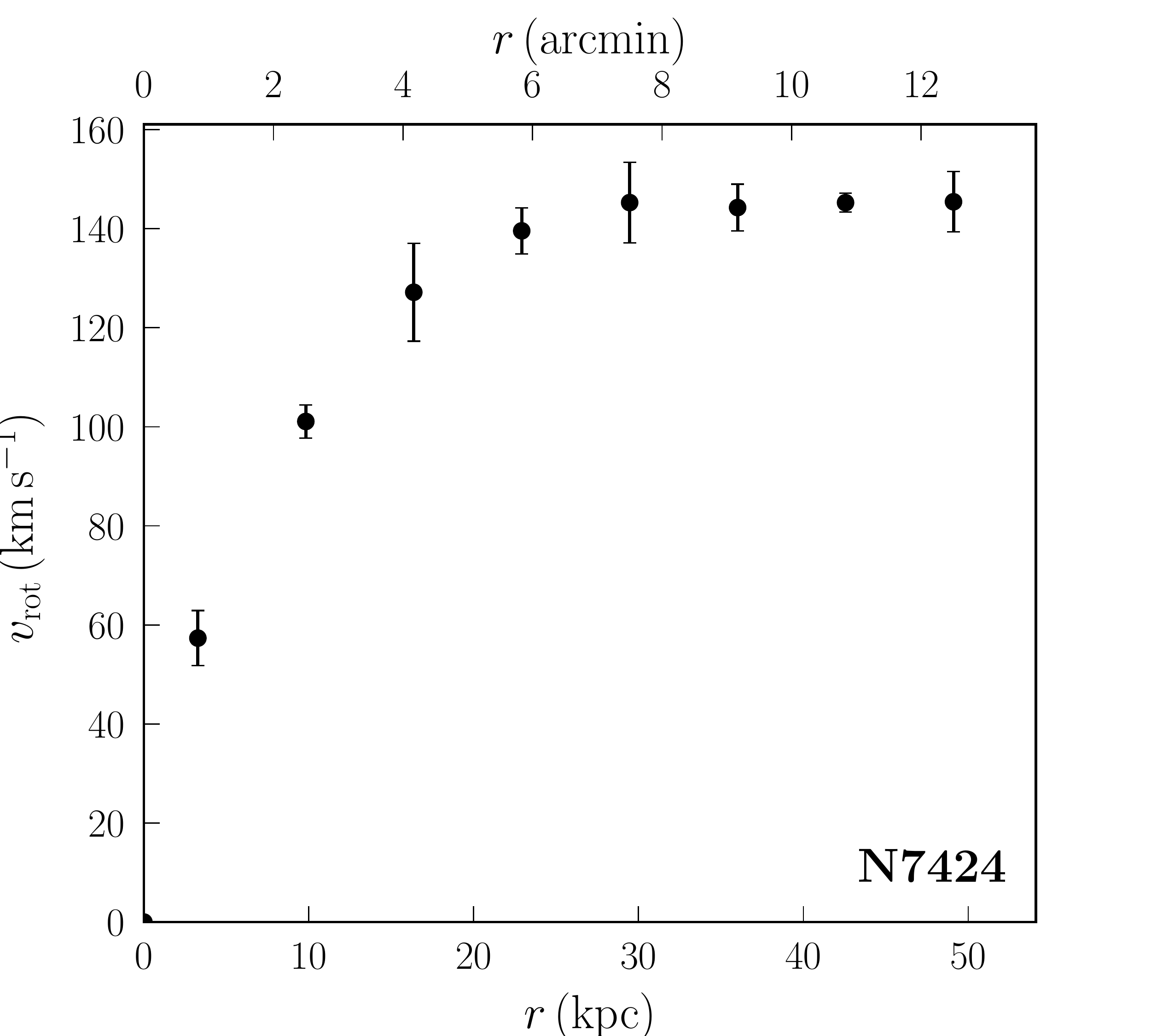}
\caption{The tilted-ring model ({\it left}) and rotation curve ({\it right}) of NGC 7424. The blue and red triangles represent respectively the data points for the approaching and receding sides, while the black stars show the points for both sides combined. The {\it dashed} horizontal lines on the left panels indicate the adopted values for the position angle and inclination, respectively.}\label{fig:vrot-n7424}
\end{figure*}

\begin{figure*}
\centering
\includegraphics[width=\columnwidth]{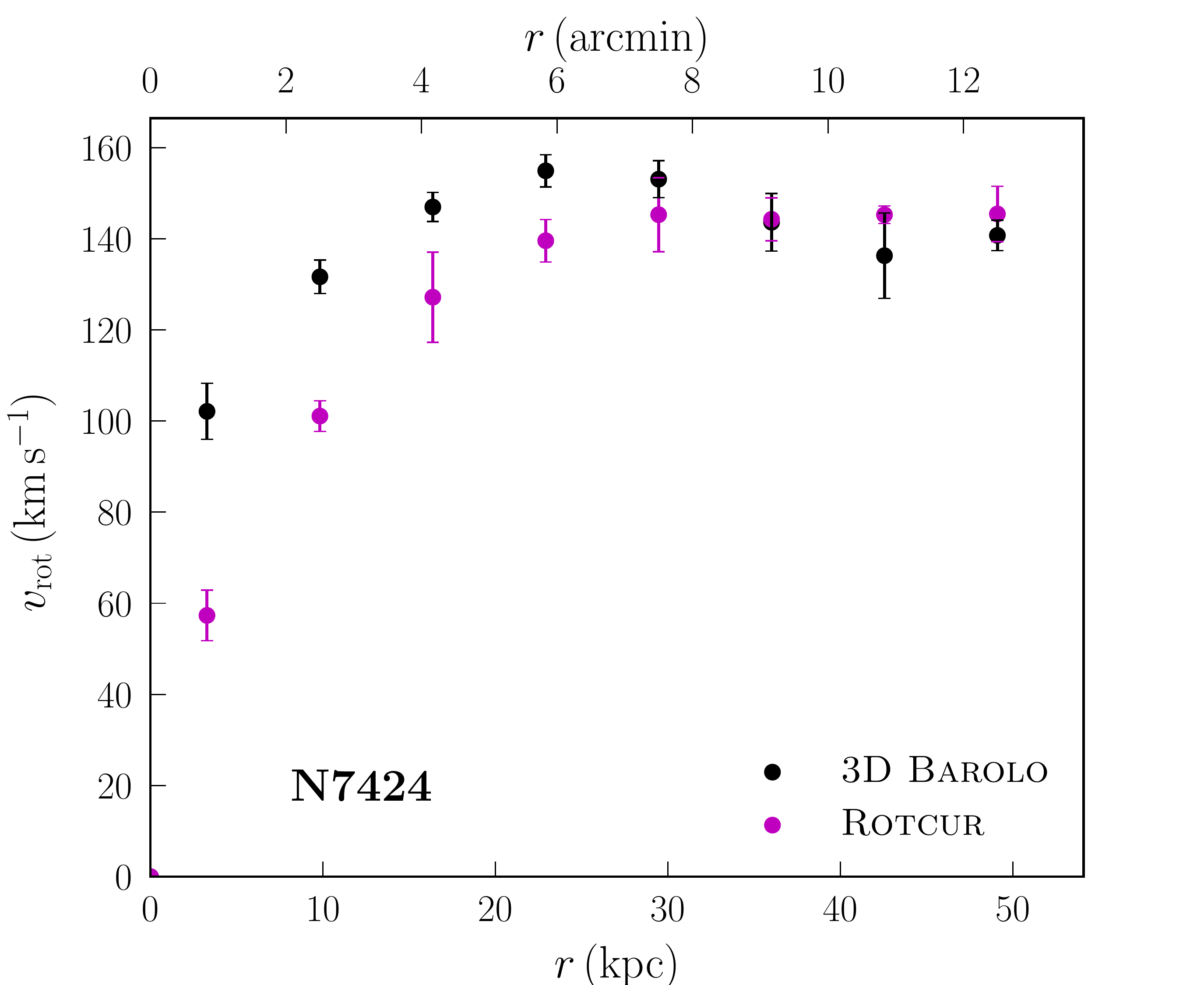}
\includegraphics[width=\columnwidth]{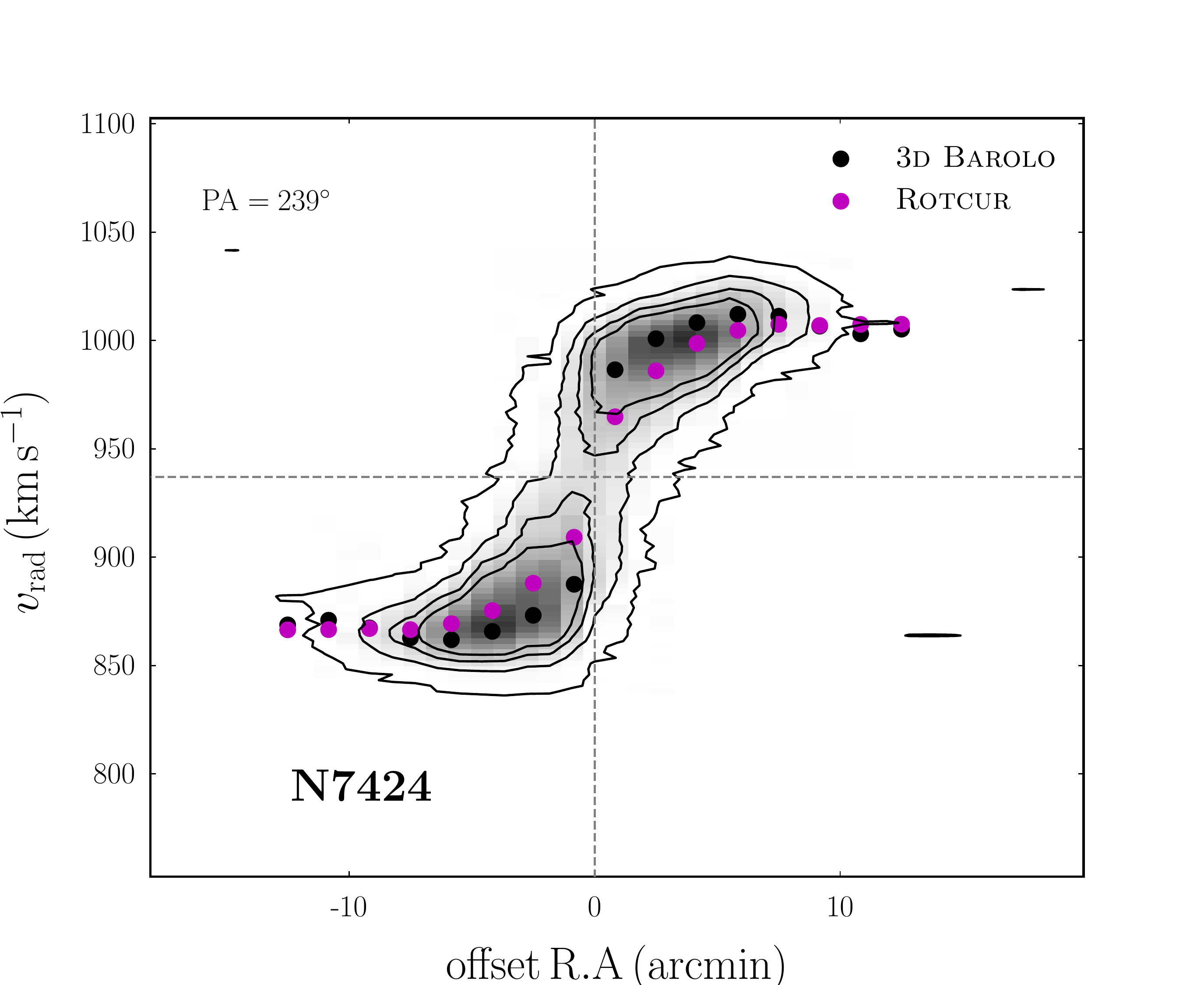}
\caption{Rotation curves of NGC 7424. {\it Left:} comparison of the {\sc Rotcur} and {\sc 3d Barolo} rotation curves. {\it Right:} overlay of the derived rotation curves on the PV diagram of the galaxy taken along its major axis.}\label{fig:n7424-barolo}
\end{figure*}

\begin{figure}
\centering
\includegraphics[width=\columnwidth]{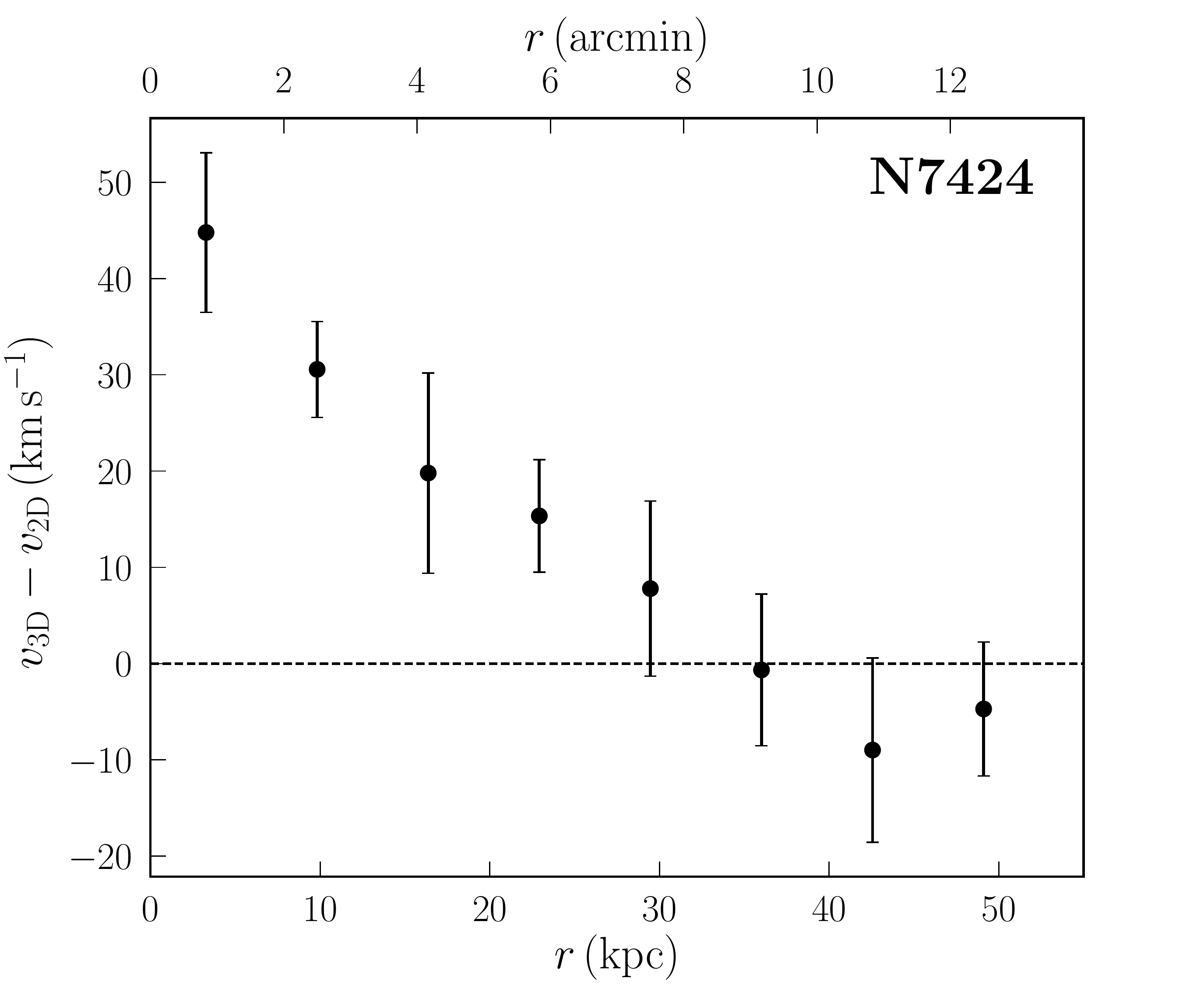}
\caption{Difference between the rotation curves of NGC 7424 obtained using {\sc rotcur} ($v_{\rm 2D}$) and {\sc 3d Barolo} ($v_{\rm 3D}$). The difference is highest at the centre and decreases to around zero in the outer regions.}\label{fig:veldiff}
\end{figure}

\subsection{Search for \hi\, clouds}
The MHONGOOSE galaxies were selected to exclude strongly interacting galaxies, and dense group and cluster environments, to minimise the gas contribution of companion galaxies. This way, the contribution of the cold gas accretion should be more apparent. 

We searched, in the whole observed sample, for \hi\, clouds associated with the galaxies that may be signs of gas accretion. However, none of the candidates present such structures down to a $3\sigma$ level (for any emission that fills the beam) of $\sim2.2\e{18}\,\cm$ over a 20 \kms\, line width, with a beam size of $9.1'$. However, the particular case of suspected cloud near the galaxy UGCA 250 caught our attention; a cloud-like structure was seen but could not be confirmed due to artefacts in which it was embedded. The structure was located south-west of the galaxy, in a region of the data that was affected by RFIs and where the noise was higher. We then re-observed the galaxy at night (to avoid solar interferences) and the analysis of the new data showed that the structure is in fact an artefact, as we did not detect anything above $3\sigma$ at its position (see \Cref{app:ugca250}).

\begin{table*}
	\center
	\begin{tabular}{l c c c c c c c c c c}
	\hline
	\hline
	Galaxy & $v_{\rm sys}$ & $W_{50}$ & $i$ & $S_\HI$ & $S_{\textsc{hipass}}$ & $M_\HI$ & $d_\HI$ & $d_{25}$ & $d_\HI/d_{25}$ & Tel.\\
 & (\kms) & (\kms) & (deg) & (Jy \kms) & (Jy \kms) & ($10^8\,\Mo$) & (arcmin) & (arcmin) &  & (K/G)\\
 (1) & (2) & (3) & (4) & (5) & (6) & (7) & (8) & (9) & (10) & (11)\\
	\hline
	ESO 300-G014\dotfill & 955.0 & 127.7 & 61.2 & $25.8\pm0.6$ & 25.6 & $10.1\pm0.2$ & 12.10 & 5.01 & 2.41 & K\\
	            \dotfill & 954.5 & 126.9 & \dots & $25.9\pm0.5$ & \dots & $10.1\pm0.2$ & 15.75& \dots & 3.14 & G\\
	ESO 300-G016\dotfill & 710.1 & 30.2 & 35.6$^a$ & $2.9\pm0.2$ & 4.3 & $0.6\pm0.1$ & $<9.0$ & 0.71$^a$ & <12.68 & G\\ 
	ESO 302-G014\dotfill & 871.4 & 70.0 & 27.6 & $10.4\pm0.4$ & 11.1 & $3.3\pm0.1$ & 12.38 & 1.70 & 7.28 & G\\
	ESO 302-G009$^b$\dotfill & 989.8 & 146.0 & 78.9 & $11.8\pm0.6$ & 13.9 & $10.7\pm0.5$ & 11.80 & 2.34 & 5.04 & G\\
	ESO 357-G007\dotfill & 1113.8 &  123.2 & 72.0 & $8.9\pm0.3$ & 11.0 & $6.4\pm0.2$ & 12.25 & 2.34 & 5.23 & G\\
	IC 4951\dotfill  & 813.0 & 119.2 & 90.0 & $26.1\pm0.4$ & 24.6 & $7.8\pm0.1$ & 11.81 & 2.75 & 4.29 & K\\
	KK98-195\dotfill & 570.1 & 25.2 & 55.7$^a$ & $6.0\pm0.2$ & 5.6 & $0.4\pm0.0$ & 10.21 & 0.45$^a$ & 22.69 & G\\
	KKS2000-23\dotfill & 1039.0 & 78.0 & 90.0$^a$ & $11.1\pm0.4$ & 11.1 & $4.2\pm0.1$ & 15.05 & 3.47$^a$ & 4.34 & G\\
	NGC 0625\dotfill & 396.0 & 71.6 & 74.4 & $34.5\pm0.4$ & 31.2 & $1.3\pm0.0$ & 13.71 & 5.75 & 2.38 & K\\
	NGC 1371\dotfill & 1460.5 & 382.6 & 47.5 & $71.2\pm1.0$ & 69.9 & $69.5\pm1.0$ & 22.28 & 5.62 & 3.96 & G\\
	NGC 1385$^b$\dotfill & 1493.6 & 182.4 & 55.6 & $27.4\pm0.6$ & 29.3 & $4.0\pm0.1$ & 15.75 & 3.39 & 4.64 & G\\
	NGC 1592\dotfill & 943.8 & 50.8 & 64.4 & $2.7\pm0.2$ & 5.9 & $1.1\pm0.1$ & $<9.0$ & 1.41 & <6.38 & G\\ 
	NGC 3511\dotfill & 1117.5 & 257.9 & 72.6 & $69.7\pm1.0$ & 87.5 & $33.1\pm0.5$ & 19.25 & 5.75 & 3.35 & G\\
	NGC 3513$^{c}$\dotfill & 1194.5 & 92.1 & 38.3 & $7.1\pm0.3$ & 29.3 & $1.0\pm0.1$ & $<9.0$ & 2.82 & <3.19 & G\\ 
	NGC 5068\dotfill & 667.0 & 69.3 & 30.1 & $177.0\pm0.7$ & 129.4 & $19.8\pm0.1$ & 24.88 & 7.24 & 3.44 & G\\
	NGC 5170\dotfill & 1497.6 & 508.9 & 90.0$^a$ & $96.9\pm1.2$ & 74.1 & $222.2\pm2.7$ & 23.55 & 8.32 & 2.83 & G\\
	NGC 5253$^d$\dotfill & 406.6 & 62.6 & 70.1 & $49.2\pm0.6$ & 43.9& $1.5\pm0.0$ & 20.48 & 5.01 & 4.09 & G\\
	NGC 7424\dotfill & 939.0 & 155.1 & 32.4 & $269.0\pm1.3$ & 250.2 & $116.5\pm0.6$ & 16.36 & 9.55 & 1.71 & K\\
	NGC 7793\dotfill & 227.0 & 176.0 & 48.8 & $274.2\pm1.2$ & 280.5 & $9.9\pm0.1$ & 20.85 & 9.33 & 2.23 & K\\
	UGCA 015\dotfill & 295.5 & 31.9 & 67.4 & $3.3\pm0.2$ & 3.7 & $0.1\pm0.0$ & $<9.0$ & 1.70 & <5.29 & G\\ 
	NGC 0247$^b$\dotfill & 156.2 & 199.2 & 74.9 & $741.4\pm1.1$ & \dots & $19.0\pm0.0$ & 40.57 & 21.38 & 1.90 & G\\
	UGCA 250\dotfill & 1703.1 & 274.7 & 90.0$^a$ & $86.9\pm1.1$ & 70.9 & $121.8\pm1.5$ & 21.66 & 4.36 & 4.97 & G\\
	UGCA 307\dotfill & 822.8 & 72.3 & 62.0 & $26.9\pm0.4$ & 26.6 & $4.7\pm0.1$ & 17.50 & 2.00 & 8.75 & G\\
	UGCA 320\dotfill & 740.9 & 109.2 & 90.0$^a$ & $122.7\pm0.7$ & 93.9 & $17.3\pm0.1$ & 22.82 & 5.62 & 4.06 & G\\
	UGCA 319$^{bc}$\dotfill & 757.3 & 87.6 & 50.0 & $3.7\pm0.2$ & 27.7 & $3.4\pm0.2$ & $<9.0$ & 1.62 & <5.55 & G\\ 
	\hline
	\end{tabular}
	\caption{Properties of the galaxies in the KAT-7 and GBT samples. The parameters of NGC 3621, which is neither part of the MHONGOOSE sample nor a secondary source, are given in \Cref{sec:n3621props} instead. Column (1): Galaxy name; Column 2: Systemic velocity derived from the \hi\ global profile; Column (3): \hi\ profile width at 50\% of the peak intensity; Column (4): Optical inclination from the RC3 catalogue \citep{DeVaucouleurs1991a}; Column (5): Measured \hi\ integrated flux; Column (6): HIPASS flux \citep{Doyle2005}; Column (7): Derived \hi\ mass; Column (8): Measured \hi\ diameter. An upper limit of $9'$ was used as the \hi\, diameter of the GBT galaxies smaller than the GBT beam; Column (9): optical diameter at measured at the 25th magnitude from the RC3 catalogue; Column (10): \hi\ to optical diameter ratio; Column (11): Telescope used for observation: K=KAT-7, G=GBT.
	Notes: $^a$ Value from HyperLEDA \citep{Makarov2014}; $^b$ Source is secondary, not part of the MHONGOOSE sample; $^c$ HIPASS flux from \citet{Koribalski2004}; $^d$ HIPASS flux from \citet{Kobulnicky2008}.}\label{tb:hi-props}
\end{table*}


\section{\hi\, kinematics of NGC 7424 \& NGC 3621}\label{sec:kinematics}

\subsection{NGC 7424}
NGC 7424 is a late-type, moderately barred face-on galaxy hosting numerous ultraluminous X-ray sources \citep[e.g.][]{Soria2006} and the supernova 2001ig \citep{Chandra2002}. The \hi\, mass of the galaxy as measured from the KAT-7 data is $\log{(\Mhi/\rm M_\odot)} = 10.1$ and, using its {\it WISE} W1 photometry, we derived a stellar mass of $\log{(\Ms/\rm M_\odot)} = 9.4$ based on \citet{Cluver2014}. These values show that the galaxy is gas-dominated, in agreement with its local environment: it is a field galaxy residing in the vicinity of the IC 1459 group \citep[e.g.][]{Serra2015} where gas stripping mechanisms are not as pronounced as in, for example, dense galaxy clusters where galaxies present smaller \hi\, gas fractions \citep[see][]{Sullivan1981,Haynes1984,Solanes2001,Serra2012}. 


\subsubsection{Rotation curve}\label{sec:n7424-rotcur}
We proceeded to derive the rotation curve of NGC 7424 using the implementation of the tilted ring model in the \gipsy\, task {\sc rotcur} \citep{Begeman1989}. The method assumes that the gas component of the galaxy is in circular motion, and uses a set of concentric rings to describe its rotation. Each ring is characterised by a fixed value of the inclination $i$, the position angle $\phi$ (measured counter-clockwise from North to the receding major axis of the galaxy), the systemic velocity $v_{\rm sys}$ and the rotational velocity $v_{\rm rot}$. For a given ring, the line-of-sight velocity at any point of coordinates $(x,y)$ a distance $R$ from the centre of the galaxy can be written
\begin{equation}
v(x,y) = v_{\rm sys} + v_{\rm rot}\,\sin{i}\,\cos{\theta}
\end{equation}
where $\theta$ is the azimuthal angle in the plane of the galaxy, and can be expressed as
\begin{subequations}
\begin{eqnarray}
\cos{\theta} &=& (1/R)\left[-(x-x_0)\,\sin{\phi} + (y-y_0)\,\cos{\phi}\right],\\
\sin\theta &=& (1/R\cos{i})\left[-(x-x_0)\,\cos{\phi} - (y-y_0)\,\sin{\phi}\right].
\end{eqnarray}
\end{subequations}

We choose the size of the ring such that we have about two rings per beam width. Although this sets a constraint in the dependency of the points, it allows a better representation of the galaxy's velocity field. The first step of the procedure consisted of determining the rotation centre $(x_c,y_c)$ and the systemic velocity $v_{\rm sys}$ of the individual rings, while keeping $\phi$ and $i$ fixed to their optical values. Next, with $v_{\rm sys}$ and $(x_c,y_c)$ fixed, $i$ and $\phi$ were derived at various radii. This was done for both sides of the galaxy on one hand, and separately on its approaching and receding sides on the other hand. The variations of $\phi$ and $i$ as a function of the radius $r$ are shown in the left panel of \Cref{fig:vrot-n7424}.

The systemic velocity derived for NGC 7424 is $937\pm3$ \kms, where the error is statistical. This value is  consistent with the value of $939.0\pm1.5$ \kms\, derived from the global profile of the galaxy using the \miriad\, task {\sc mbspect}, where the error is taken to be the difference between the line centres at 20\% and 50\% respectively. Because of the low inclination of the galaxy, the variation in both the inclination and the position angle is high in certain regions, especially in the inner $2'$ and beyond $11'$ (\Cref{fig:vrot-n7424}). We then use the average values of $\phi = 239\pm8^\circ$ and $i = 29\pm3^\circ$ over the region $2'$ to $11'$ to derive the rotation curve of the galaxy. The resulting curve is given in the right panel of \Cref{fig:vrot-n7424} where, as in \citet{Carignan2013}, the error on $v_{\rm rot}$ is a combination of the dispersion $\sigma(v)$ in each ring and half the difference between the velocities of the approaching and the receding sides:
\begin{equation}
\Delta v_{\rm rot} = \sqrt{\sigma^2(v) + \left[{1\over2}(v_{\rm rec} - v_{\rm app})\right]^2}
\end{equation}

Because the obtained rotation curve is greatly affected by beam smearing -- especially in the inner regions of the galaxy -- we used the automatic {\sc 3d Barolo} package \citep{DiTeodoro2015} to re-derive the curve from the 3-dimensional data cube, instead of the 2-dimensional velocity field as is done with {\sc rotcur}. The previously obtained kinematical parameters were used as initial guess, and we adopted the two-stage fit mode of the code, which enables a second fitting stage to fit a Bezier function to the inclination and position angle. The averages of the derived kinematical parameters are $i=29.8\pm1.7\dg$ and $\phi=236.9\pm3.0\dg$ for the inclination and position angle respectively, while the new systemic velocity is now $936.2$ \kms. These values are consistent with the results of the 2D fitting with {\sc rotcur}. We compare in the left panel of \Cref{fig:n7424-barolo} the rotation curves obtained from the 2D and 3D fitting methods, respectively. The new curve is less sensitive to beam smearing affecting mostly the inner regions of the galaxy. This is further seen in the right panel of the figure where we overlay the two rotation curves on the position-velocity (PV) diagram of the galaxy. The {\sc 3d Barolo} rotation curve describes better the rotation of the galaxy, while the {\sc rotcur}-derived curve underestimates the velocities in the inner regions. We quantify that in \Cref{fig:veldiff} where we plot the difference between the two velocities of the different curves as a function of the radius. We see that the difference is largest (around 45 \kms) at the centre of the galaxy, and slowly decreases to about zero in the outskirts.

\subsection{NGC 3621}
Unlike NGC 7424, the galaxy NGC 3621 has been extensively observed in \hi\, \citep[e.g.,][]{Barnes2001,Koribalski2004,DeBlok2008,Koribalski2018} and appears as one of the brightest \hi\, sources in the southern hemisphere, at a distance of 6.6 Mpc \citep{Freedman2001}.

\subsubsection{Rotation curve}
\begin{figure}
\centering
\includegraphics[width=\columnwidth]{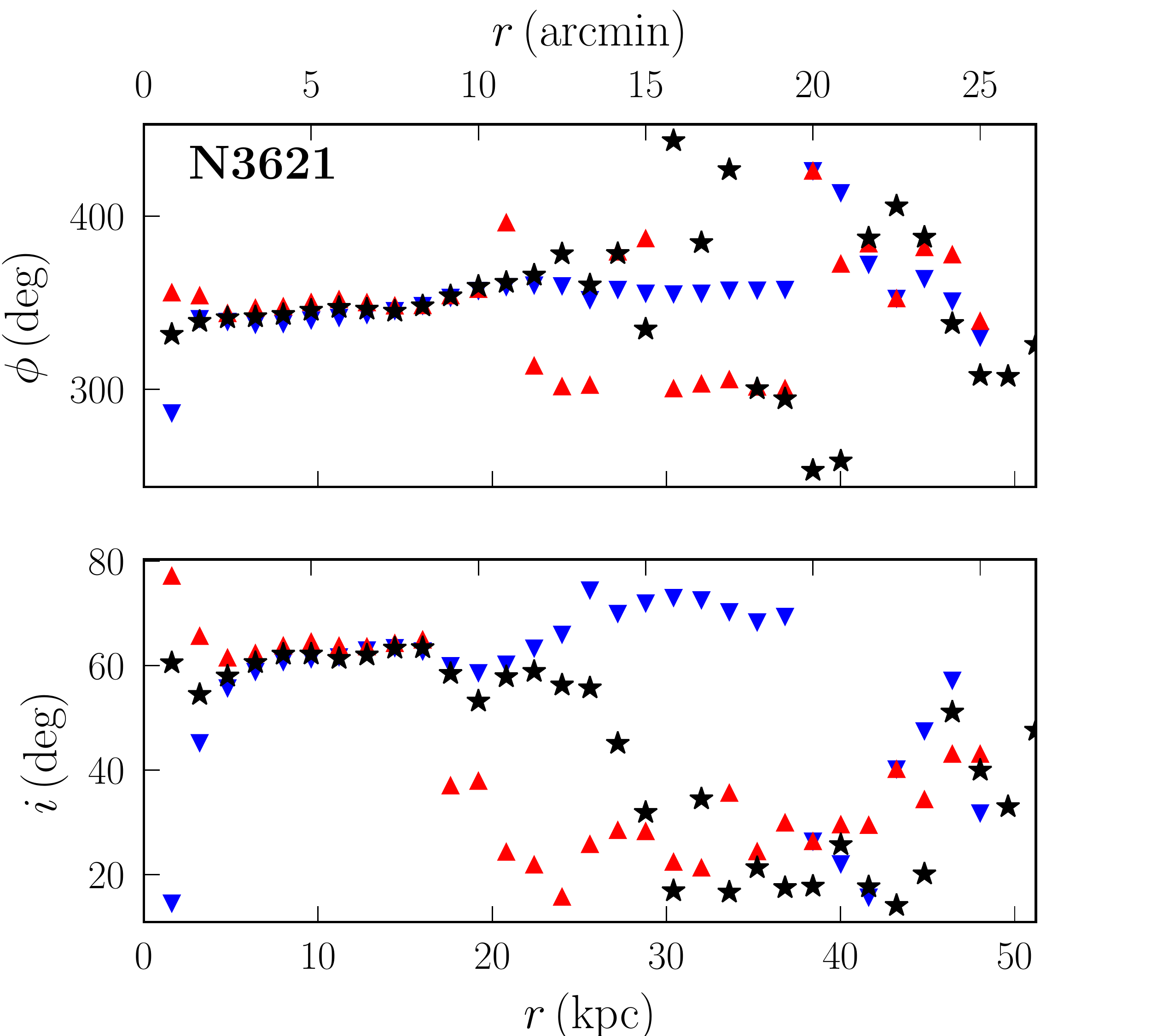}
\caption{Tilted-ring model of NGC 3621 from a {\sc Rotcur} fit of the MeerKAT data. Legends are same as in \Cref{fig:vrot-n7424}.}\label{fig:n3621-tiltedring}
\end{figure}

\begin{figure*}
\centering
\includegraphics[width=\columnwidth]{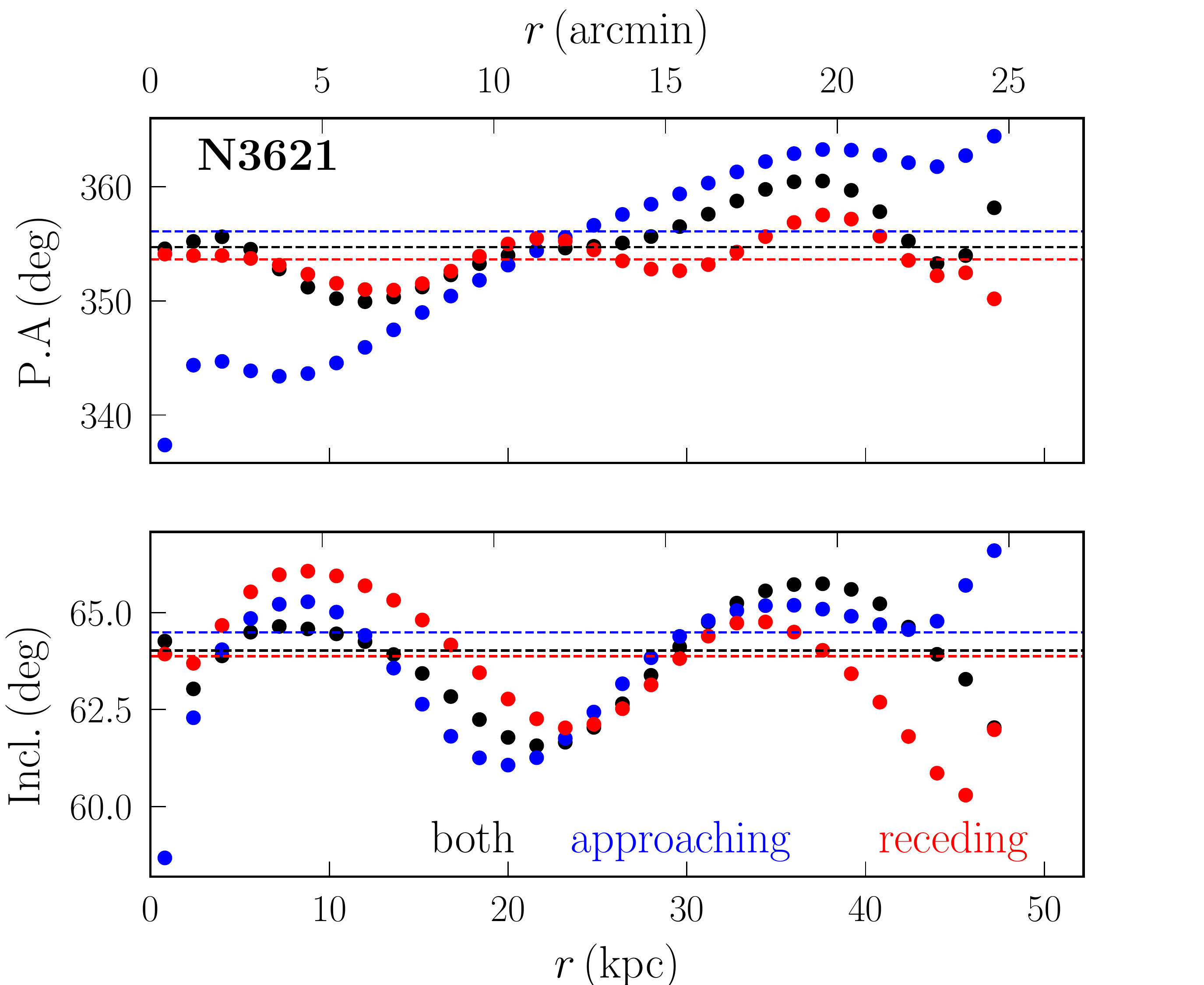}
\includegraphics[width=\columnwidth]{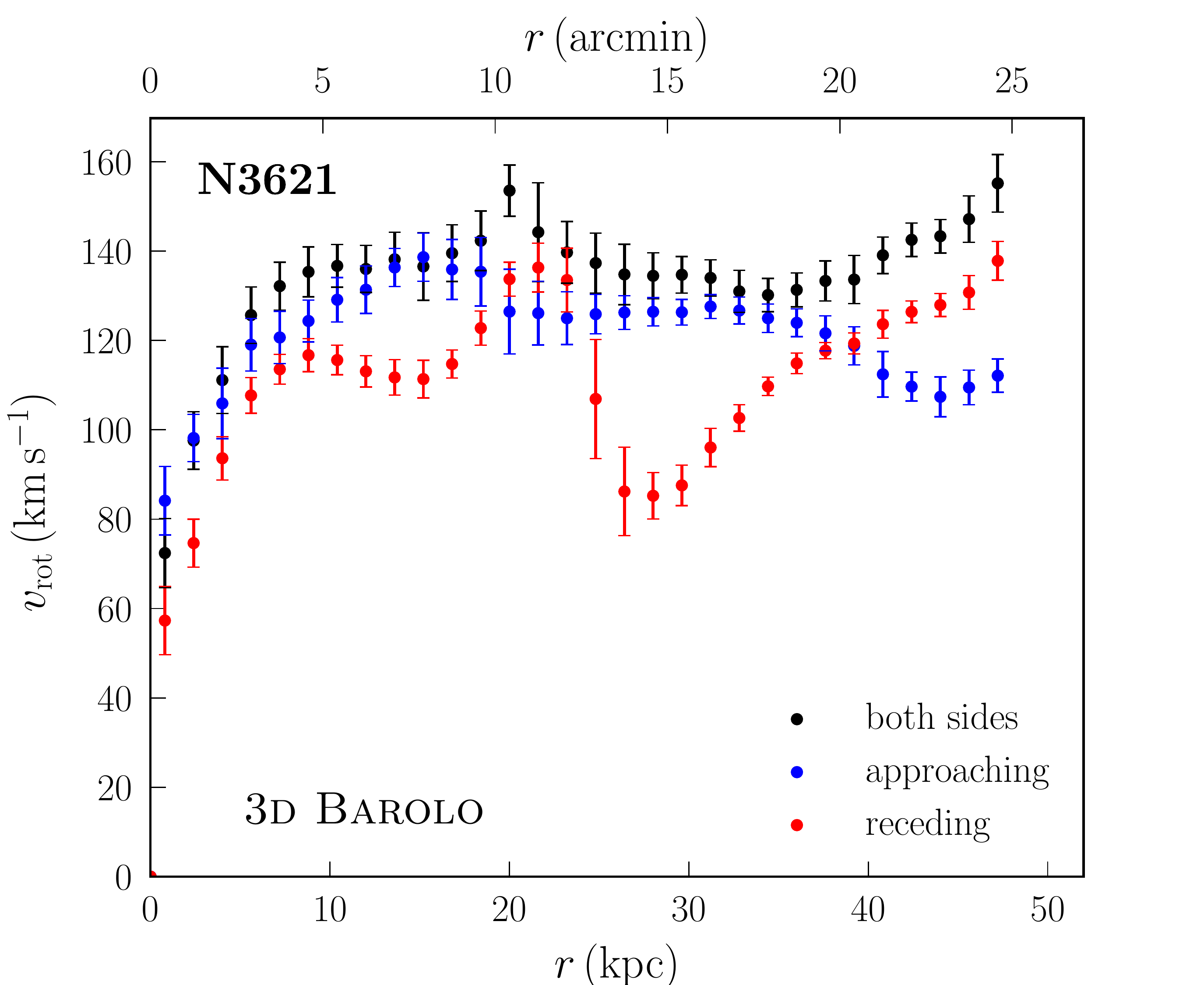}
\caption{The {\sc 3d Barolo} model of NGC 3621. {\it Left:} variations of the position angle and inclination. The average values of the parameters are shown by horizontal dotted lines. {\it Right:} the rotation curve for the approaching and receding sides, as well as both sides combined.}\label{fig:vrot-n3621}
\end{figure*}

\begin{figure*}
\centering
\includegraphics[width=\columnwidth]{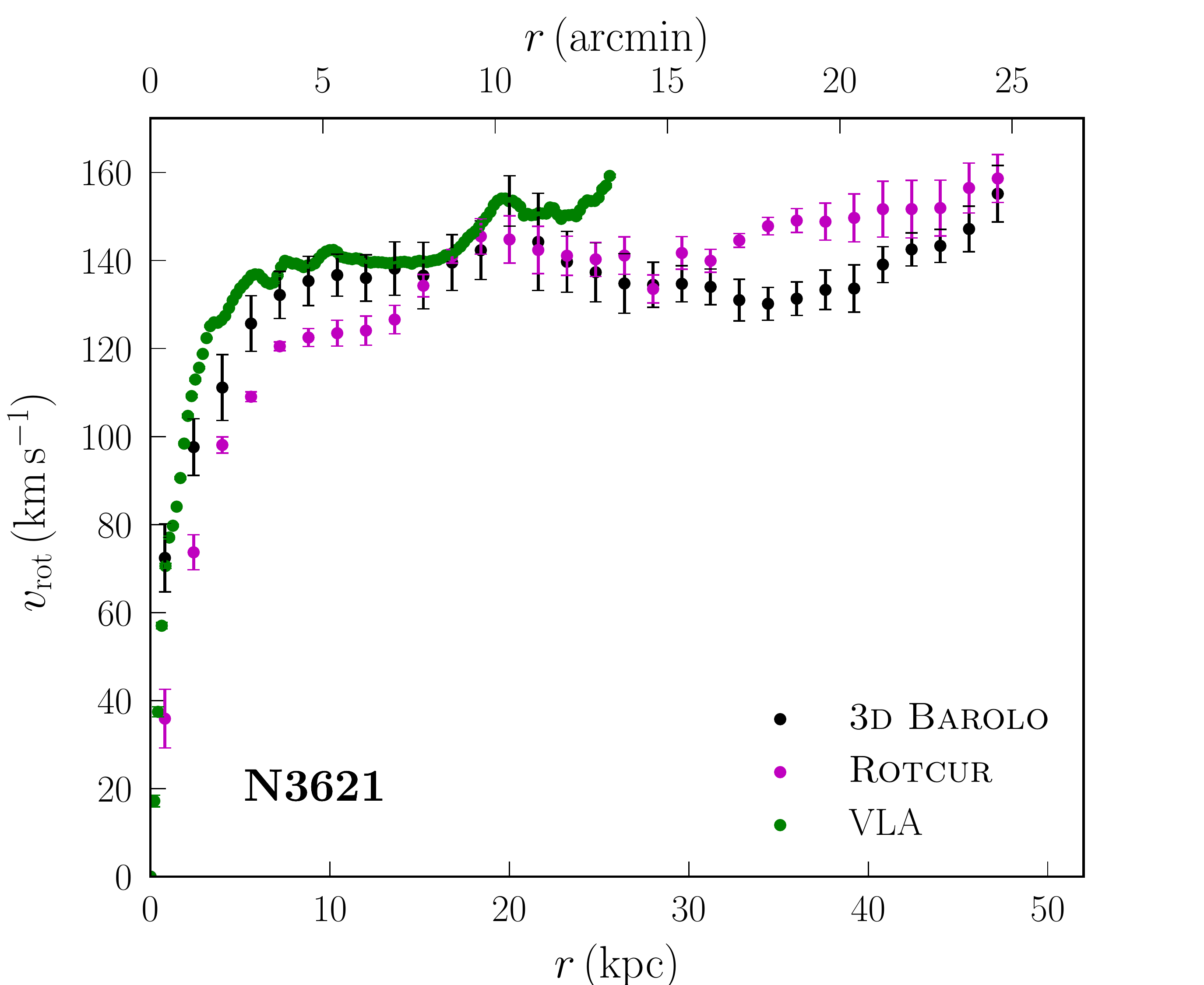}
\includegraphics[width=\columnwidth]{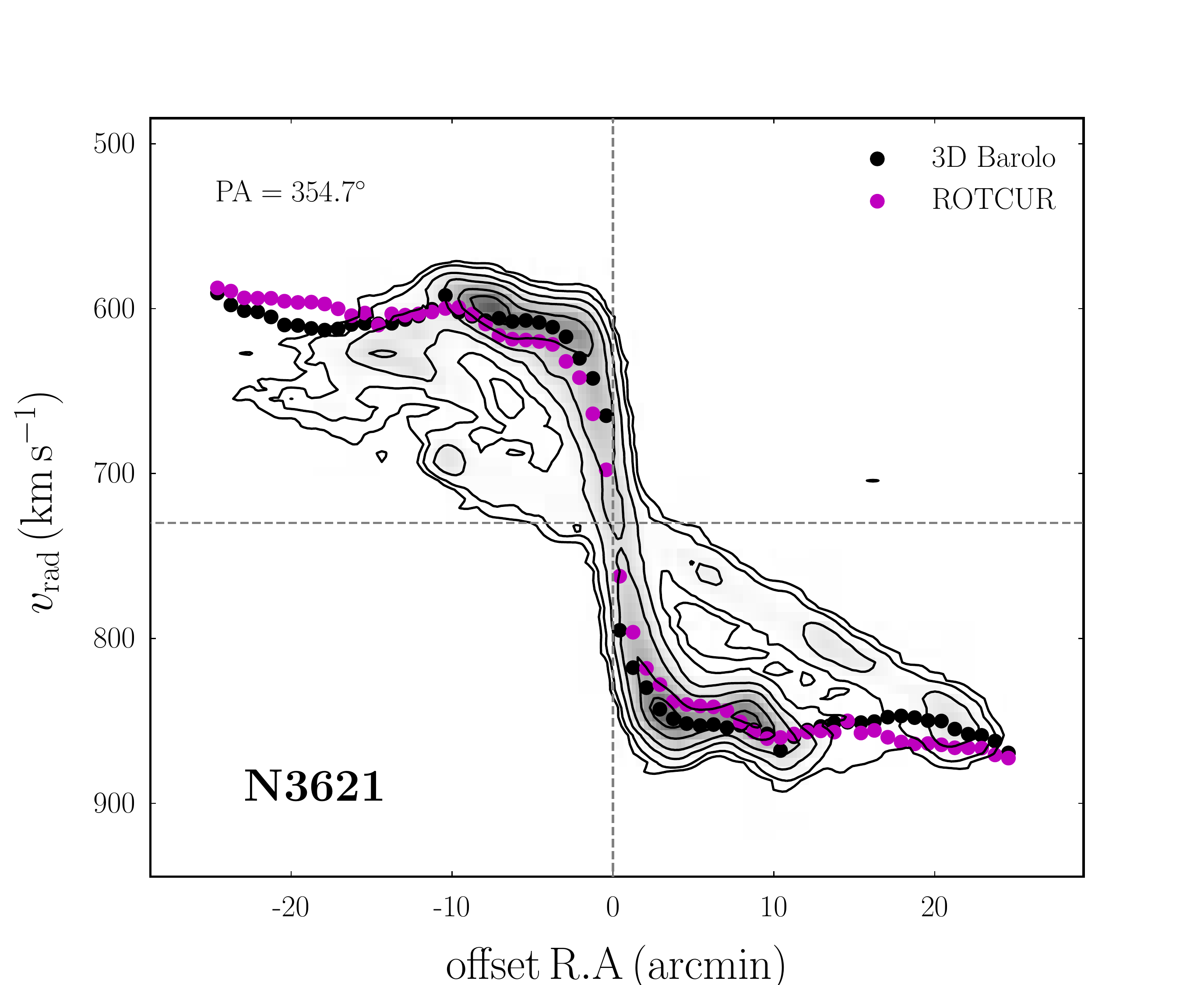}
\caption{Comparison of the derived rotation curves of NGC 3621. {\it Left:} the {\sc 3d Barolo} and {\sc Rotcur} rotation curves compared to the THINGS curve. {\it Right:} the {\sc 3d Barolo} and {\sc Rotcur} curves overlaid on the PV diagram of the galaxy.}
\label{fig:n3621-pv}
\end{figure*}

The process used to obtain the rotation curve of NGC 3621 is similar to that described in \Cref{sec:n7424-rotcur}. The systemic velocity obtained for the galaxy is $730.1\pm1.4$ \kms, consistent with the measurements of \citet[$730\pm2$ \kms]{Koribalski2004} and the value of $728.8\pm1.5$ \kms, derived from the global profile with the \miriad\, task {\sc mbspect}.
The {\sc Rotcur} fit of the galaxy revealed large scatters in the galaxy's inclination and position angle at radii larger than 19.2 kpc (\Cref{fig:n3621-tiltedring}), probably a result of the galaxy's complex kinematics. The average values of these parameters, derived in the inner regions of the galaxy, are $i=59.9\pm3.2\dg$ and $\phi=345.2\pm6.7\dg$. These values, together with the derived value of the systemic velocity, were used as initial guess for the {\sc 3d Barolo} fit. Because of the galaxy's asymmetry and disturbed velocity field, we performed additional separate fits for the approaching and receding sides. We present in \Cref{fig:vrot-n3621} the variations of the inclination and position angle from the {\sc 3d Barolo} fit, as well as the obtained rotation curves. The average inclination and position angle are respectively $64.0\pm1.3\dg$ and $354.7\pm3.0\dg$, consistent with those obtained in \citet{DeBlok2008}, although the average position angle is a little higher ($\sim3\%$) than derived by the authors. The approaching and receding curves of the galaxy are very different in shape, with the receding (northern) side being more disturbed. This asymmetry of the galaxy is likely caused by the presence of a cloud-like structure located in the receding side of the galaxy (as is seen in the velocity field in \Cref{fig:n3621-maps}). The PV diagram in \Cref{fig:n3621-pv} suggests the presence of a warp in the line-of-sight of NGC 3621, that is more pronounced in the receding side of the galaxy. This further confirms the results of \citet[][Fig. 78 therein]{DeBlok2008} regarding NGC 3621.

To compare the {\sc 3d Barolo} rotation curve to the 2D-derived curve, we ran {\sc Rotcur} a second time with the kinematical parameters derived above. In the left panel of \Cref{fig:n3621-pv} we plot the two curves together with that obtained by \citet{DeBlok2008}, where it is clear that we reach radii that are almost twice the radii probed by the VLA observations.
The velocities of the {\sc 3d Barolo} curve are higher in the inner parts of the galaxy than those of the {\sc Rotcur} curve, likely because of the beam smearing effects as in the case of NGC 7424.
The {\sc 3d Barolo} curve agrees well with the THINGS curve in the inner $\sim20$ kpc of the galaxy, although the beam smearing is not completely suppressed in the former curve. Beyond 20 kpc, our curves seem to decrease while the THINGS curve increases. Although the galaxy presents a complex velocity field affected by warps especially in its receding side, the rotation curves derived in this work do not seem to be affected by these warps (see variations of the inclination and position angle in \Cref{fig:vrot-n3621}). A meaningful interpretation of the difference between our rotation curves and that of \citet{DeBlok2008} beyond $\sim20$ kpc is hard to provide since the variations (with the radius) of the kinematical parameters obtained by the authors is not available. The difference in the adopted kinematical parameters, as well as the aforementioned stream of cloud that resides in the northern side of the galaxy, may affect the shape of the rotation curve especially in the outer regions.
In the right panel of \Cref{fig:n3621-pv} we present a PV diagram of the galaxy on which we overlay the {\sc Rotcur} and {\sc 3d Barolo} rotation curves. Like in the case of NGC 7424, the rotation curve of NGC 3621 derived with {\sc 3d Barolo} describes better the PV diagram of the galaxy, especially in the inner regions. The PV diagram moreover shows that the velocities in the outskirts of the approaching side are lower than the receding side velocities, in agreement with the rotation curves in \Cref{fig:vrot-n3621}.

\section{Mass models of NGC 7424 \& NGC 3621}\label{sec:massmodel}
\subsection{NGC 7424}\label{sec:n7424-massmodel}
\begin{figure*}
\includegraphics[width=\columnwidth]{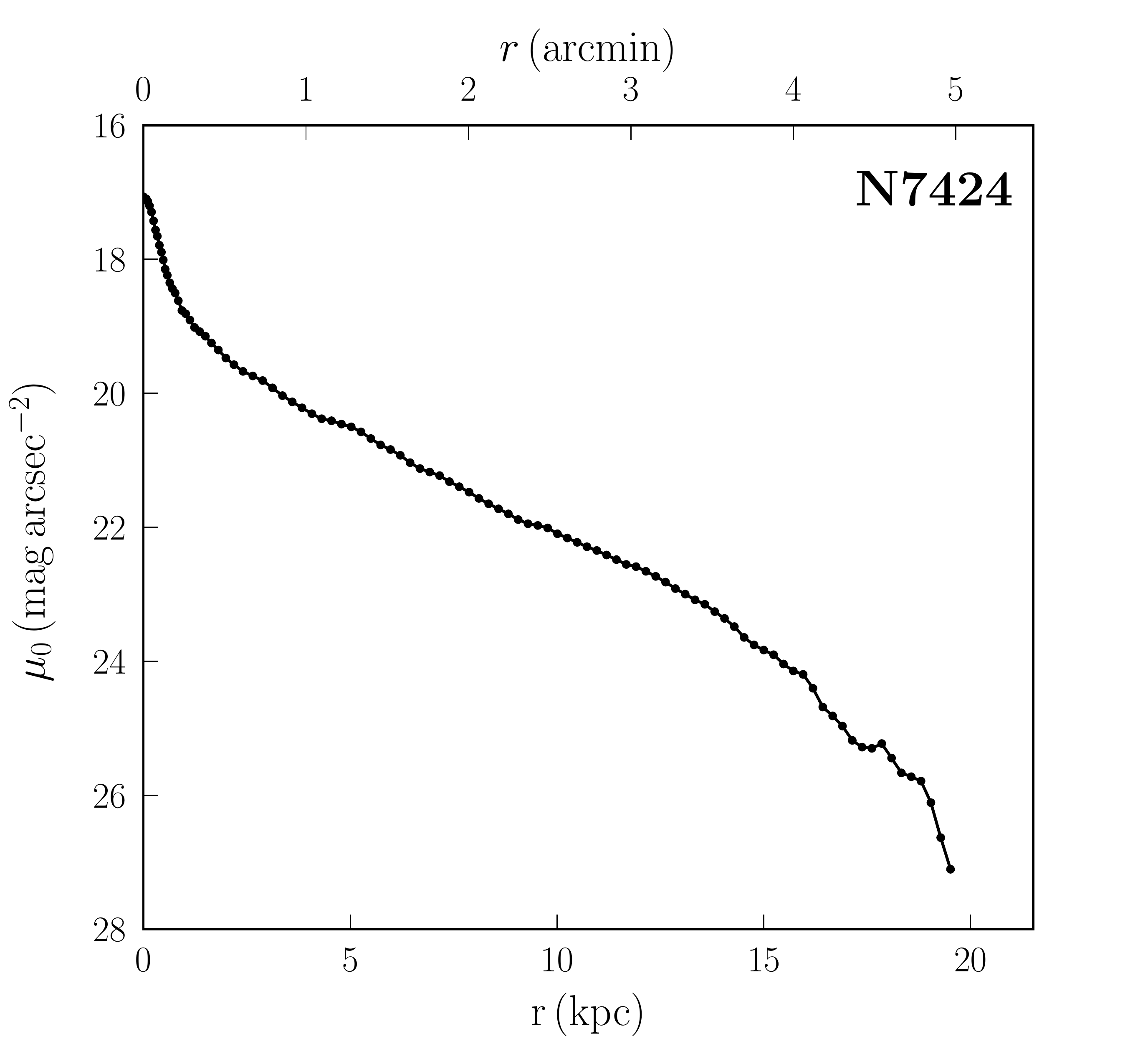}
\includegraphics[width=\columnwidth]{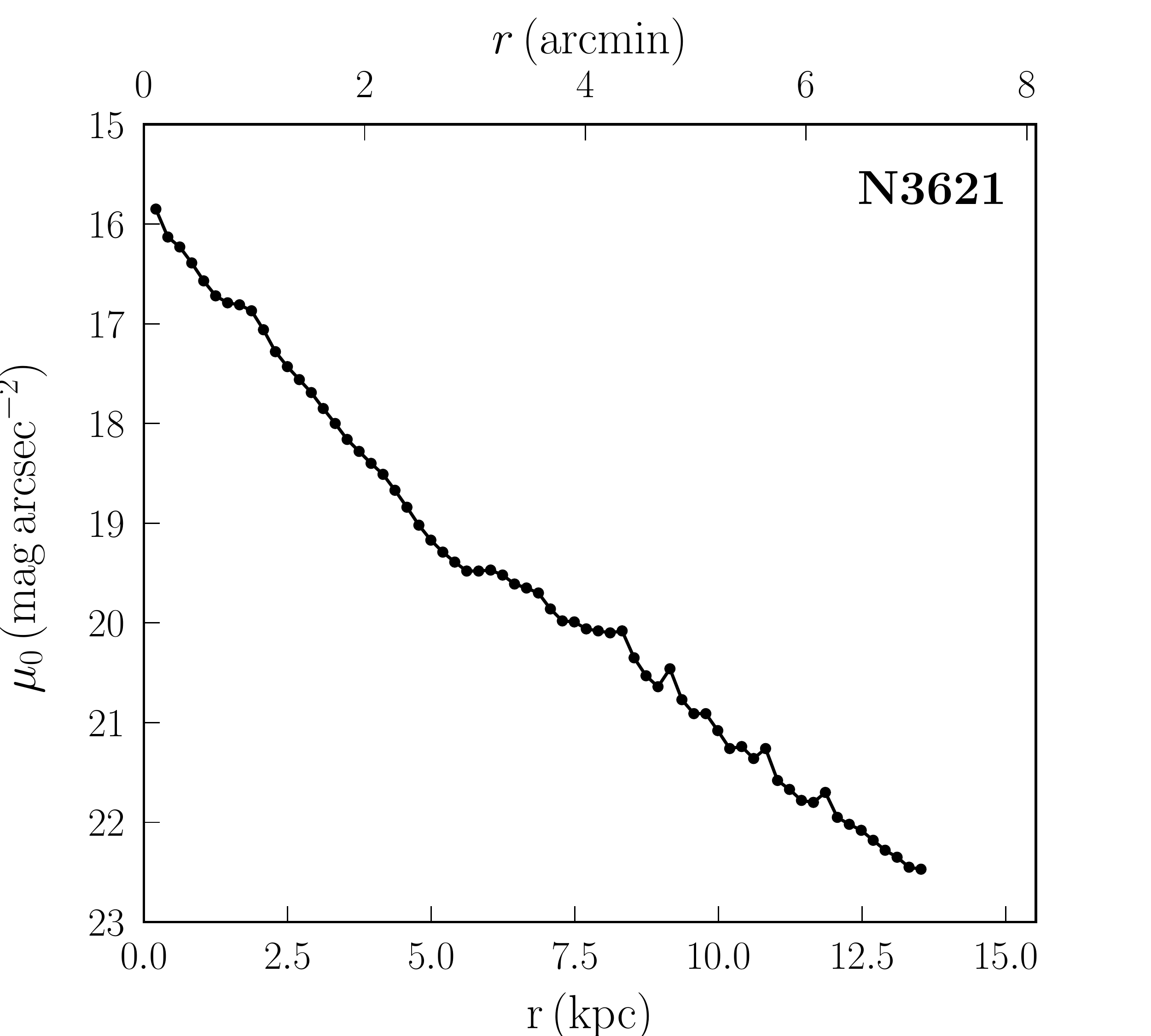}
\caption{{\it WISE} W1 light profile of NGC 7424 ({\it left}) and Spitzer $3.6\mu m$ of NGC 3621 ({\it right}.)}\label{fig:light-profile}
\end{figure*}

The mass models for NGC 7424 were derived using its {\sc 3d Barolo} rotation curve and {\it WISE} \citep{Wright2010} W1 ($\rm 3.4\mu m$) band photometry, sensitive to the old stellar populations of galaxies. The W1 light profile of the galaxy presented in \Cref{fig:light-profile}, along with the galaxy's optical image, suggests the absence of a prominent bulge. An attempt to decompose the light profile in the left panel of \Cref{fig:light-profile} further confirmed a negligible bulge. We therefore consider the stellar disk and the gas disk as the only baryonic components contributing to the observed rotation curve. We can write
\begin{equation}
  v_{\rm baryonic} = \sqrt{|v_{\rm gas}|\,v_{\rm gas} + \Upsilon_\star\,|v_{\rm disk}|\,v_{\rm disk}}
\end{equation}
where $v_{\rm gas}$ is the gas velocity, $v_{\rm disk}$ the disk velocity and $\Upsilon_\star$ the disk mass-to-light ratio ({\it M/L}). $v_{\rm gas}$ is derived from the \hi\, surface density, assuming a thin disk composed of neutral hydrogen and helium (the \hi\, surface densities have been multiplied by 1.4 to take into account the helium and heavier elements). Also, using the W1-W2 ($\rm 4.6\mu m$) colour and the calibration from \citet[][Eq. 1]{Cluver2014}, we derived an inferred disk mass-to-light ratio of $\Upsilon_\star=0.25$ \ml. In fitting the rotation curve, we fixed $\Upsilon_\star$ to the theoretical value on one hand, and let it free on the other hand (best fit model). Three different models were used in each case: the pseudo-isothermal (ISO) and the Navarro-Frenk-White (NFW) dark matter models, and the MOdified Newtonian Dynamics (MOND) model.

\subsubsection{The pseudo-ISO model}
Besides the baryonic components of the galaxy, an important fraction of galaxies' mass is contained in their dark matter (DM) halo. The DM halo contributes to the total rotational velocity of a galaxy, and we can therefore write that
\begin{equation}
  v_{\rm rot} = \sqrt{v_{\rm baryonic}^2 + v_{\rm halo}^2}
\end{equation}
where $v_{\rm halo}$ is the DM halo contribution.

The density and velocity distributions of the model as a function of the distance $r$ from the halo centre are respectively given by
\begin{eqnarray}
  \rho_{\rm ISO}(r) &=& {\rho_0 \over 1+(r/r_0)^2},\\
  v_{\rm ISO}(r) &=& \sqrt{4\pi G\rho_0 r_0^2\left[1-(r/r_0)\arctan{(r/r_0)}\right]}
\end{eqnarray}
where $\rho_0$ and $r_0$ are respectively the central density and the scaling radius of the halo. We used the model implemented in the \gipsy's task \rotmas, with NGC 7424. As stated above, the stellar disk's {\it M/L} was first fixed to the {\it WISE} inferred value, then let free. \Cref{tb:massmod-n7424} summarises the different values derived. The {\it M/L} value obtained in the best fit is $0.18\pm0.57$ \ml, which is consistent with the {\it WISE} inferred value, and shows that the two fits are in agreement. We present in the top panel of \Cref{fig:massmod-n7424} the pseudo-ISO fit with the {\it M/L} fixed by the {\it WISE} colour.

\subsubsection{The Navarro-Frenk-White model}
The DM halo profile can also be described with the Navarro-Frenk-White \citep[NFW,][]{Navarro1997} profile derived from N-body simulations. The model describes the density profile by
\begin{equation}
\rho_{\rm NFW}(r) = {\rho_i\over r/r_{\rm S}(1 + r/r_{\rm S})^2},
\end{equation}
where $r_{\rm S}$ is the scale radius of the halo, and $\rho_i$ is related to the critical density, i.e, the density of the universe at which the DM halo begins to collapse. The resulting rotation velocity can be written
\begin{equation}
v_{\rm NFW} = v_{200}\sqrt{{\ln{(1+cx)} - cx/(1+cx)\over x\left[\ln{(1+c)} - c/(1+c)\right]}}
\end{equation}
where $r_{200}$ and $v_{200}$ are respectively the radius and density at which the density of the halo exceeds 200 times the critical density of the universe \citep{Navarro1996}. The ratio $x=r/r_{200}$ represents the radius in units of virial radius, and $c=r_{200}/r_{\rm S}$ the concentration of the halo. The NFW mass density profile is cuspy in the inner regions of the halo and can be represented by $\rho\sim r^\alpha$ , where $\alpha=-1$.

The fit of the NFW model to NGC 7424's observed rotation curve provides a negative {\it M/L} value for the stellar disk when the parameter is not fixed to the {\it WISE} inferred value of 0.25 \ml, which is physically unacceptable. Similarly, \citet{Carignan2013} obtained a non-physical stellar {\it M/L} of $\Upsilon_\star=0$ for NGC 3109. However, when the stellar {\it M/L} is fixed to the {\it WISE} inferred value, the model provides a good fit to the observed rotation. The NFW model of NGC 7424 is presented in the middle panel of \Cref{fig:massmod-n7424}.


\subsubsection{The MOND model}
First proposed by \citet{Milgrom1983a,Milgrom1983b} as an alternative to DM, the MOND theory has been widely discussed in the last two decades and has proved successful in accurately reproducing the rotation curves of numerous galaxies \citep[e.g.,][]{Begeman1991,Sanders1998,Gentile2010}. The model postulates that the classical Newtonian dynamics breaks in regimes where the acceleration is much smaller than the universal constant acceleration $a_0$,  and the law of gravity is modified. The gravitational acceleration of a particle is then given by
\begin{equation}
  \mu(x=g/a_0)g = g_N
\end{equation}
where $g$ is the gravitational acceleration, $g_N$ the Newtonian acceleration and $\mu(x)$ the MOND simple interpolating function \citep{Zhao2006}.

\begin{figure}
\centering
\includegraphics[width=\columnwidth]{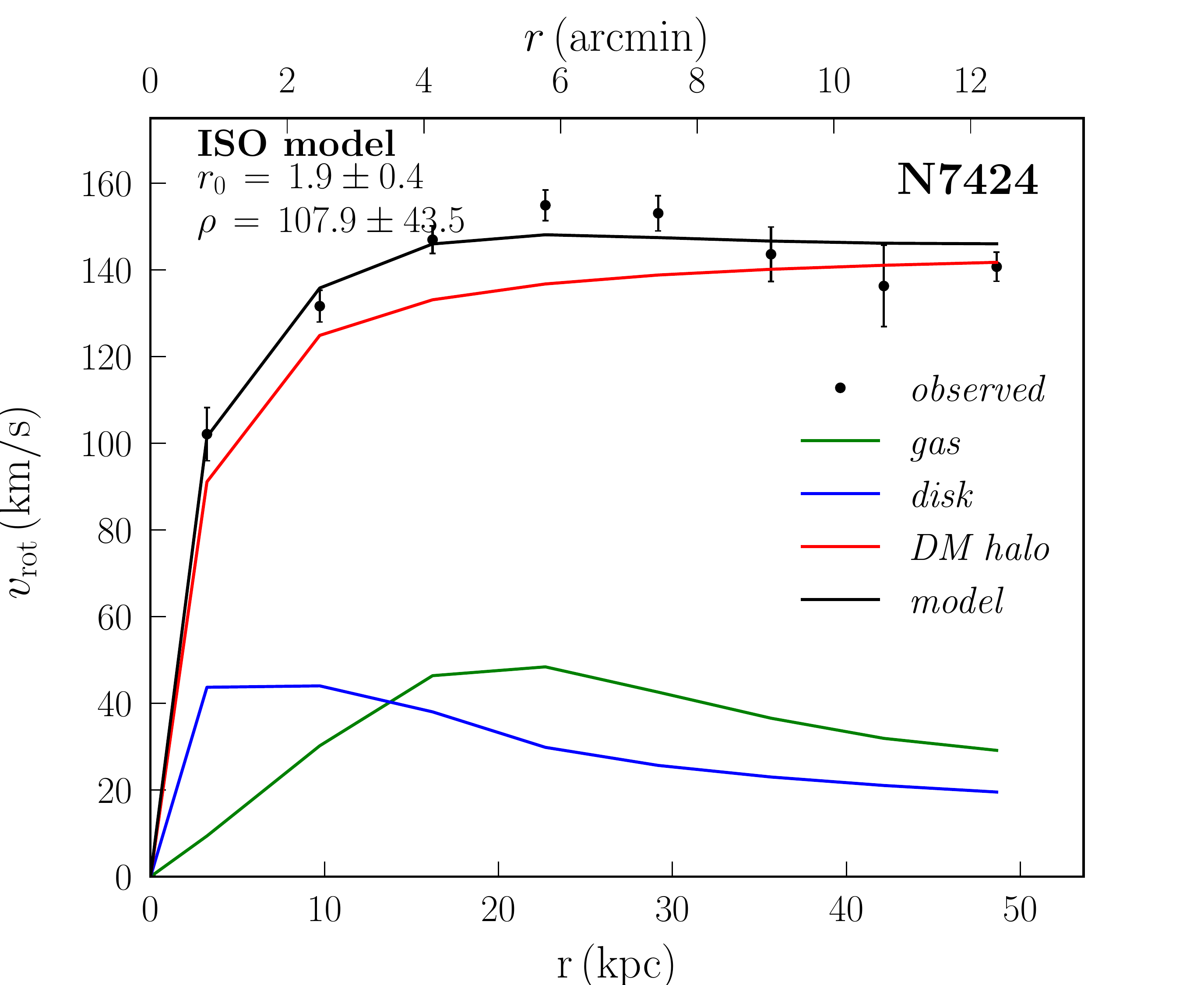}
\includegraphics[width=\columnwidth]{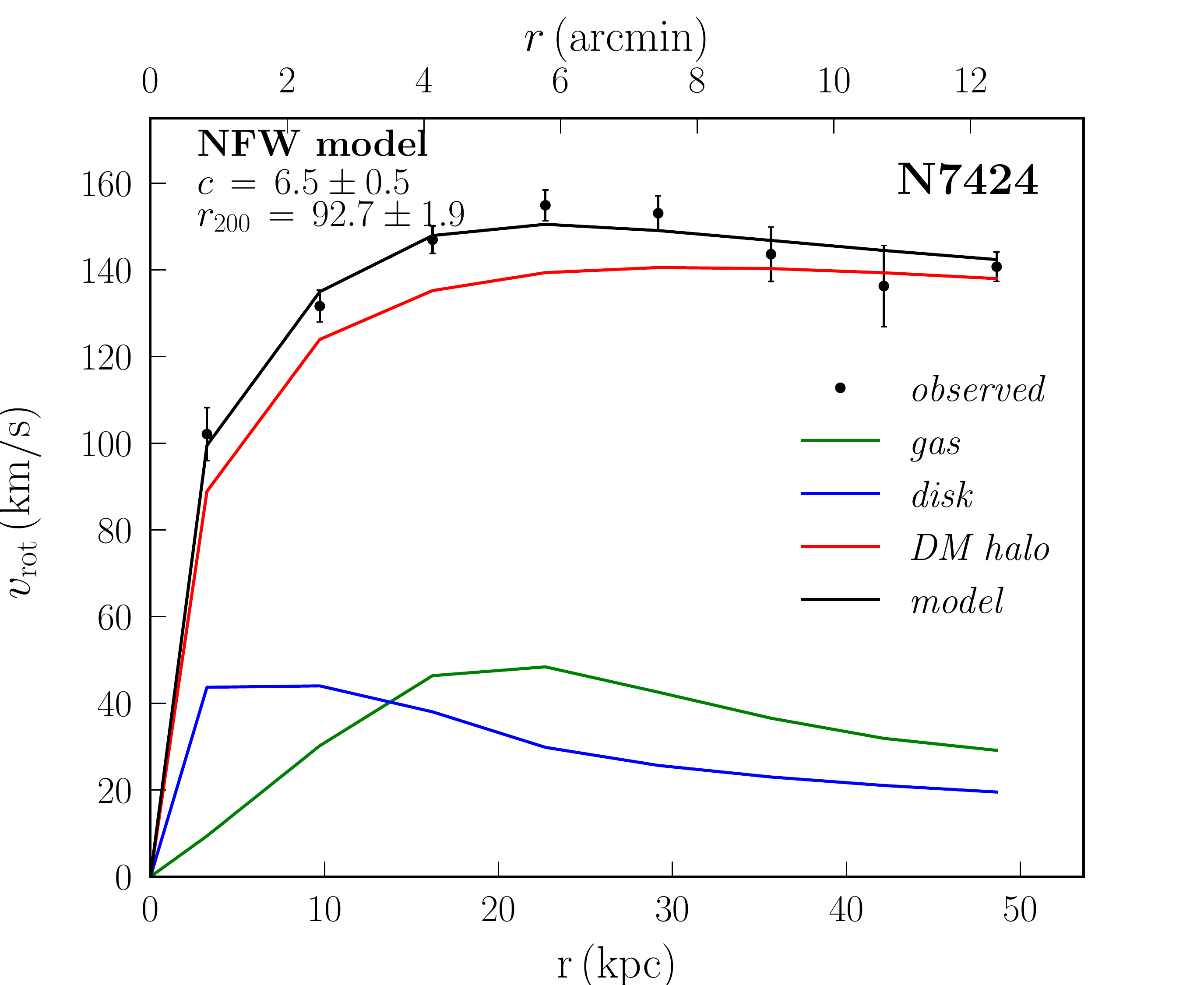}
\includegraphics[width=\columnwidth]{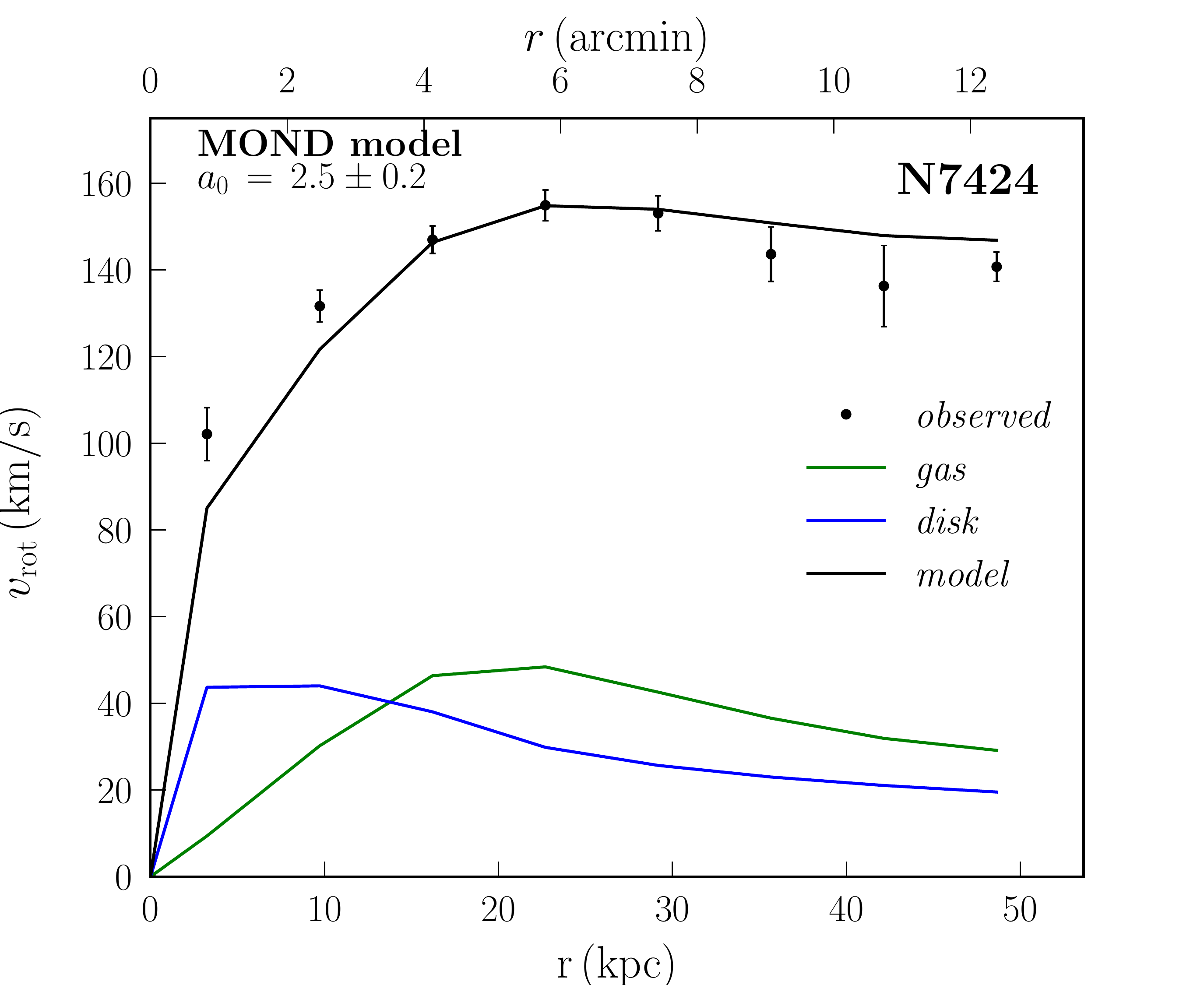}
\caption{The mass models of NGC 7424 for the pseudo-ISO ({\it top panel}), NFW ({\it middle panel}) and MOND ({\it bottom panel}) models.}\label{fig:massmod-n7424}
\end{figure}

As for the pseudo-ISO model, we fit the model to the rotation curve of the galaxy for both a fixed and a free {\it M/L} parameter. Moreover, we fixed the parameter $a_0$ to the generally accepted value of $1.21\e{-8}\,\rm cm\,s^{-2}$ \citep{Begeman1991} on one hand, and let it free to vary on the other. The results of the fits are summarised in \Cref{tb:massmod-n7424}. 
The value of $a_0$ obtained when $\Upsilon_\star$ is fixed to the {\it WISE} colour is two times higher than the generally accepted value. On the other hand, when $a_0$ is fixed to the \citet{Begeman1991}'s value, we obtain a $\Upsilon_\star$ value of 0.92, i.e more than three times higher than the {\it WISE} inferred value. These may hint that the MOND model does not provide a good fit to the galaxy's rotation curve, although the obtained $\chi^2$ values suggest otherwise.
This is further seen in the bottom panel of \Cref{fig:massmod-n7424} where we present the MOND fits to the galaxy's rotation curve with fixed {\it M/L} and free $a_0$: the model tends to underestimate the galaxy's velocities in the inner regions.

\begin{table}
  \begin{tabular}{lcclcc}
    \hline\hline
    \multirow{2}{*}{Model} & & \multicolumn{2}{c}{\multirow{2}{*}{Parameter}} & \multicolumn{2}{c}{Values} \\
    \cmidrule{5-6}
          &              & & & $\Upsilon_\star$ fixed & $\Upsilon_\star$ free \\
    \hline
    \multirow{4}{*}{ISO} & & \multicolumn{2}{c}{$\Upsilon_\star$} & 0.25  & $0.18\pm0.57$ \\
                         & & \multicolumn{2}{c}{$r_0$}      & $1.9\pm0.4$   & $1.8\pm1.2$\\
                         & & \multicolumn{2}{c}{$\rho_0$}   & $107.9\pm43.5$ & $124.7\pm178.5$\\
                         & & \multicolumn{2}{c}{$\chi_{\rm red}^2$} & 1.8 & 2.1\\
    \hline
    \multirow{4}{*}{NFW} & & \multicolumn{2}{c}{$\Upsilon_\star$} & 0.25 & \multirow{4}{*}{---}\\
                         & & \multicolumn{2}{c}{$r_{200}$} & $92.7\pm1.9$ & \\
                         & & \multicolumn{2}{c}{$c$} & $6.5\pm0.5$ & \\
                         & & \multicolumn{2}{c}{$\chi_{\rm red}^2$} & 4.8 & \\
    \hline
    \multirow{6}{*}{MOND} & \multirow{3}{*}{$a_0$ free} & & $\Upsilon_\star$ & 0.25 & $0.69\pm0.14$\\
                       & & & $a_0$ & $2.5\pm0.2$ & $1.4\pm0.2$\\
                       & & & $\chi_{\rm red}^2$ & 3.0 & 0.8\\
    \cmidrule{2-6}
     & \multirow{3}{*}{$a_0$ fixed} & & $\Upsilon_\star$ & \multirow{3}{*}{---} & $0.92\pm0.05$\\
                       & & & $a_0$  & & $1.21$\\
                       & & & $\chi_{\rm red}^2$ & & 0.9\\
    \hline
  \end{tabular}
  \caption{Results of the NGC 7424's mass models fits. The $\Upsilon_\star$ ratio is expressed in units of \ml, and the radii and densities have units of kpc and $10^{-3}\Mpcc$, respectively. The units of $a_0$ are $\rm 10^{-8}\,cm\,s^{-2}$.}
  \label{tb:massmod-n7424}
\end{table}

%
%
%

\subsection{NGC 3621}\label{sec:n3621-massmodel}
\begin{figure}
\centering
\includegraphics[width=\columnwidth]{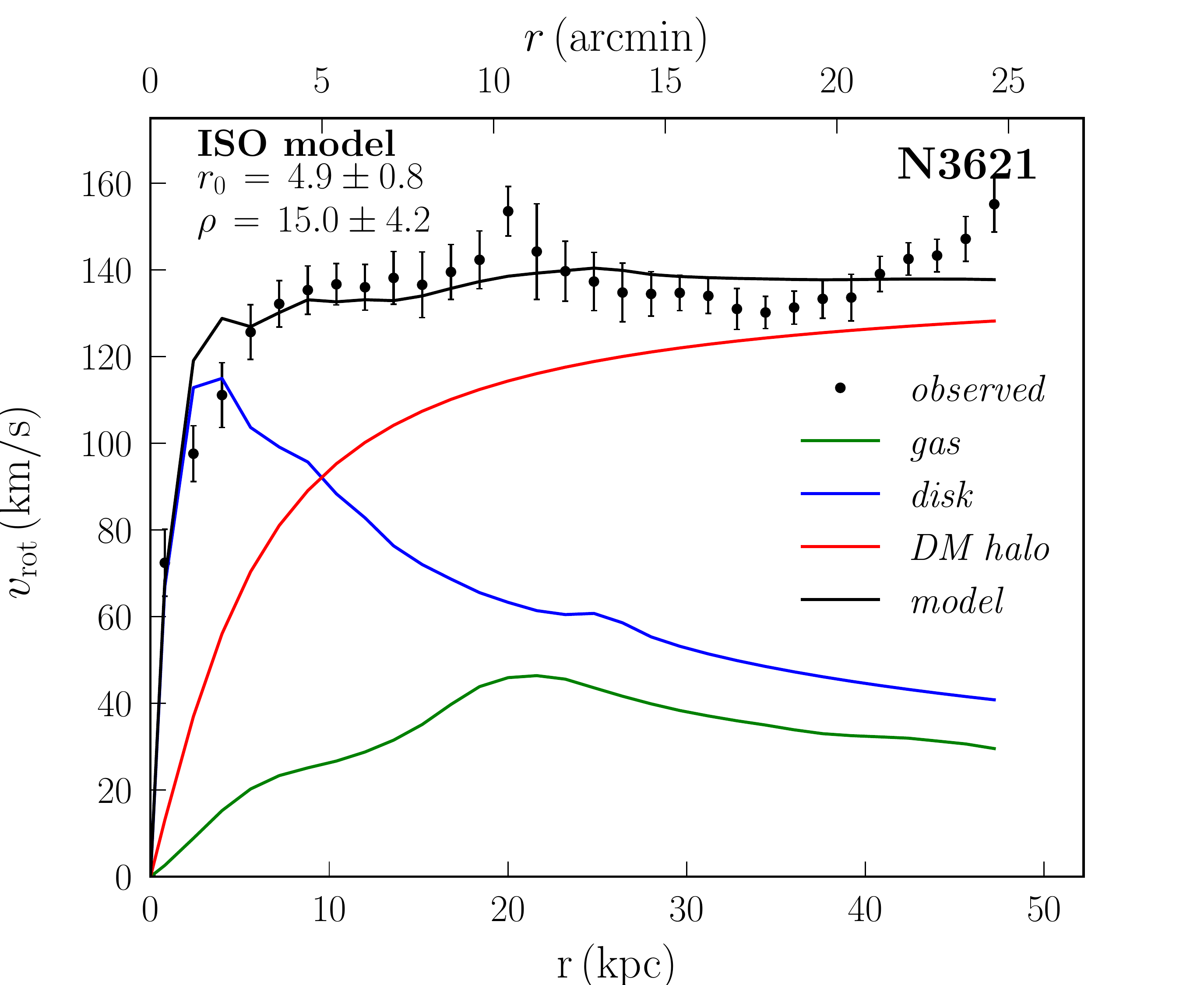}
\includegraphics[width=\columnwidth]{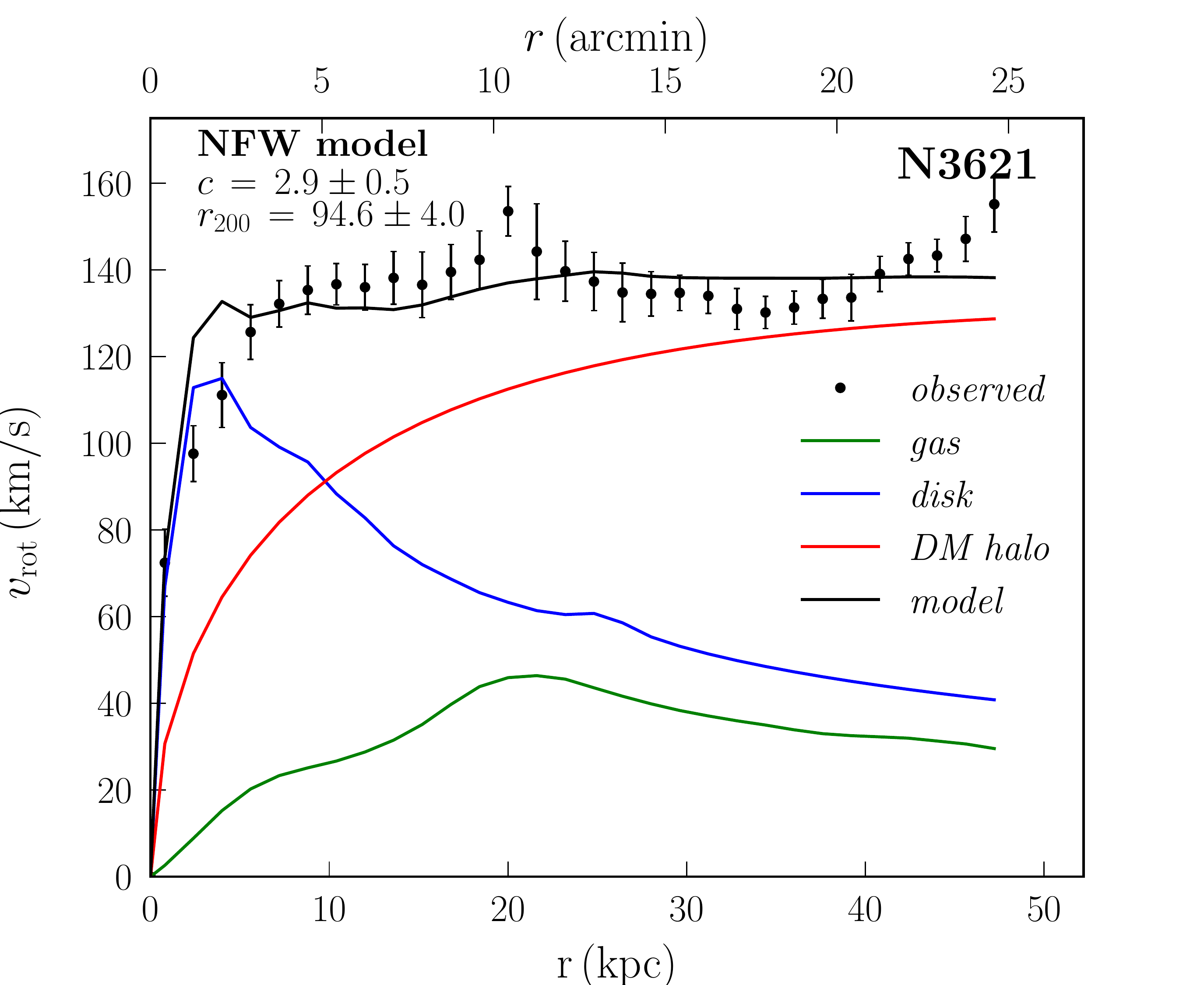}
\includegraphics[width=\columnwidth]{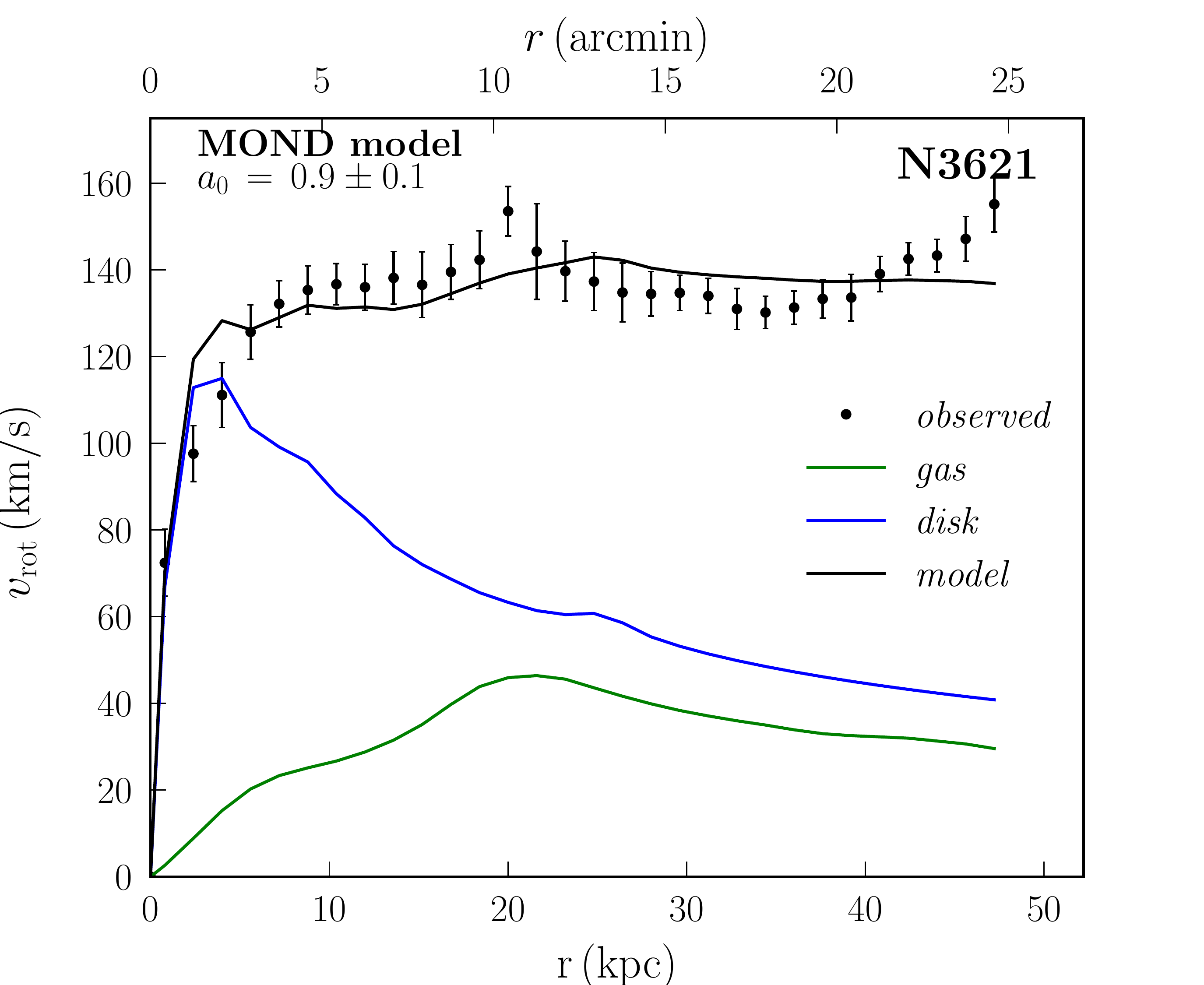}
\caption{The mass models of NGC 3621 for the ISO ({\it top panel}), NFW ({\it middle panel}) and MOND ({\it bottom panel}) models.}\label{fig:massmod-n3621}
\end{figure}

The {\it M/L} of NGC 3621's stellar disk, as derived from the galaxy's {\it WISE} inferred colour, is $\Upsilon_\star=0.50$ \ml. This value is in agreement with the value of 0.59 \ml\, derived in \citet{DeBlok2008} who used the Spitzer $3.6\rm\mu m$ band photometry to describe the optical component of the galaxy, and also similar to the $3.6\rm\mu m$ value suggested by \citet{Lelli2016}. Using the method described in \Cref{sec:n7424-massmodel}, we constructed its mass models for the ISO, NFW and MOND models. However, for a more rigorous comparison of the mass models with those obtained in \citet{DeBlok2008}, we used, instead of the {\it WISE} W1 profile, the Spitzer $3.6\mu m$ light profile of the galaxy as in \citet{DeBlok2008}. Like in the case of NGC 7424, the galaxy's light profile does not present a bulge (right panel of \Cref{fig:light-profile}), making us consider its stellar and gas disks as the only baryonic components to account for. Similarly to NGC 7424, we derived both the best fit and colour-fixed {\it M/L} models (see \Cref{tb:massmod-n3621}). The stellar {\it M/L} derived from the ISO best fit is, like in the case of NGC 7424, consistent with the value derived from the {\it WISE} colour while it agrees less for the NFW and MOND models. \Cref{fig:massmod-n3621} presents the different fits for the fixed {\it M/L} models.

\begin{table}
  \begin{tabular}{lcclcc}
    \hline\hline
    \multirow{2}{*}{Model} & & \multicolumn{2}{c}{\multirow{2}{*}{Parameter}} & \multicolumn{2}{c}{Values} \\
    \cmidrule{5-6}
          &              & & & $\Upsilon_\star$ fixed & $\Upsilon_\star$ free \\
    \hline
    \multirow{4}{*}{ISO} & & \multicolumn{2}{c}{$\Upsilon_\star$} & 0.50  & $0.47\pm0.16$ \\
                         & & \multicolumn{2}{c}{$r_0$}      & $4.8\pm0.8$   & $4.4\pm1.7$\\
                         & & \multicolumn{2}{c}{$\rho_0$}   & $14.9\pm4.2$ & $18.0\pm1.0$\\
                         & & \multicolumn{2}{c}{$\chi_{\rm red}^2$} & 2.0 & 1.7\\
    \hline
    \multirow{4}{*}{NFW} & & \multicolumn{2}{c}{$\Upsilon_\star$} & 0.50 & $0.36\pm0.09$\\
                         & & \multicolumn{2}{c}{$r_{200}$} & $162.7\pm4.0$ & $87.4\pm2.5$\\
                         & & \multicolumn{2}{c}{$c$} & $5.3\pm0.5$ & $8.0\pm0.1$\\
                         & & \multicolumn{2}{c}{$\chi_{\rm red}^2$} & 2.4 & 2.0\\
    \hline
    \multirow{6}{*}{MOND} & \multirow{3}{*}{$a_0$ free} & & $\Upsilon_\star$ & 0.50 & $0.43\pm0.06$\\
                       & & & $a_0$ & $0.9\pm0.1$ & $0.9\pm0.1$\\
                       & & & $\chi_{\rm red}^2$ & 2.1 & 2.1\\
   \cmidrule{2-6}
     & \multirow{3}{*}{$a_0$ fixed} & & $\Upsilon_\star$& \multirow{3}{*}{---} & $0.36\pm0.02$\\
                       & & & $a_0$  & & 1.21\\
                       & & & $\chi_{\rm red}^2$ & & 2.1\\
    \hline
  \end{tabular}
  \caption{Results of the NGC 3621's mass models fits. The units are same as in \Cref{tb:massmod-n7424}.}
  \label{tb:massmod-n3621}
\end{table}



\section{Summary \& Discussions}\label{sec:discussion}
The test observations of the MHONGOOSE galaxies allowed to identify the best calibrators for each of the five galaxies observed. Three different flux calibrators (PKS 1934-638, 0407-658 and 3C138) were used in total, two of which were used per individual galaxy. All the flux calibrators performed well overall, with the flux being quite constant. However, due to the low observing time that the sessions with 3C138 have received, we cannot fully assess the quality of the calibrator. As for the phase calibrators, we successfully identified the best of three candidate sources for each galaxy. The most extreme case of source quality variation was observed between the phase calibrators 0039-445 and 0201-440, where the quality difference of the respective data cubes of the target galaxy was very significant.

The calibrations applied to the KAT-7 observations proved to slightly improve the quality of the data after a self-calibration of the calibrators was performed, as illustrated in \Cref{fig:phase-0039}. However, an estimation of the SNR of the resulting data cubes suggests that the improvement is a function of the size and the brightness of the target galaxy. In fact, the SNR is defined as the ratio between the strongest signal in a particular channel and the channel-to-channel rms:
\begin{equation}
\rm SNR = {S_\nu \over \sigma_{rms}}.
\end{equation}
Since the change in the rms is small compared to a galaxy's signal, and given that the rms in the data does not vary with the galaxy's size and brightness, the change in the SNR will be more noticeable for fainter galaxies than for brighter galaxies. This is what we obtain when comparing the SNR of the five KAT-7 galaxies. For example, the SNR of NGC 7424 before and after self-calibration is about 55, while that of NGC 625 goes from about 26 to about 35, that is, a $\sim30\%$ increase in the quality of the data.

The search for gas clouds around the candidate galaxies -- that might hint at ongoing gas accretion -- was unsuccessful despite the high sensitivity of the GBT. Nonetheless, this quest allows us to put an upper limit on the column density levels to which one can expect to detect signs of gas accretion in these galaxies. Indeed, for the galaxies observed with the GBT, any eventual search for gas clouds should be conducted at column density levels below a $3\sigma$ detection level of $\sim2.2\e{18}\,\rm cm^{-2}$, for a 20 \kms\, line width. However, if the low column density gas is in discrete clouds, small compared to the GBT beam, we may have missed them. In effect, the physical size of the GBT beam at the distance of the galaxies range from about 9 to 64 kpc with an average of about 30 kpc, which is the range of sizes at which the clouds are expected to be found. This confirms that like sensitivity, the resolution of \hi\, observations is important to detect low-column density emission, as was clearly shown for the inter-galactic \hi\, between M31 and M33 \citep{Wolfe2013}, and for some HALOGAS galaxies \citep{Pingel2018}. This is exactly what MeerKAT will provide us with: the ability to detect low column density \hi\, at high spatial resolutions.

The ISO and NFW Dark Matter models of NGC 7424 show a DM-dominated galaxy at all radii (see \Cref{fig:massmod-n7424}). This is not surprising given the morphological type of the galaxy (late-type spiral), and is in agreement with previous studies on similar galaxies \citep[e.g.,][]{Cote1991, Carignan2013}. The free {\it M/L} fit of the galaxy's rotation curve with the ISO model yields a stellar {\it M/L} value of $0.18\pm0.57$ \ml, consistent with with the {\it WISE} inferred value of 0.25 \ml, although the error associated with the fitted value is large. This large error may hint that using the best fit method is not necessarily the best way to derive the disk {\it M/L} ratio of galaxies, and measuring the ratio from the photometry (in this case in the {\it WISE} W1 band) beforehand can help to better constrain the mass of the stellar component.
As for the NFW model, we obtain a negative, therefore physically unacceptable value, making the model less suited than the ISO model to fit the rotation curve of NGC 7424 with free {\it M/L}. However, when the stellar {\it M/L} is fixed to the value derived from the {\it WISE} colour, the two models are both consistent with the rotation curve of the galaxy as is shown in the first two panels of \Cref{fig:massmod-n7424}.
For the MOND model, we found that the fits provided values that are inconsistent with the {\it WISE} colour of the galaxy, and when the disk {\it M/L} is fixed, we obtain a value of the parameter $a_0$ that is not in agreement with the generally accepted value.

The MeerKAT commissioning observations of NGC 3621 with 16 antennas allowed to reach column density sensitivities of $10^{19}\,\cm$, similar to the LVHIS observations with the ATCA. The rotation curve of the galaxy derived in this work extends to about 50 kpc, two times more extended than the THINGS survey previously reached with the VLA. The galaxy's total \hi\ intensity map and velocity field suggest that NGC 3621 is asymmetric, and presents a warp in the line-of-sight (see \Cref{fig:n3621-pv}). Its rotation curves for both its approaching and receding sides have very different shapes, confirming the complex kinematics of the galaxy. When overlaid on the PV diagram of the galaxy (\Cref{fig:n3621-pv}), the rotation curve obtained by averaging both sides fails to accurately describe the approaching side of the galaxy at regions around $10'$, most likely due to the effect of the receding side on the averaged curve.
Unlike NGC 7424 whose mass models suggest a DM-dominated galaxy at all radii, the models for NGC 3621 reveal a maximum stellar disk in the inner regions of the galaxy. 
The fits of the ISO, NFW and MOND models to the galaxy's rotation curve provide more or less the same accuracy, with the ISO model providing the closest {\it M/L} value to the {\it WISE} inferred value. This result, together with the result of NGC 7424 fits, allows us to conclude that the ISO model is to be preferred over the other two models for these galaxies.








\section*{Acknowledgments}
The KAT-7 and MeerKAT telescopes are operated by the South African Radio Astronomy Observatory (SARAO), which is a facility of the National Research Foundation (NRF), an agency of the Department of Science and Technology (DST).
We thank the West Virginia University Research Office for its support of the operations of the GBT.
We also thank Tom Jarrett who provided the {\it WISE} photometric data of the galaxies NGC 7424 and NGC 3621 used in this work, and Danielle Lucero who helped with the KAT-7 observations.
The work of CC is based upon research supported by the South African Research Chairs Initiative (SARChI) of the Department of Science and Technology (DST), SARAO and NRF. The research of AS and MK has been supported by SARChI fellowship. DJP, NMP, and Amy Sardone acknowledge partial support by NSF CAREER grant AST-1149491.

\bibliographystyle{mnras}
\bibliography{library}

\FloatBarrier

\appendix
\section{\hi\, maps of the GBT detections}\label{sec:gbtmaps}
Below we present the GBT \hi\, maps and global spectra of the 16 MHONGOOSE galaxies, as well as the secondary sources listed in \Cref{tb:hi-props}. \Cref{app:eso300-g014,app:eso300-g016,app:eso302-g014,app:eso357-g007,app:kk98-195,app:kks2000-23,app:ngc1371,app:ngc1592,app:ngc3511,app:ngc5068,app:ngc5170,app:ngc5253,app:ugca015,app:ugca250,app:ugca307,app:ugca320} show the column density map on the left panel and the global spectrum on the right. 
The \hi\, column density levels are $0.5, 1.0, 2.0, 4.0, ..., \times10^{19}\,\rm cm^{-2}$, overlaid on optical {\it WISE} W1 grayscale images. The global profiles of the four secondary sources are given in \Cref{app:profiles}. For each profile, the systemic velocity is marked by a vertical upward arrow.
\input{mapsgbt}

\bsp	
\label{lastpage}
\end{document}

%% file: mapsgbt.tex
\begin{figure*}
	\includegraphics[width=\columnwidth]{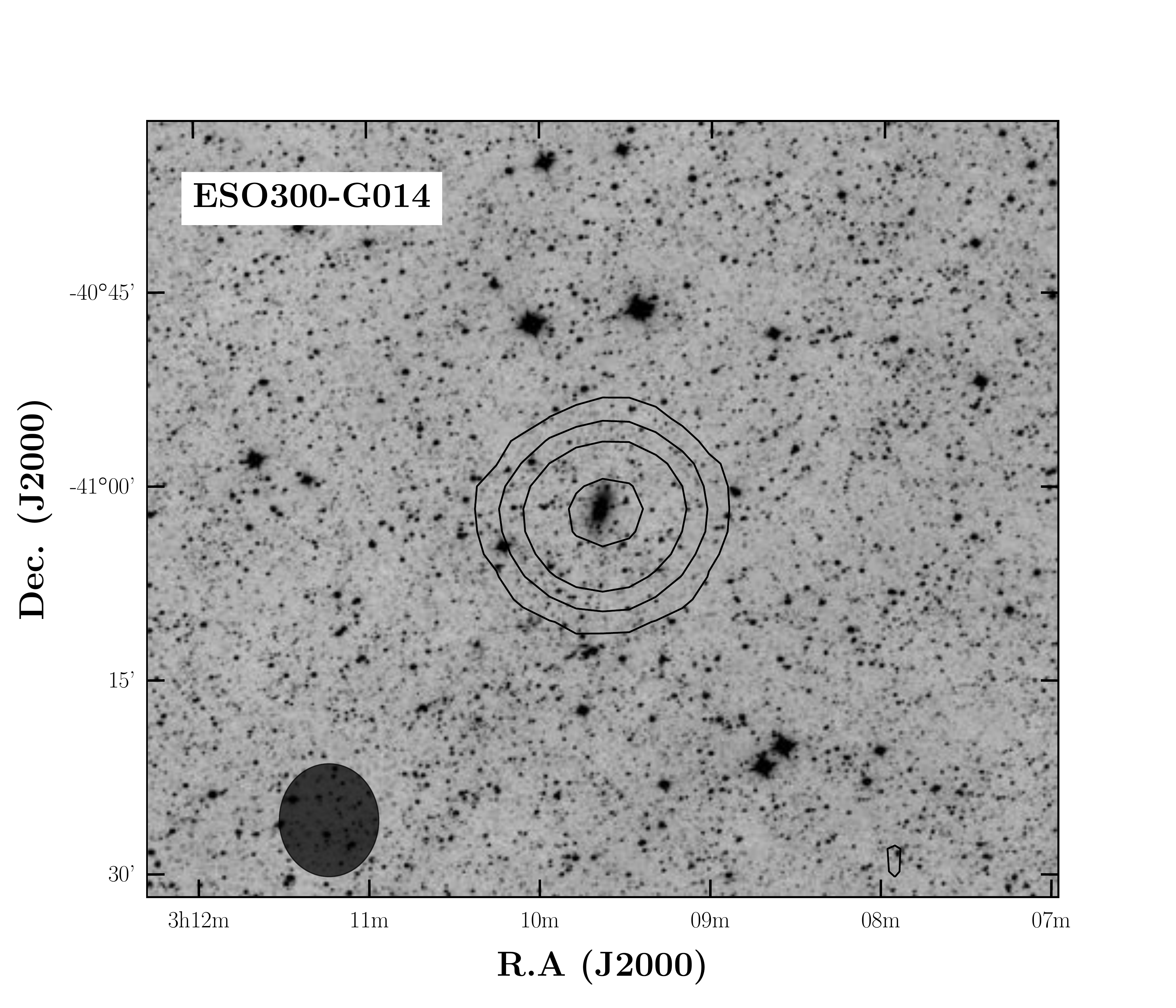}
	\includegraphics[width=\columnwidth]{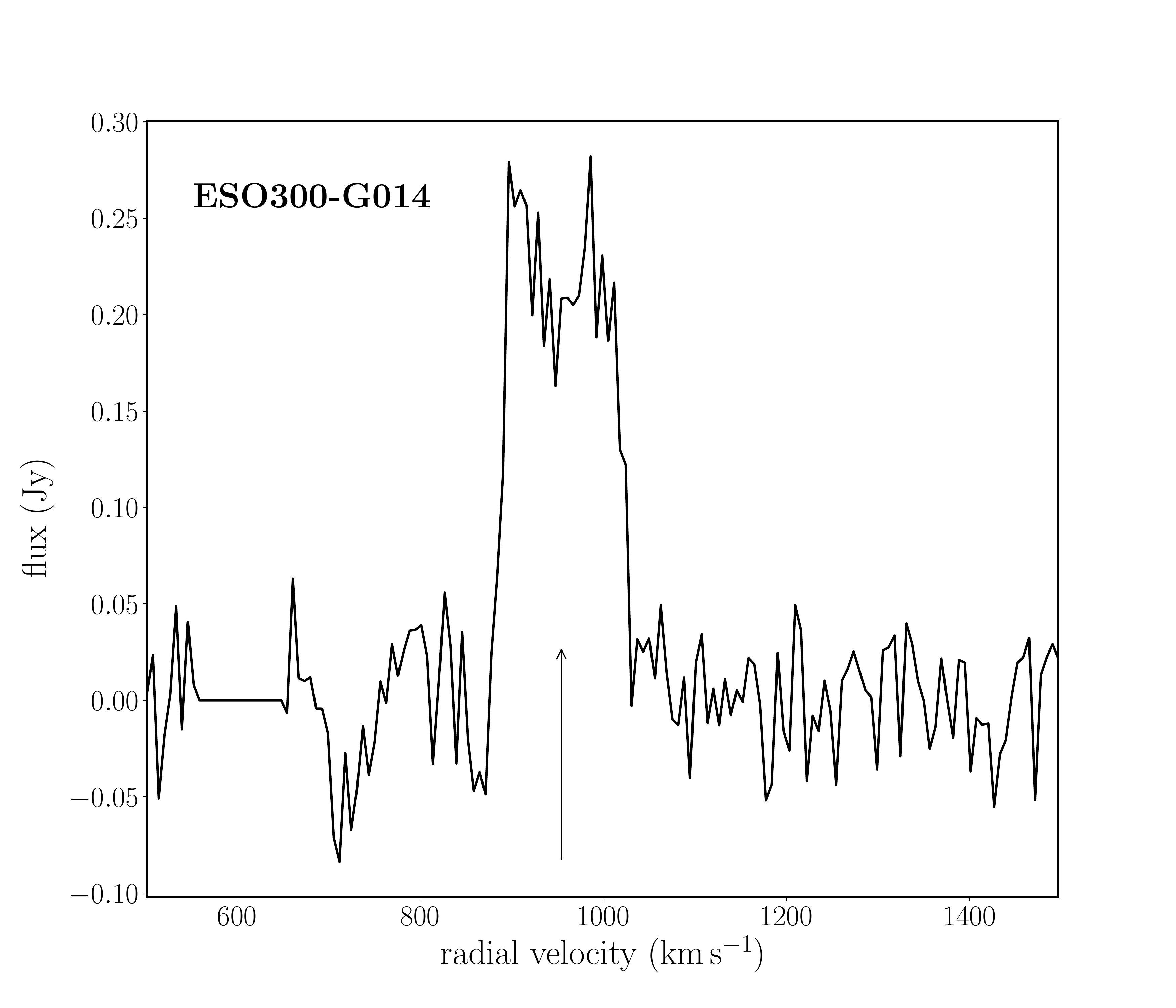}
    \caption{ESO 300-G014}
    \label{app:eso300-g014}
\end{figure*}
\begin{figure*}
	\includegraphics[width=\columnwidth]{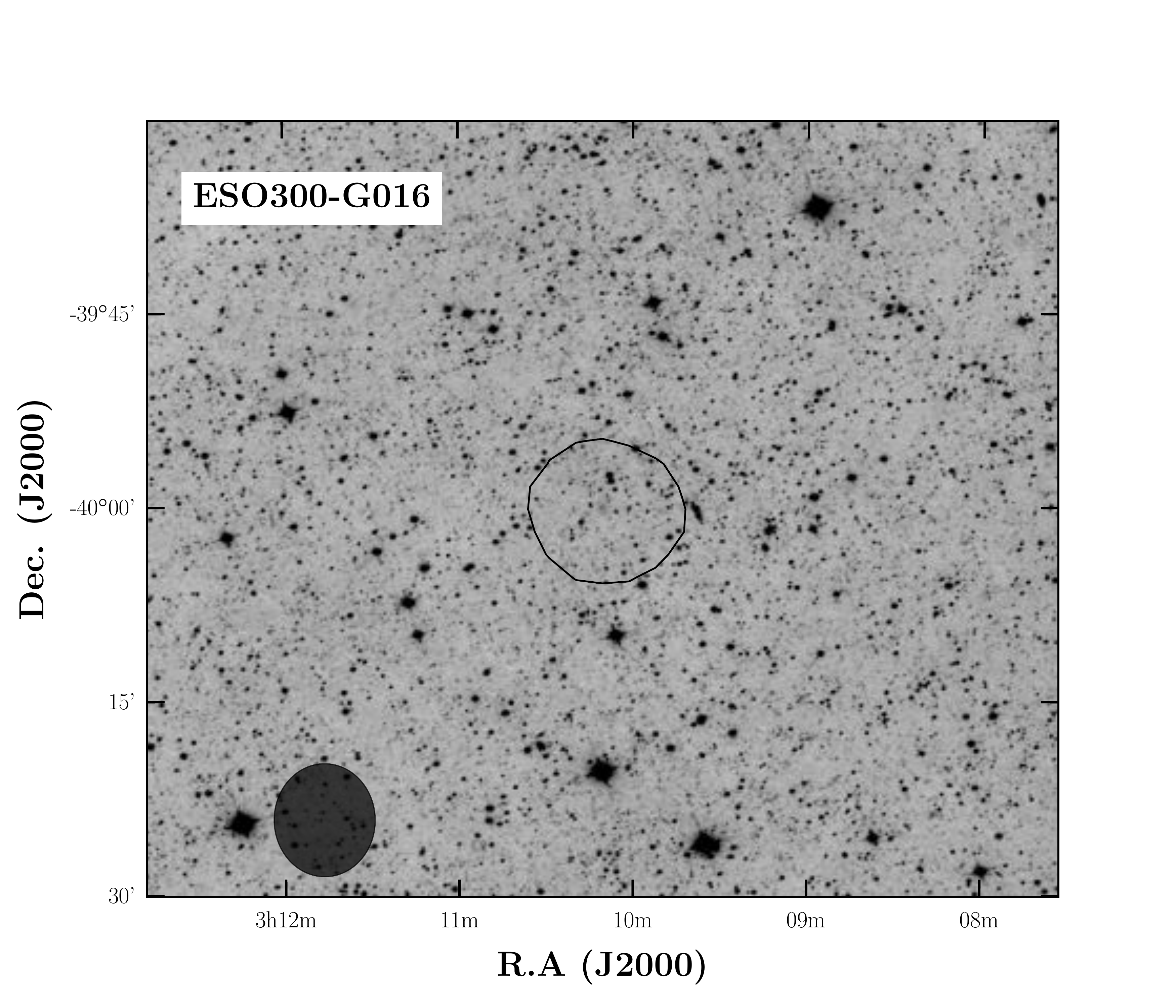}
	\includegraphics[width=\columnwidth]{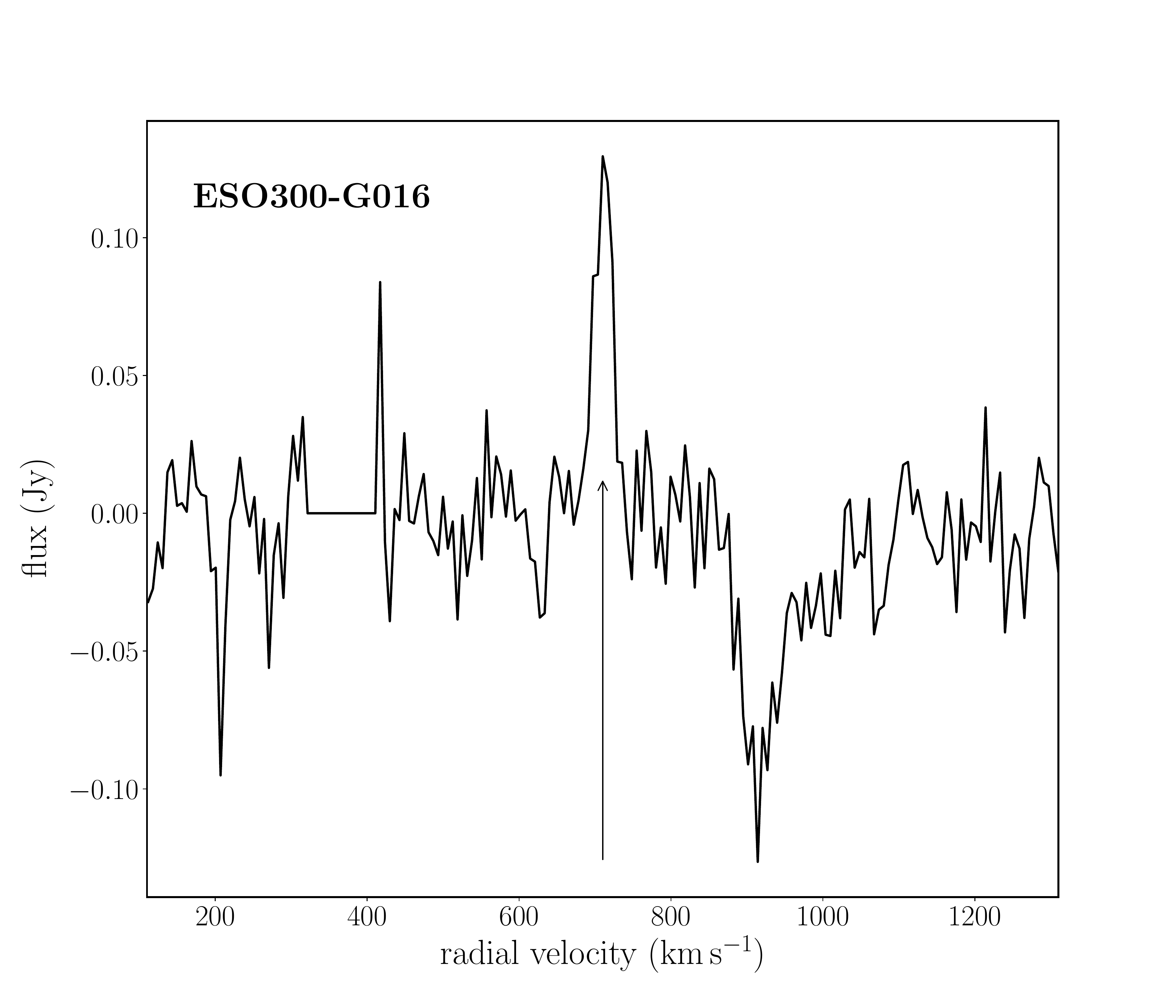}
    \caption{ESO 300-G016}
    \label{app:eso300-g016}
\end{figure*}
\begin{figure*}
	\includegraphics[width=\columnwidth]{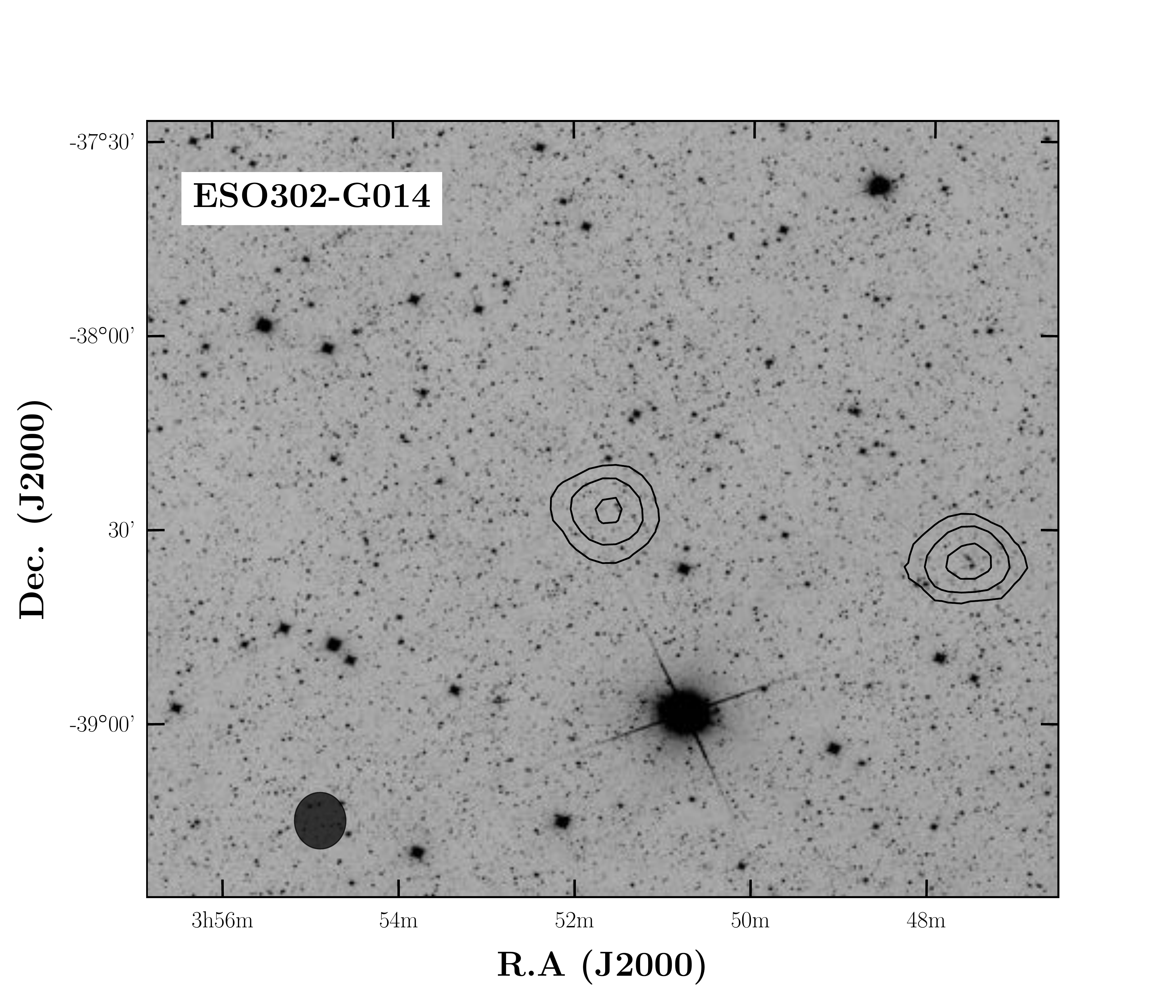}
	\includegraphics[width=\columnwidth]{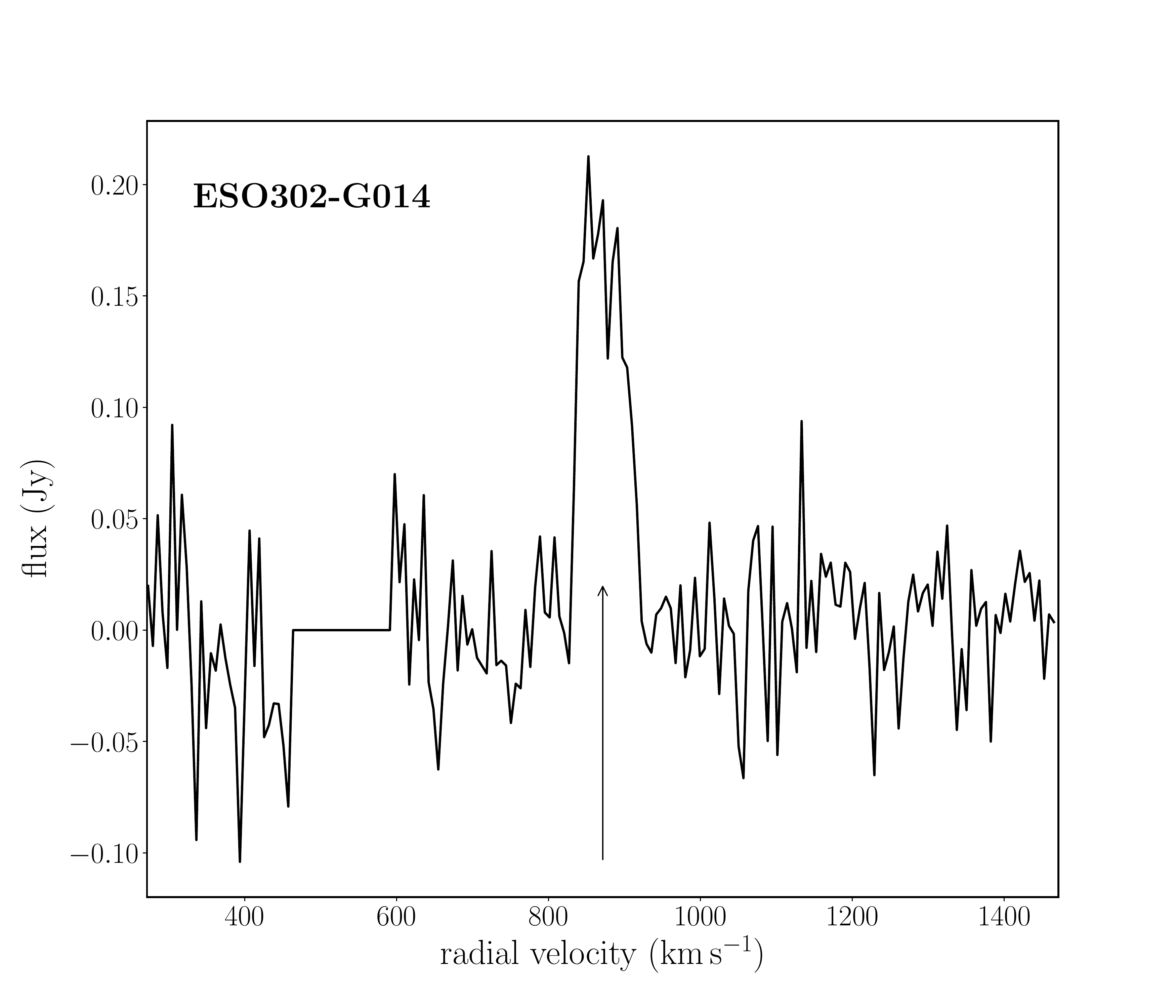}
    \caption{ESO 302-G014 ({\it centre}) and secondary source ESO 302-G009 ({\it right}).}
    \label{app:eso302-g014}
\end{figure*}
\begin{figure*}
	\includegraphics[width=\columnwidth]{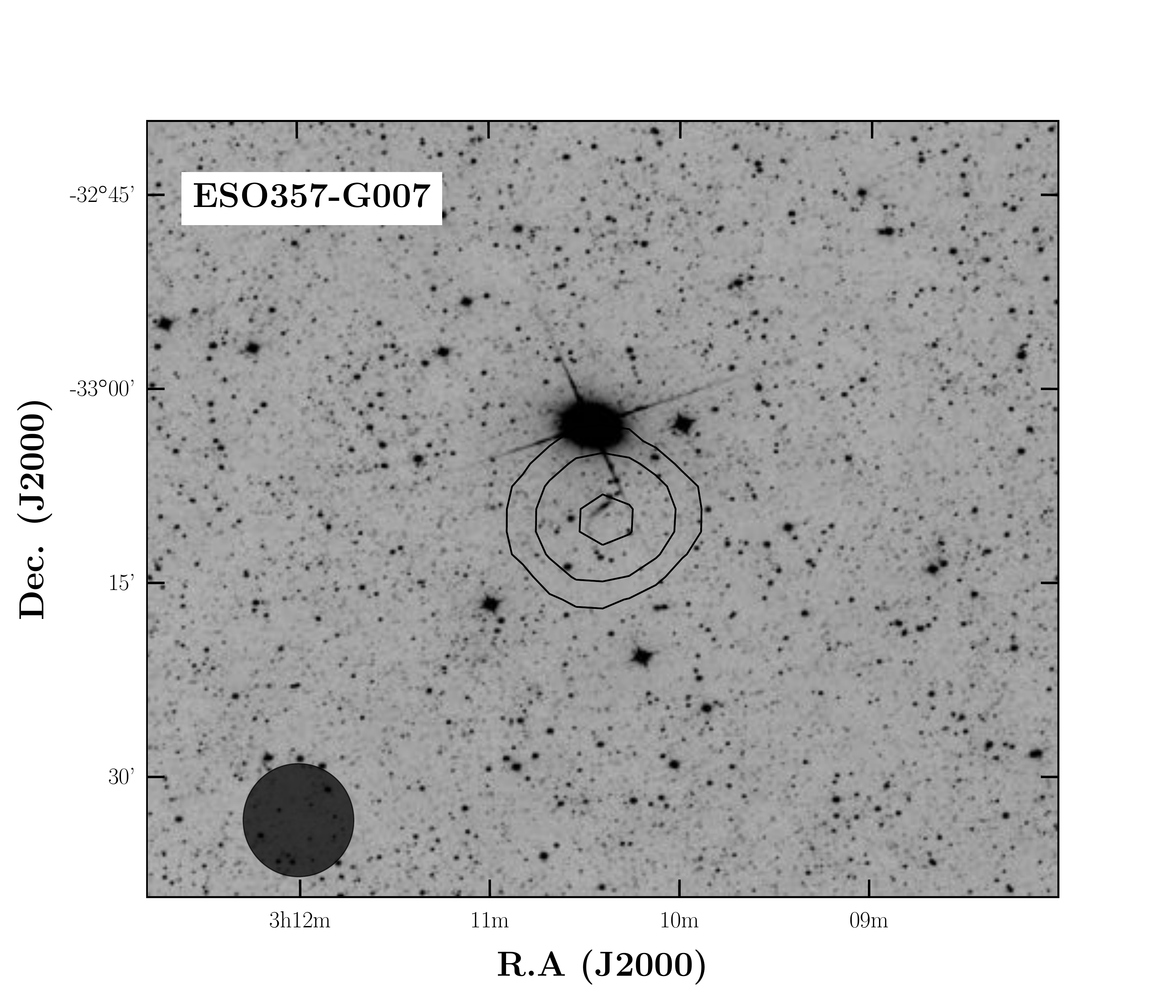}
	\includegraphics[width=\columnwidth]{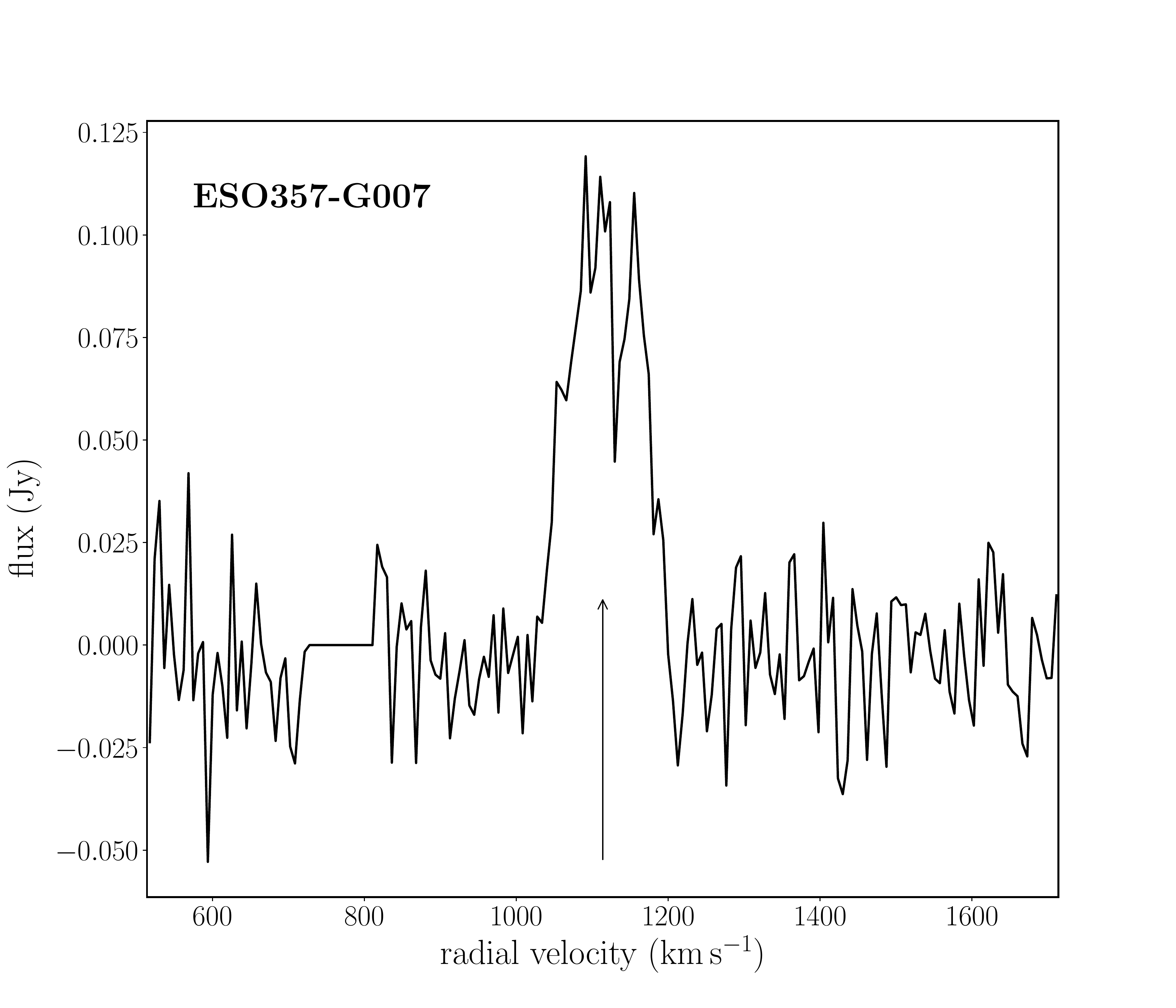}
    \caption{ESO 357-G007}
    \label{app:eso357-g007}
\end{figure*}
\begin{figure*}
	\includegraphics[width=\columnwidth]{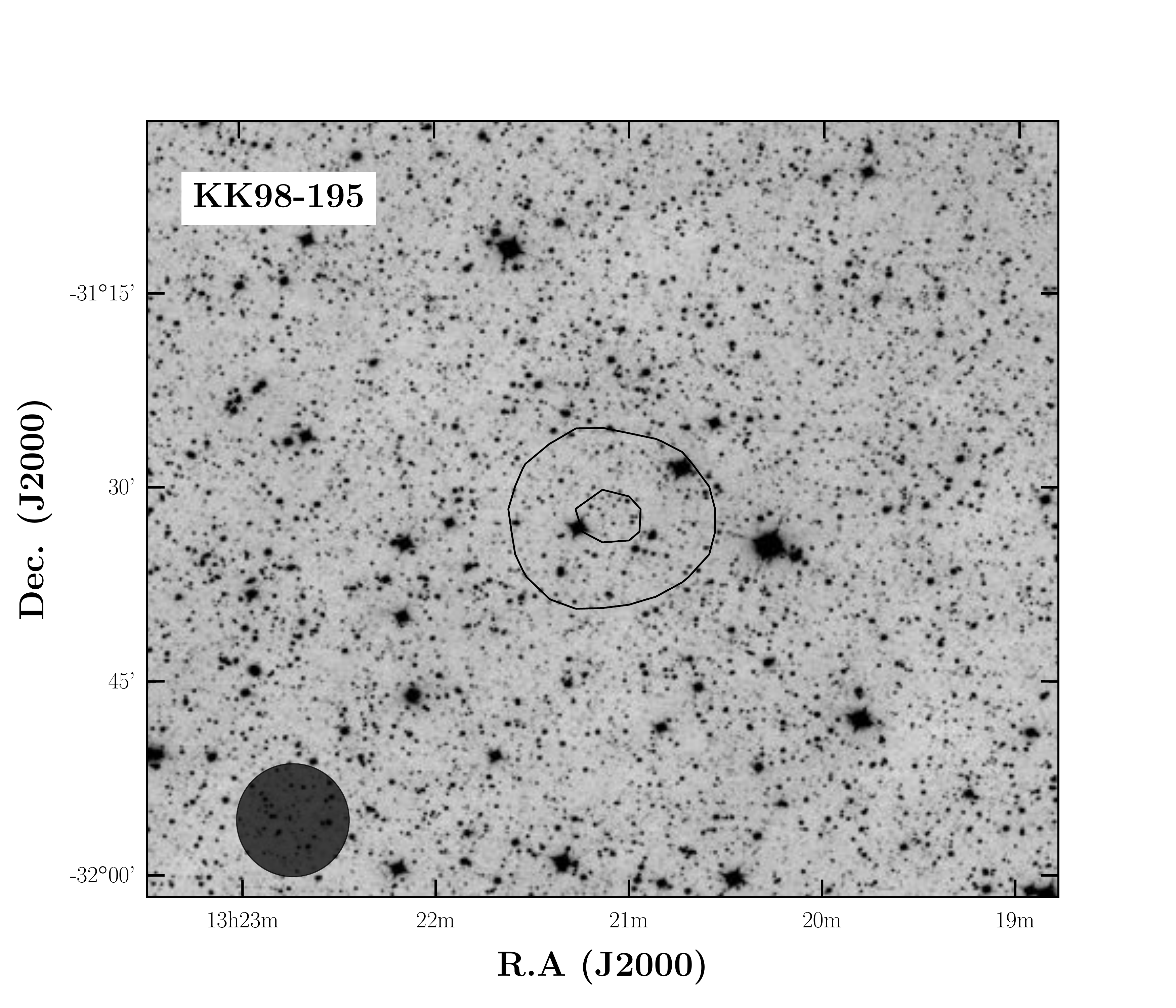}
	\includegraphics[width=\columnwidth]{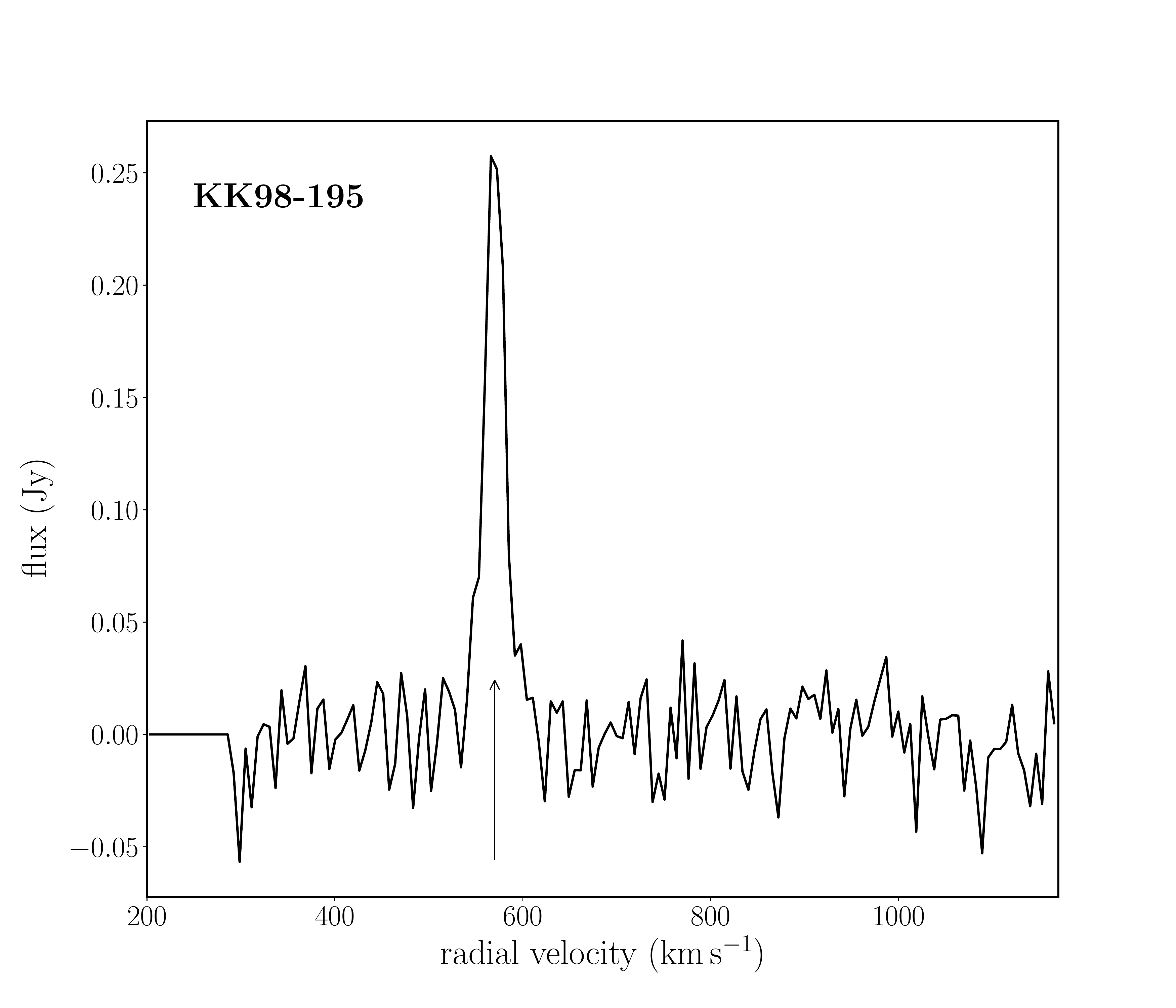}
    \caption{KK 98-195}
    \label{app:kk98-195}
\end{figure*}
\begin{figure*}
	\includegraphics[width=\columnwidth]{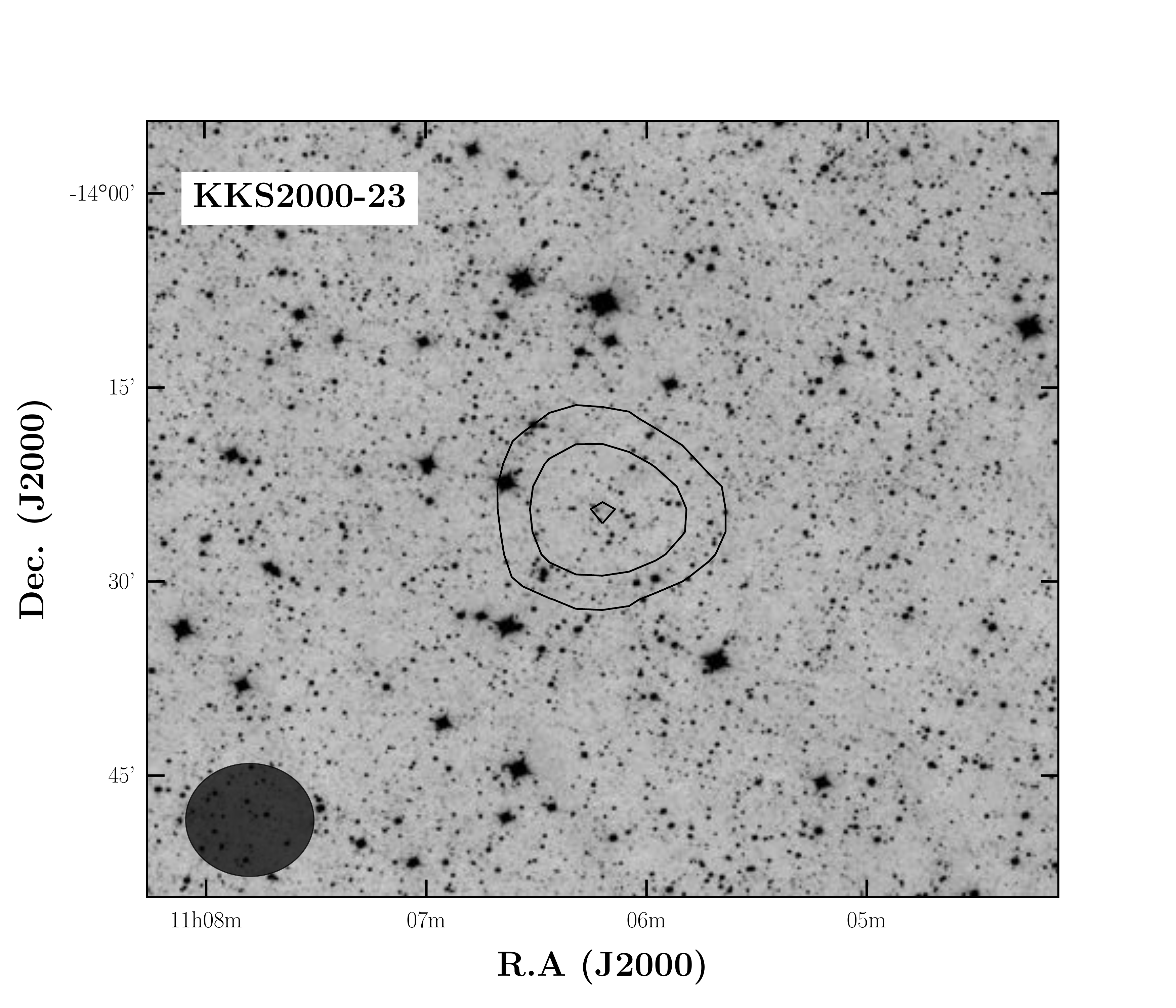}
	\includegraphics[width=\columnwidth]{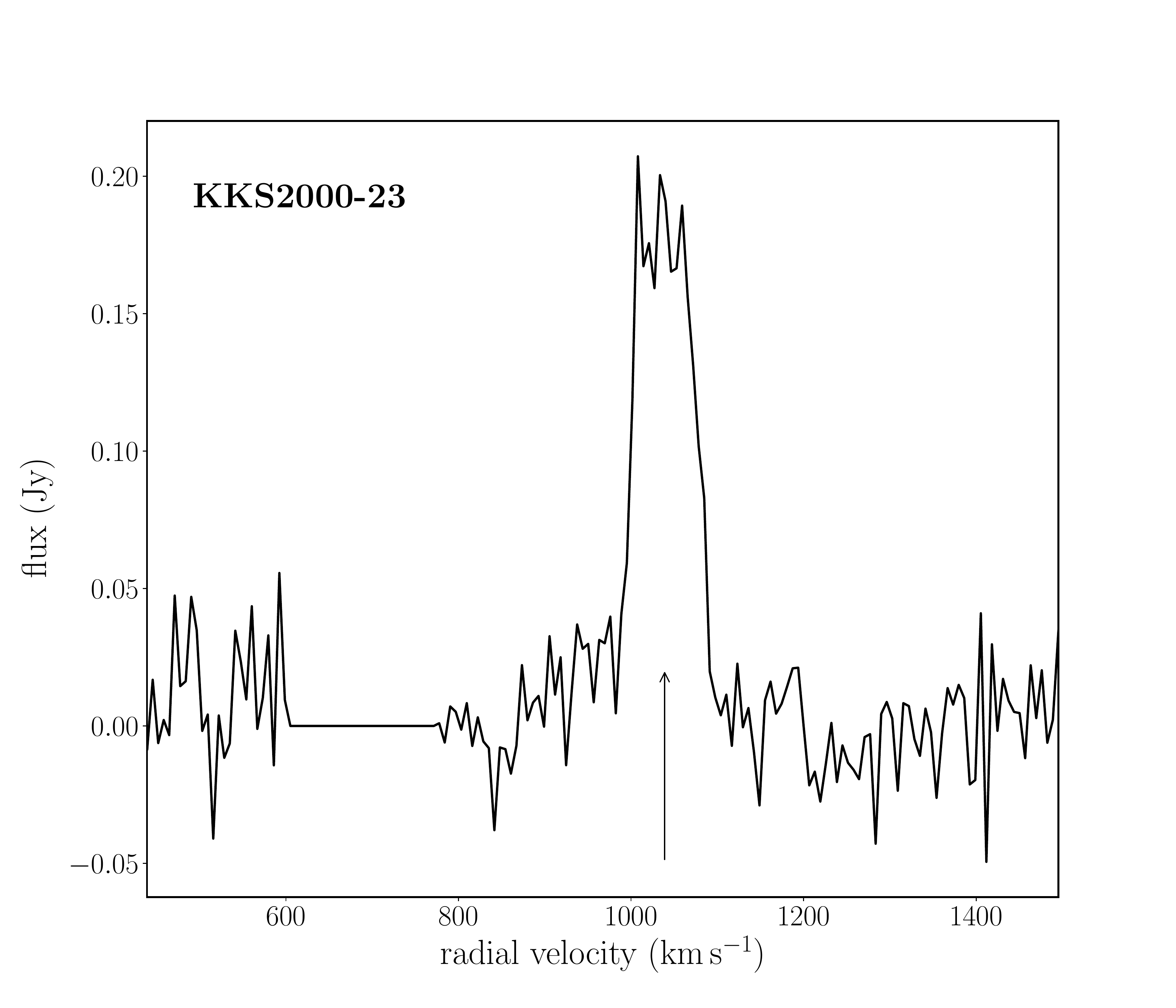}
    \caption{KKS 2000-23}
    \label{app:kks2000-23}
\end{figure*}
\begin{figure*}
	\includegraphics[width=\columnwidth]{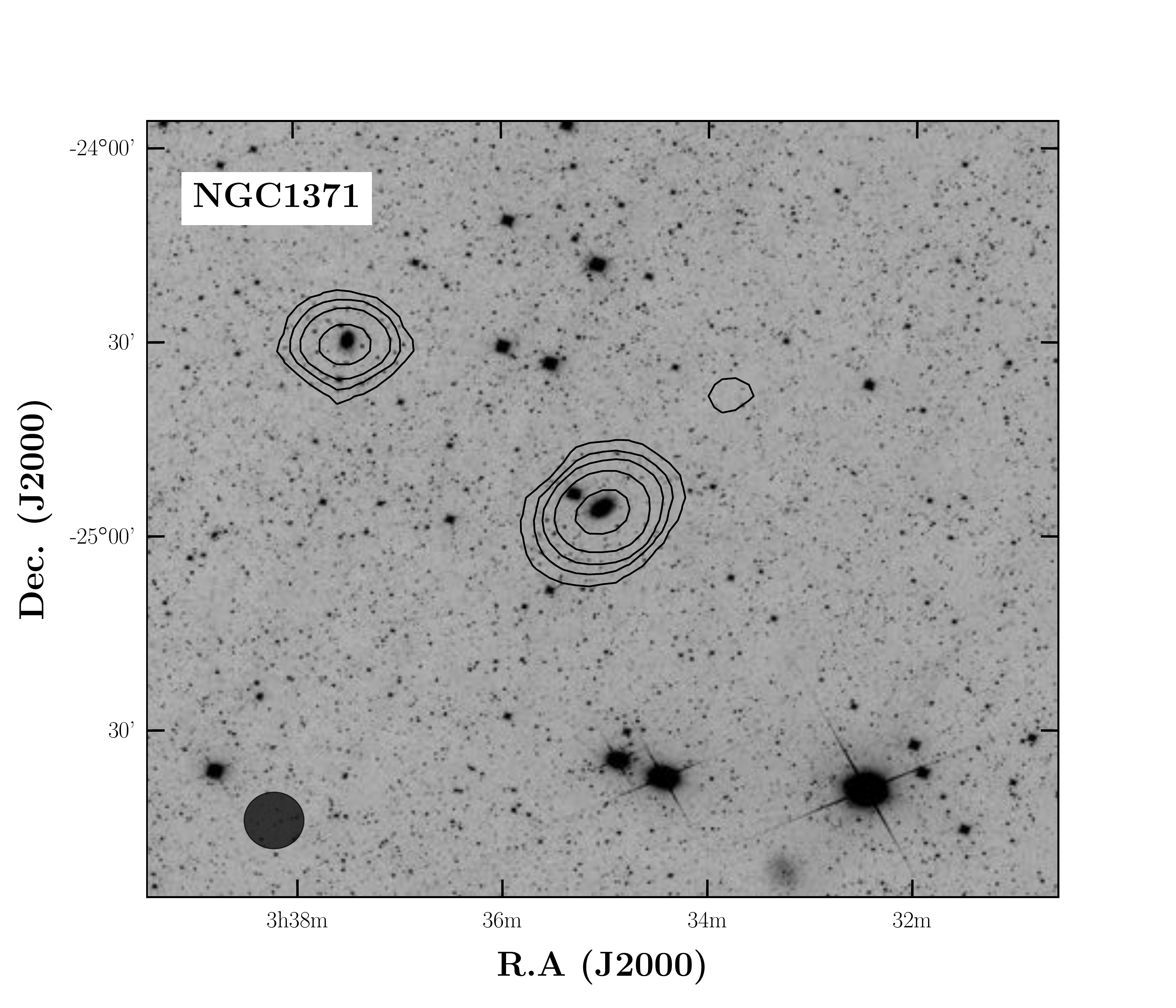}
	\includegraphics[width=\columnwidth]{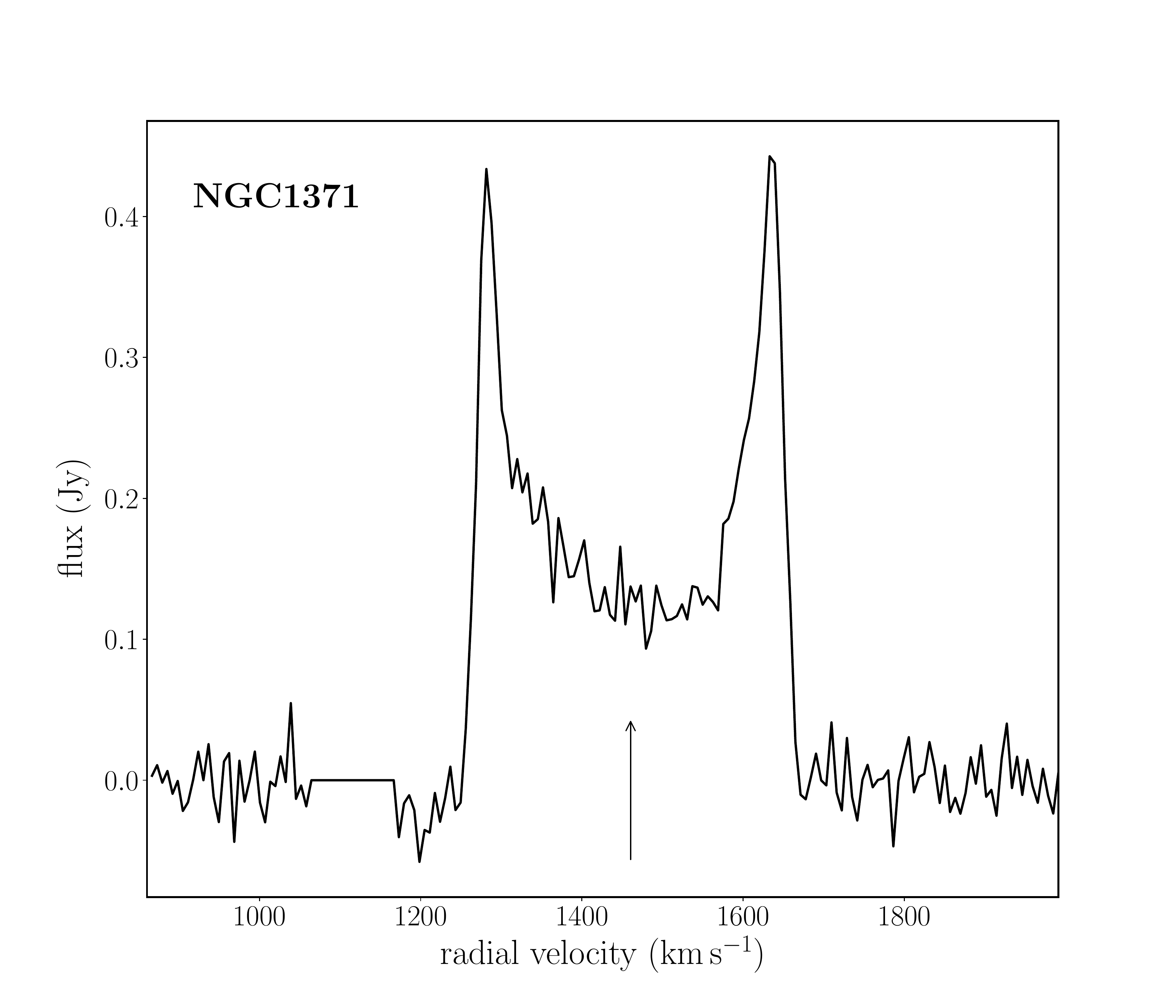}
    \caption{NGC 1371 ({\it centre}) and secondary source NGC 1385 ({\it upper left contours}).}
    \label{app:ngc1371}
\end{figure*}
\begin{figure*}
	\includegraphics[width=\columnwidth]{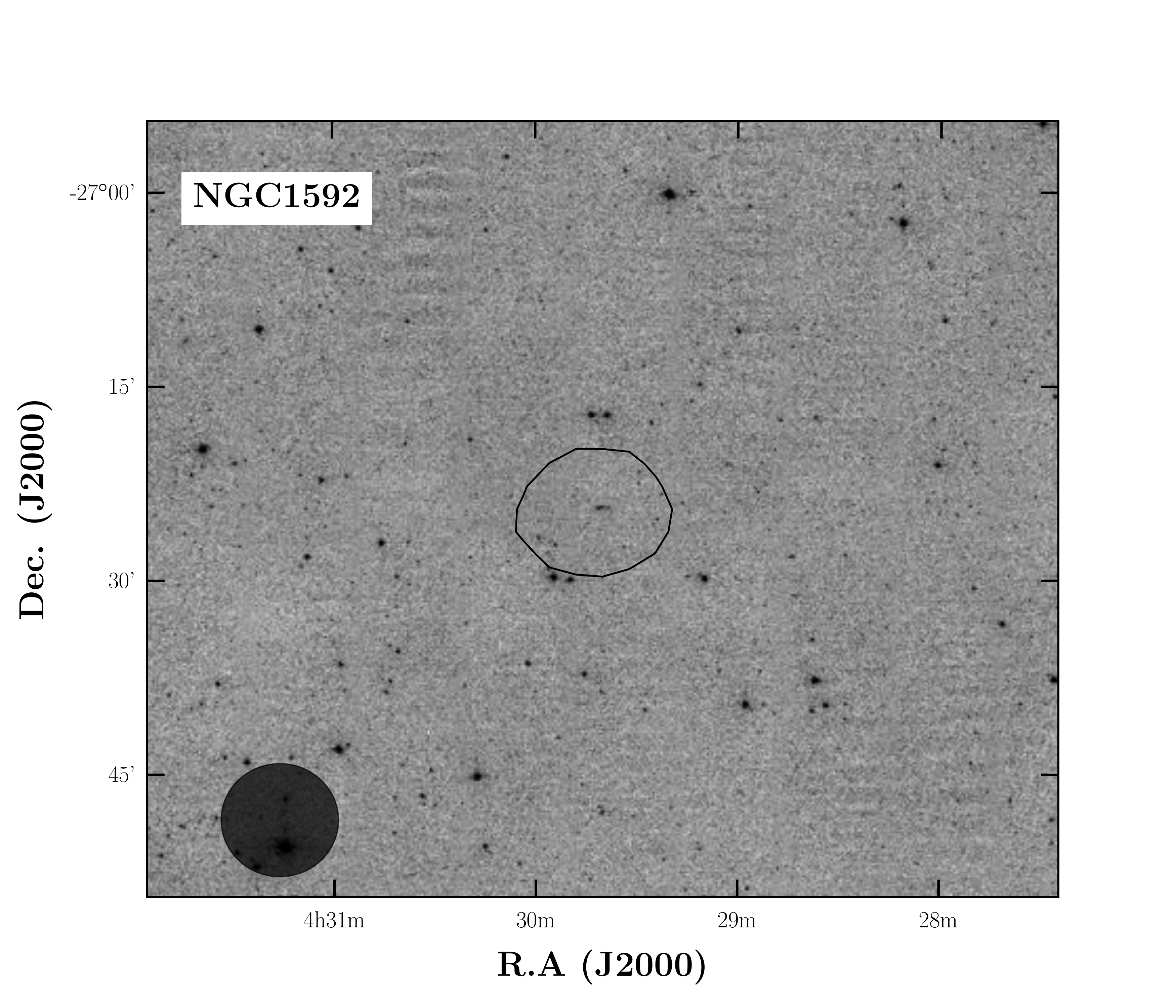}
	\includegraphics[width=\columnwidth]{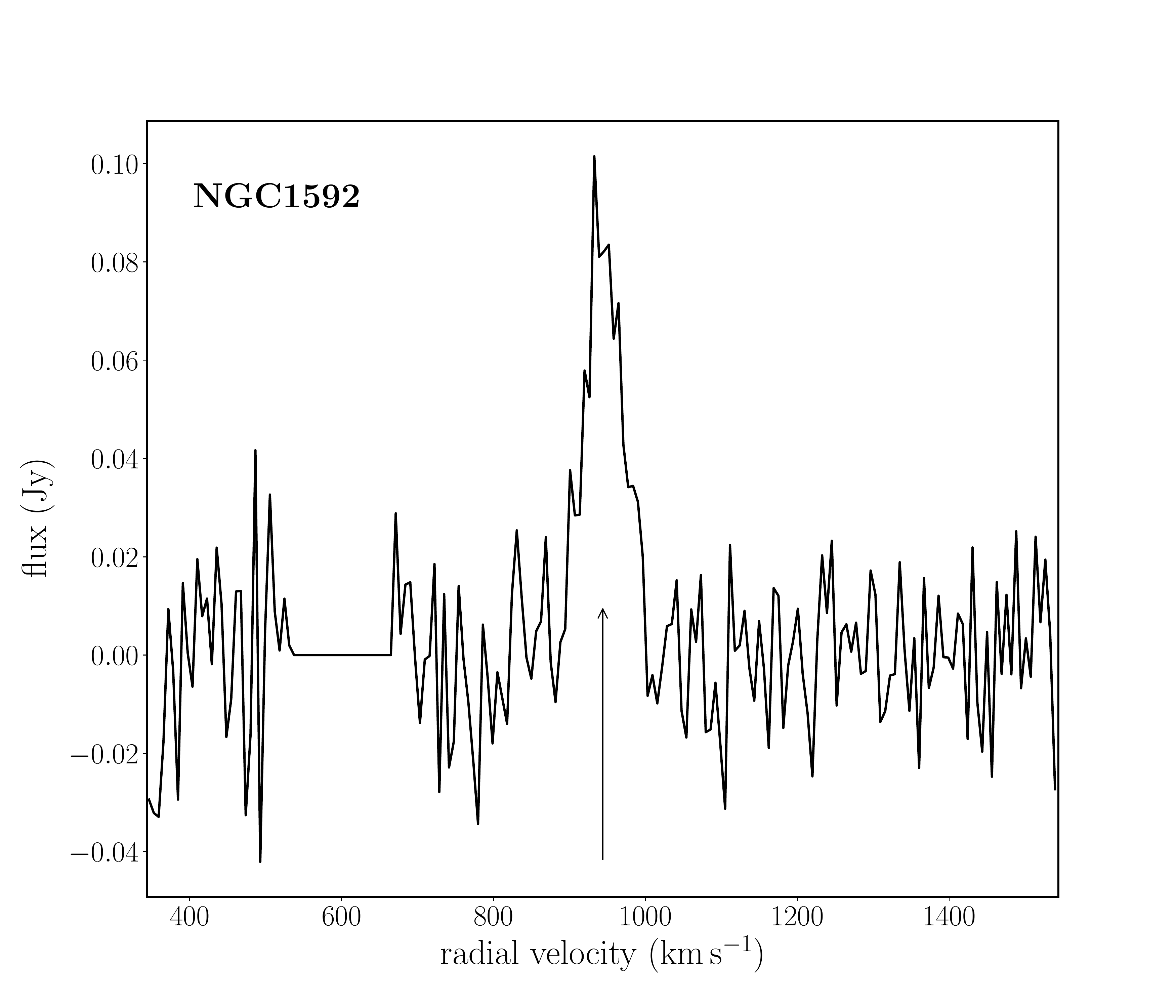}
    \caption{NGC 1592}
    \label{app:ngc1592}
\end{figure*}
\begin{figure*}
	\includegraphics[width=\columnwidth]{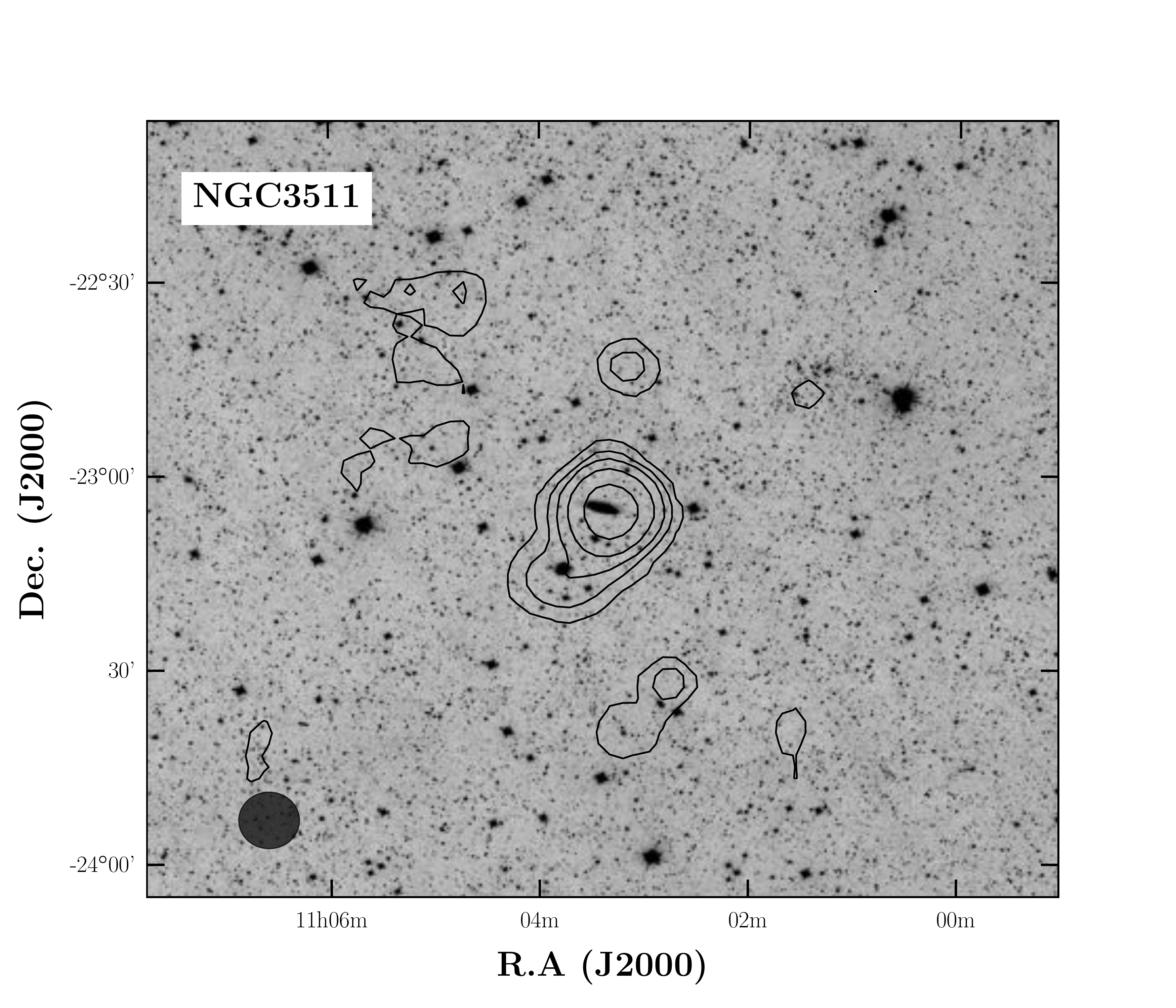}
	\includegraphics[width=\columnwidth]{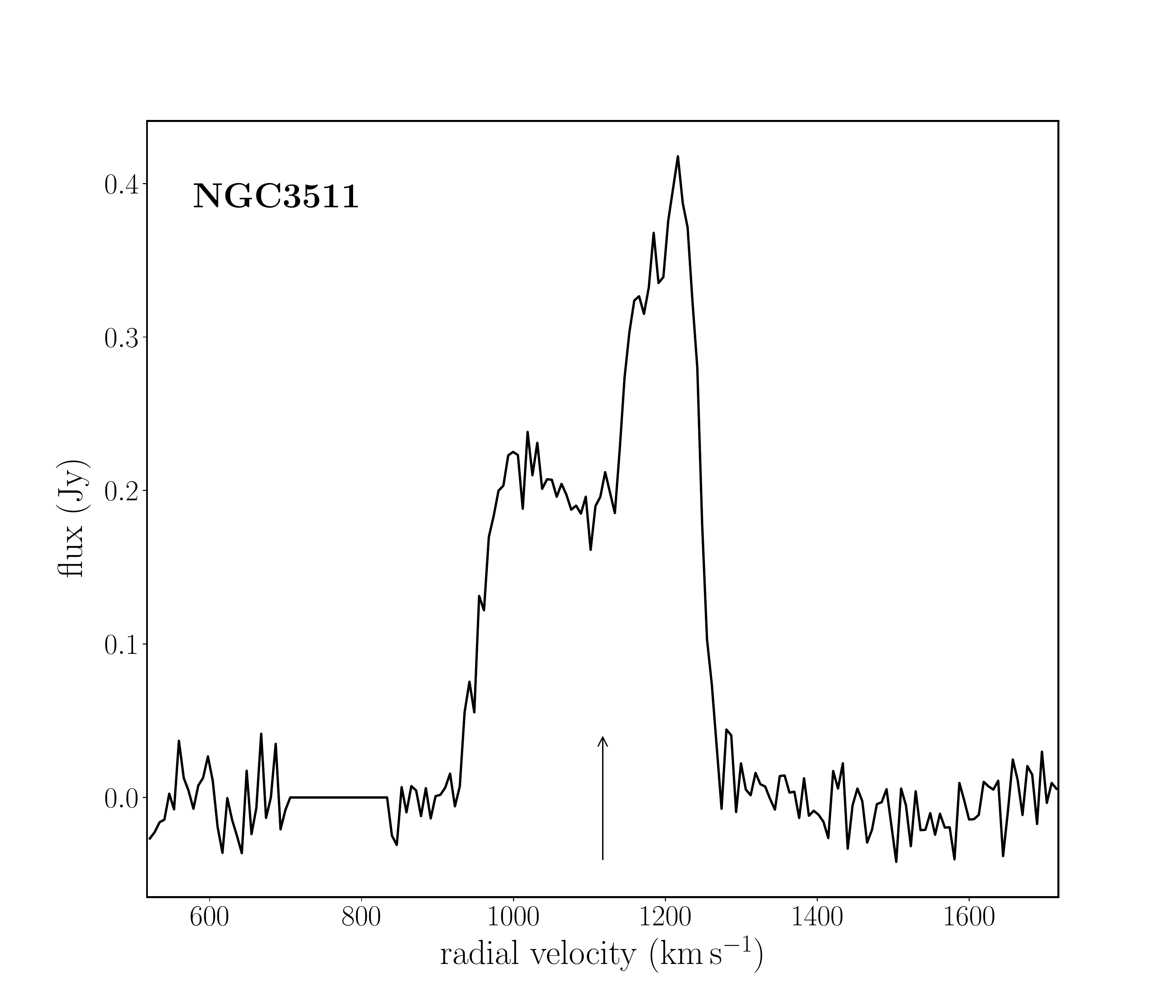}
    \caption{The GBT-confused sources NGC 3511 ({\it centre}) and NC 3513 ({\it lower left}). The contours in the outskirts do not have reported optical counterparts.}
    \label{app:ngc3511}
\end{figure*}
\begin{figure*}
	\includegraphics[width=\columnwidth]{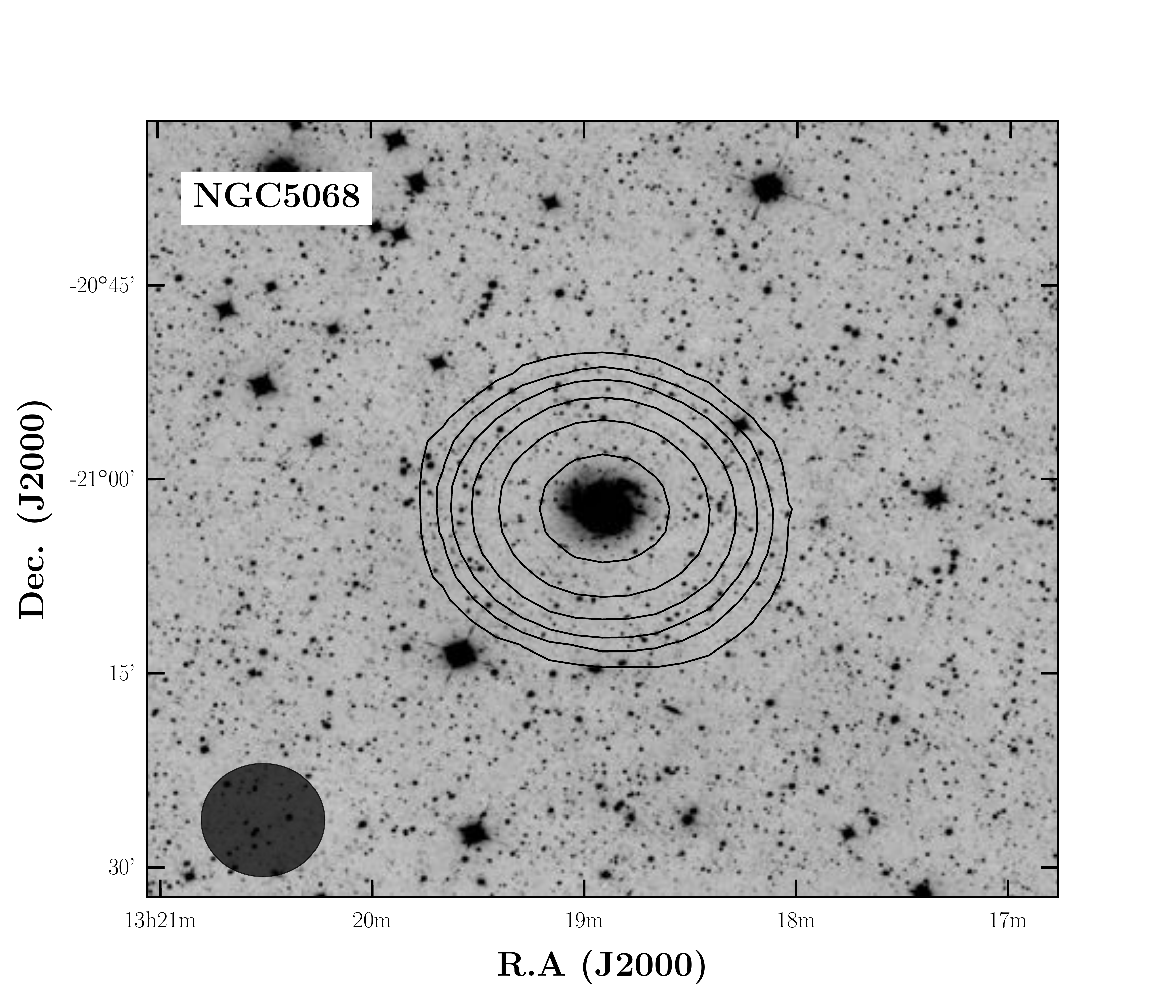}
	\includegraphics[width=\columnwidth]{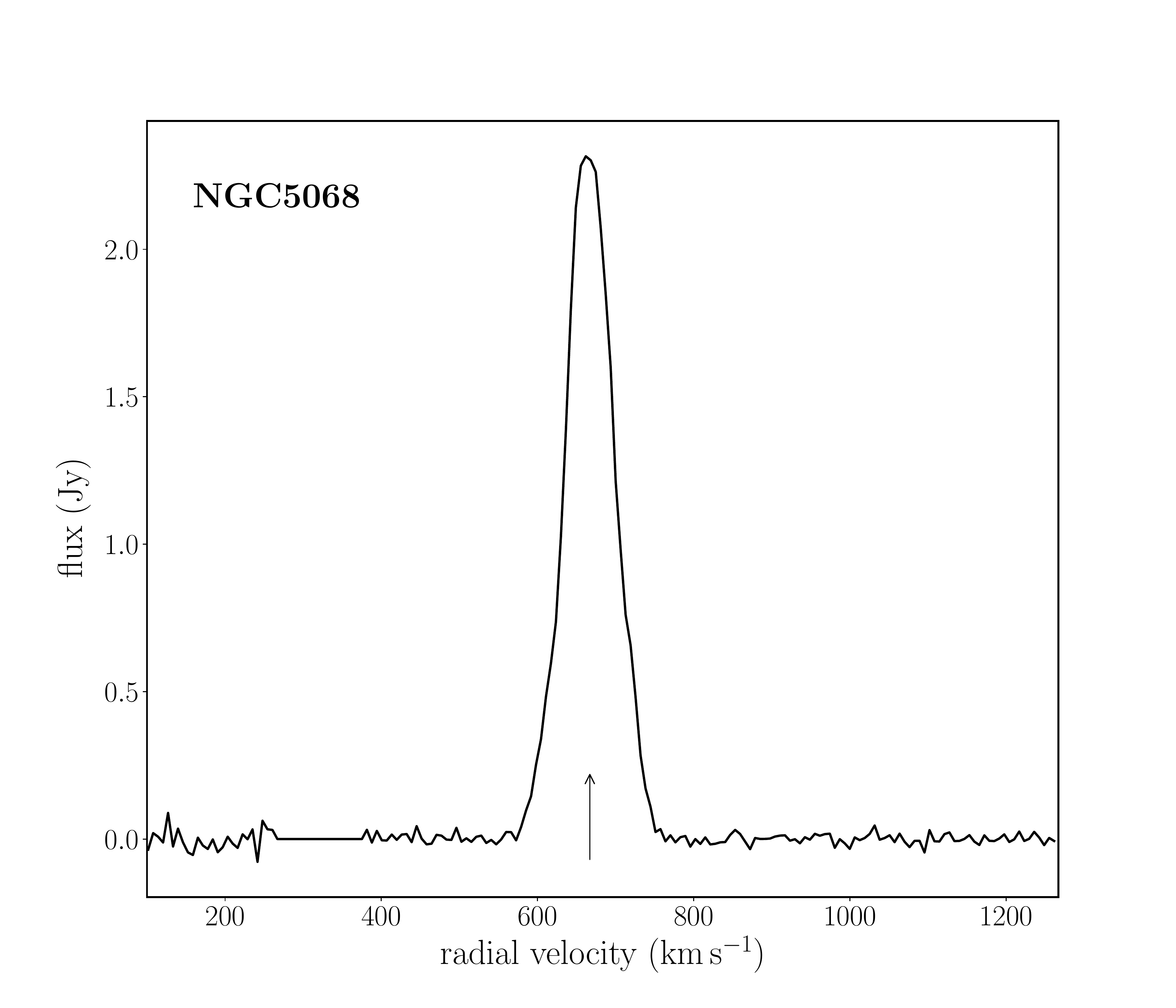}
    \caption{NGC 5068}
    \label{app:ngc5068}
\end{figure*}
\begin{figure*}
	\includegraphics[width=\columnwidth]{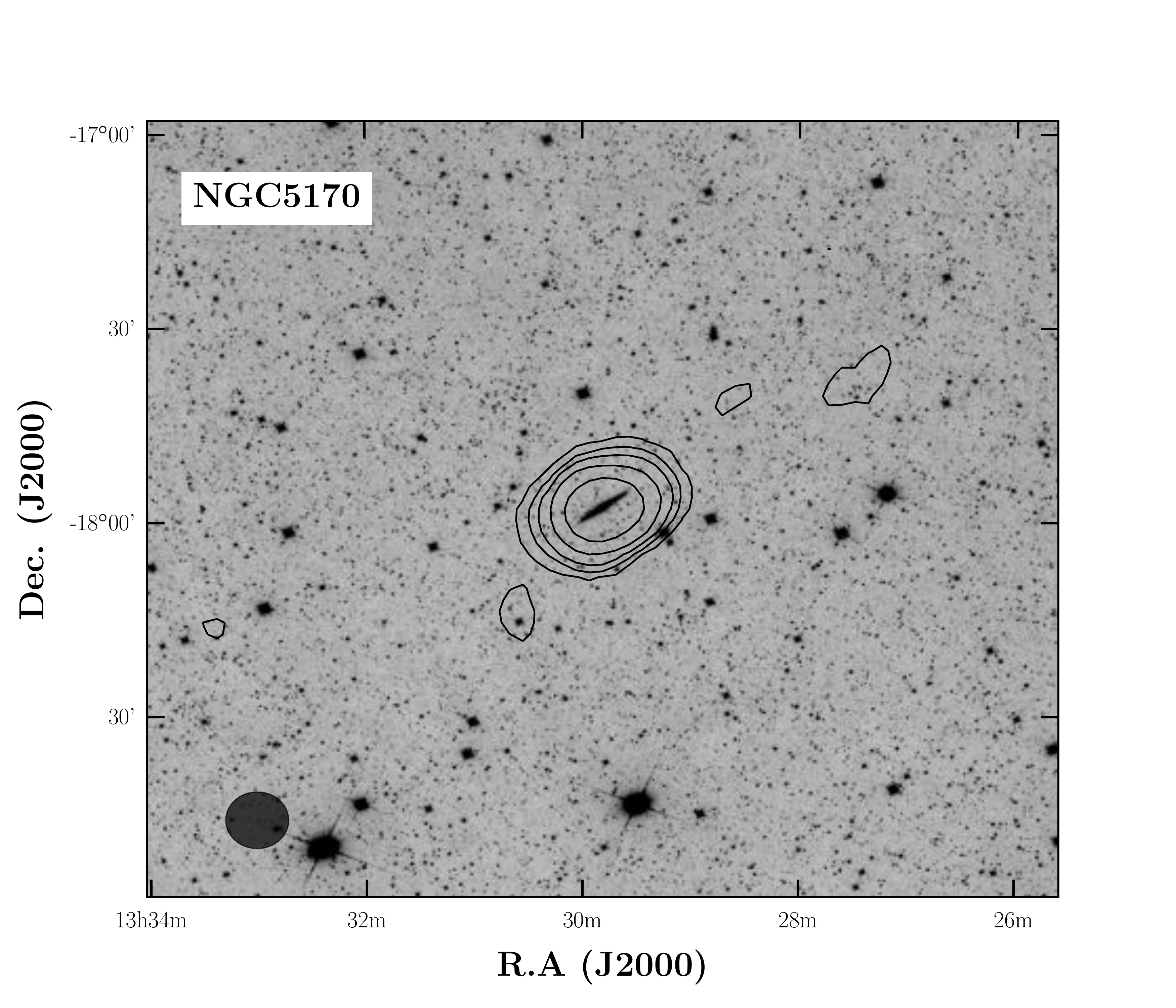}
	\includegraphics[width=\columnwidth]{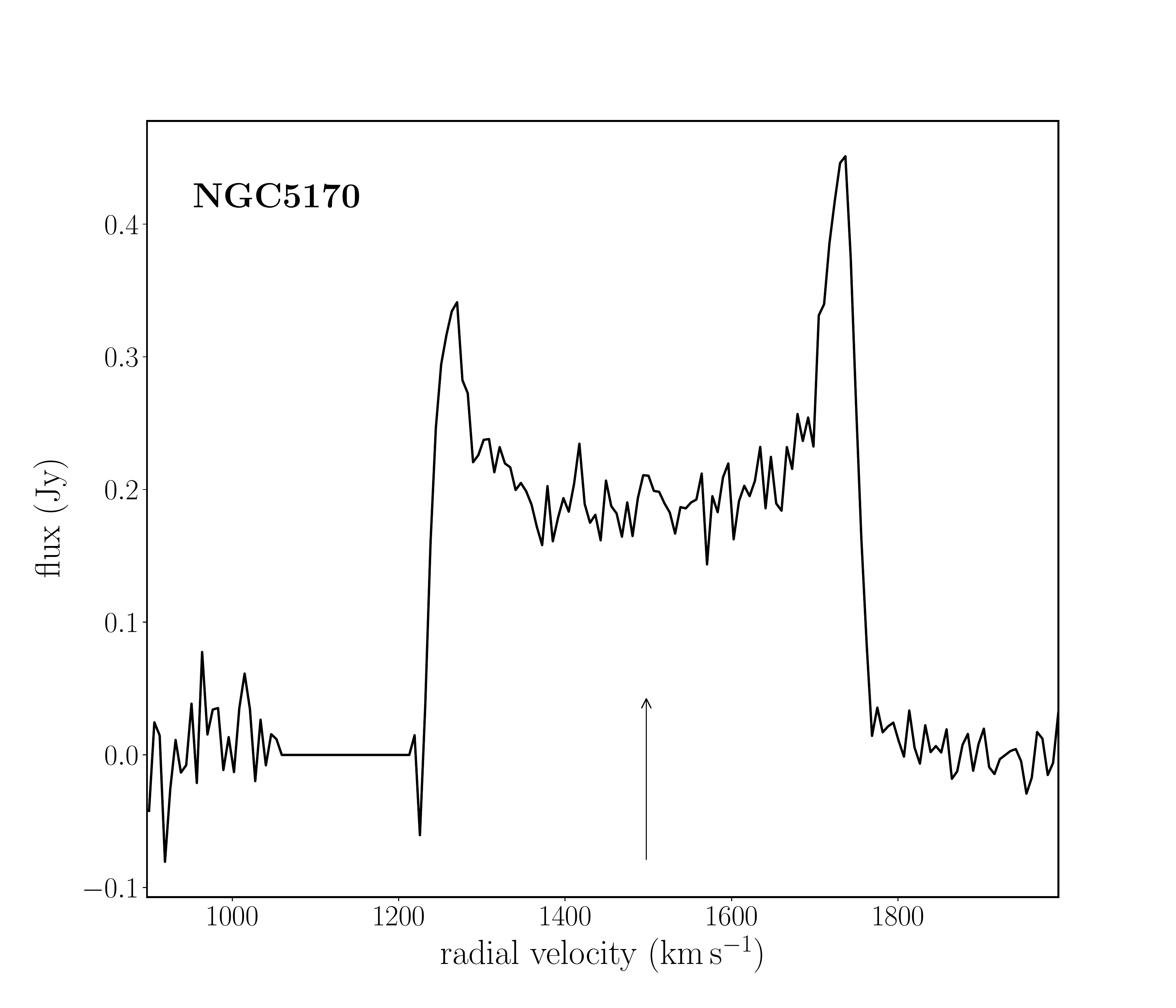}
    \caption{NGC 5170}
    \label{app:ngc5170}
\end{figure*}
\begin{figure*}
	\includegraphics[width=\columnwidth]{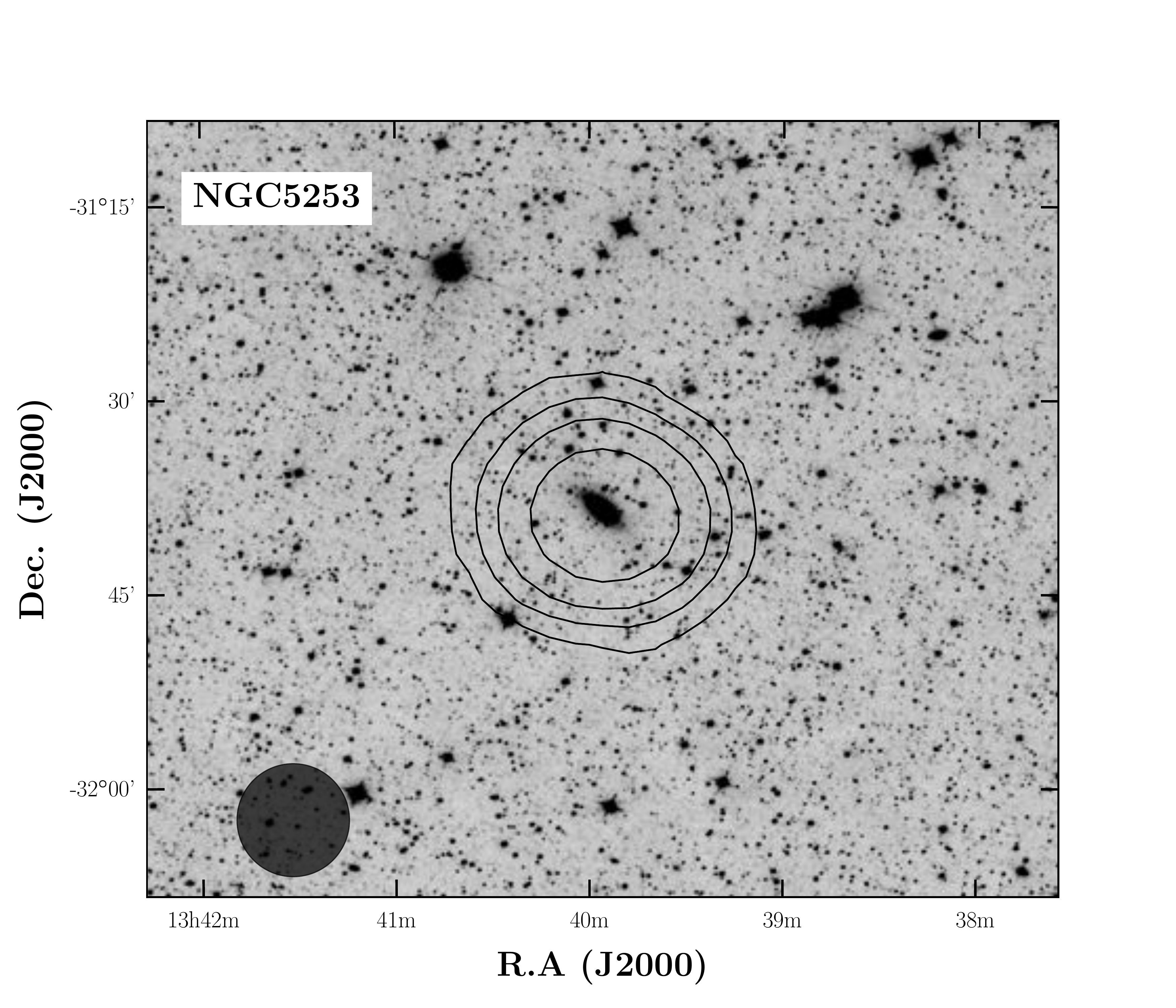}
	\includegraphics[width=\columnwidth]{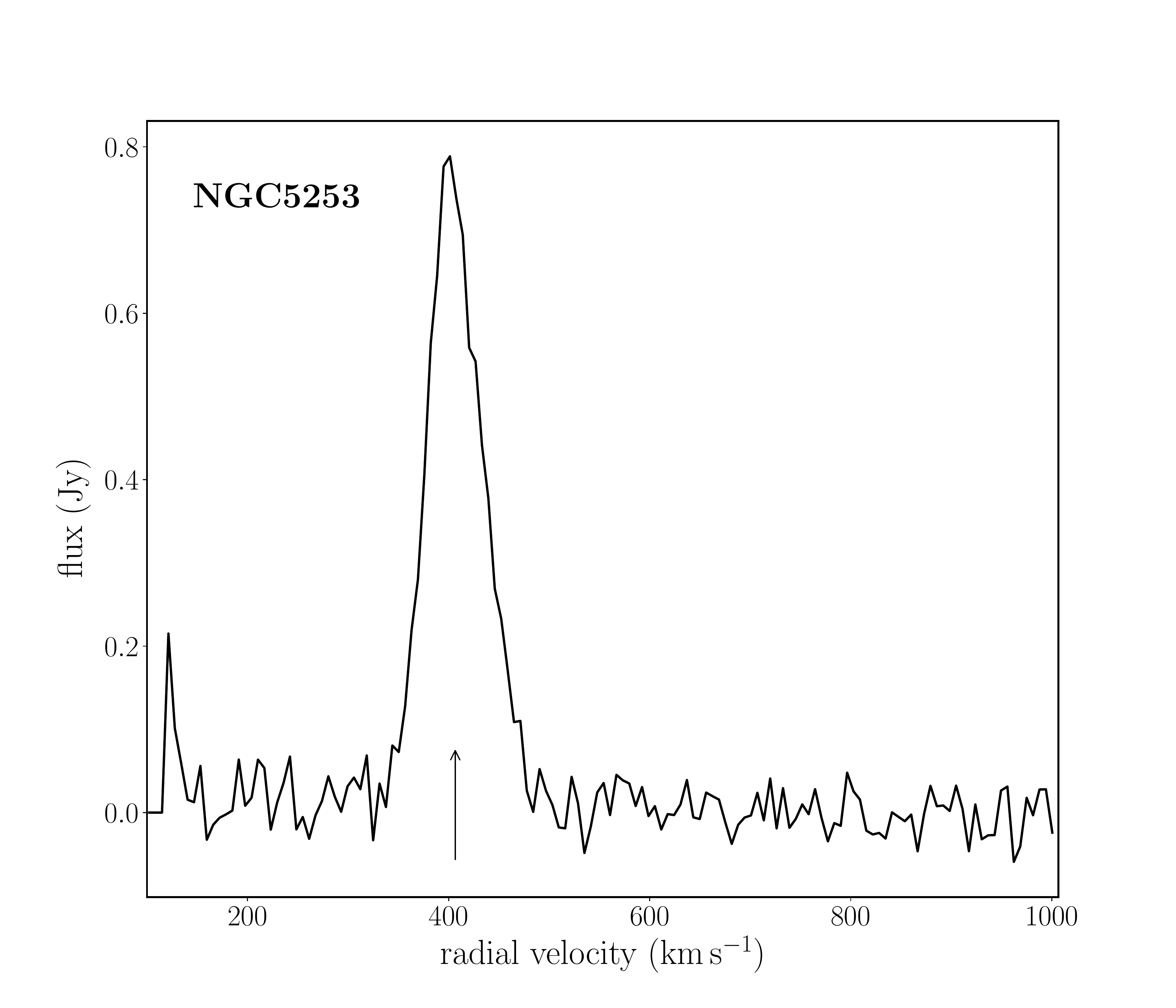}
    \caption{NGC 5253}
    \label{app:ngc5253}
\end{figure*}
\begin{figure*}
	\includegraphics[width=\columnwidth]{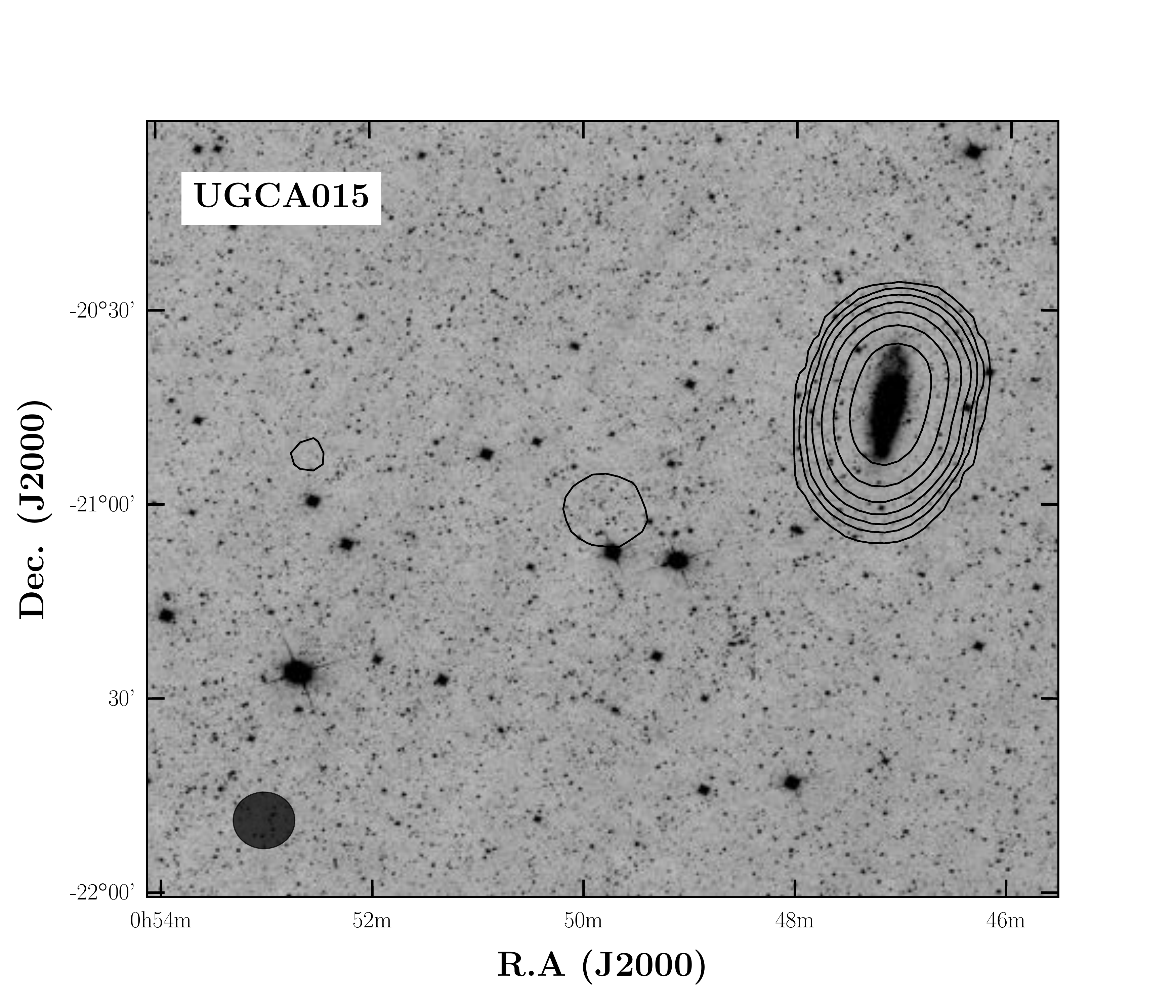}
	\includegraphics[width=\columnwidth]{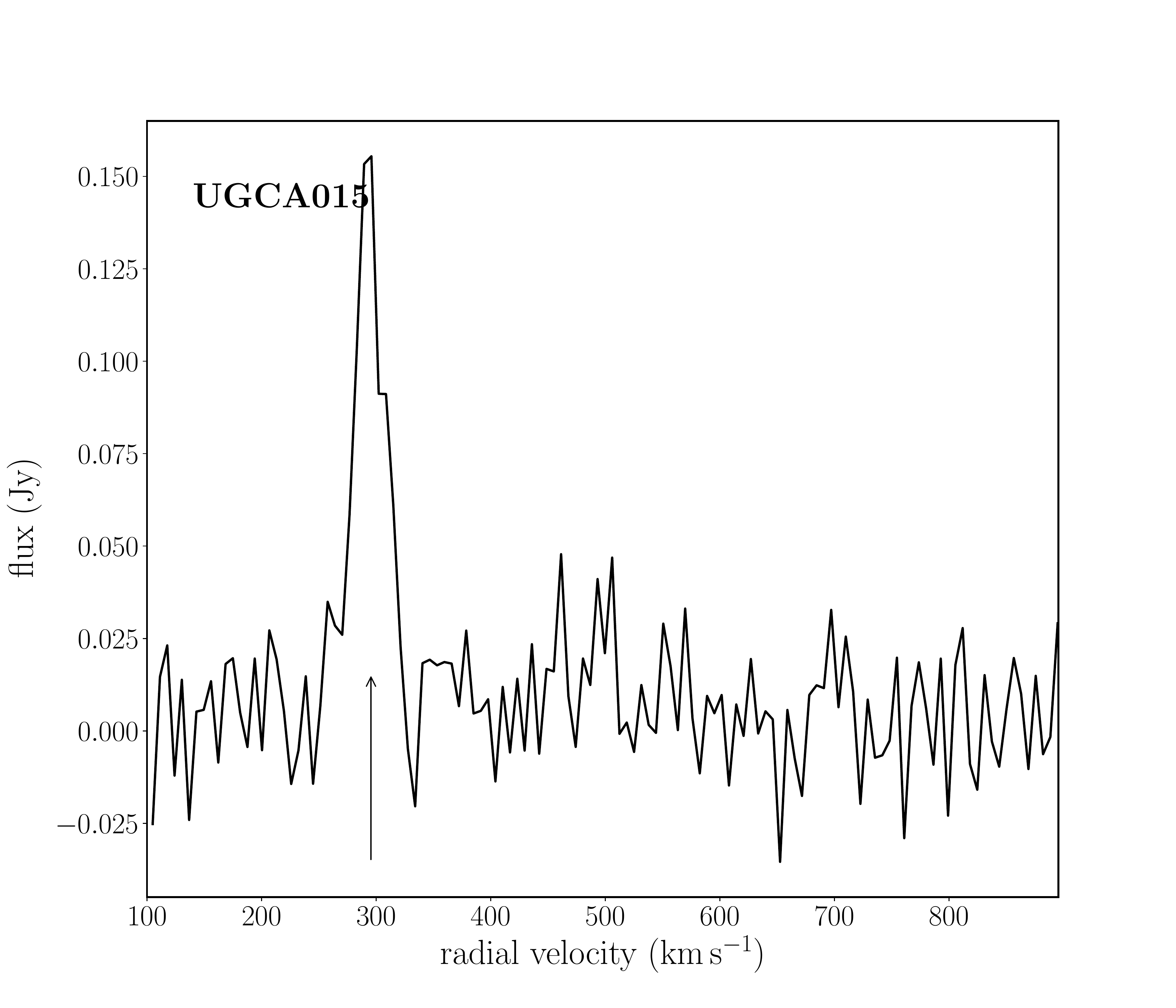}
    \caption{UGCA 015 ({\it centre}) and the secondary source NGC 247 ({\it right}).}
    \label{app:ugca015}
\end{figure*}
\begin{figure*}
	\includegraphics[width=\columnwidth]{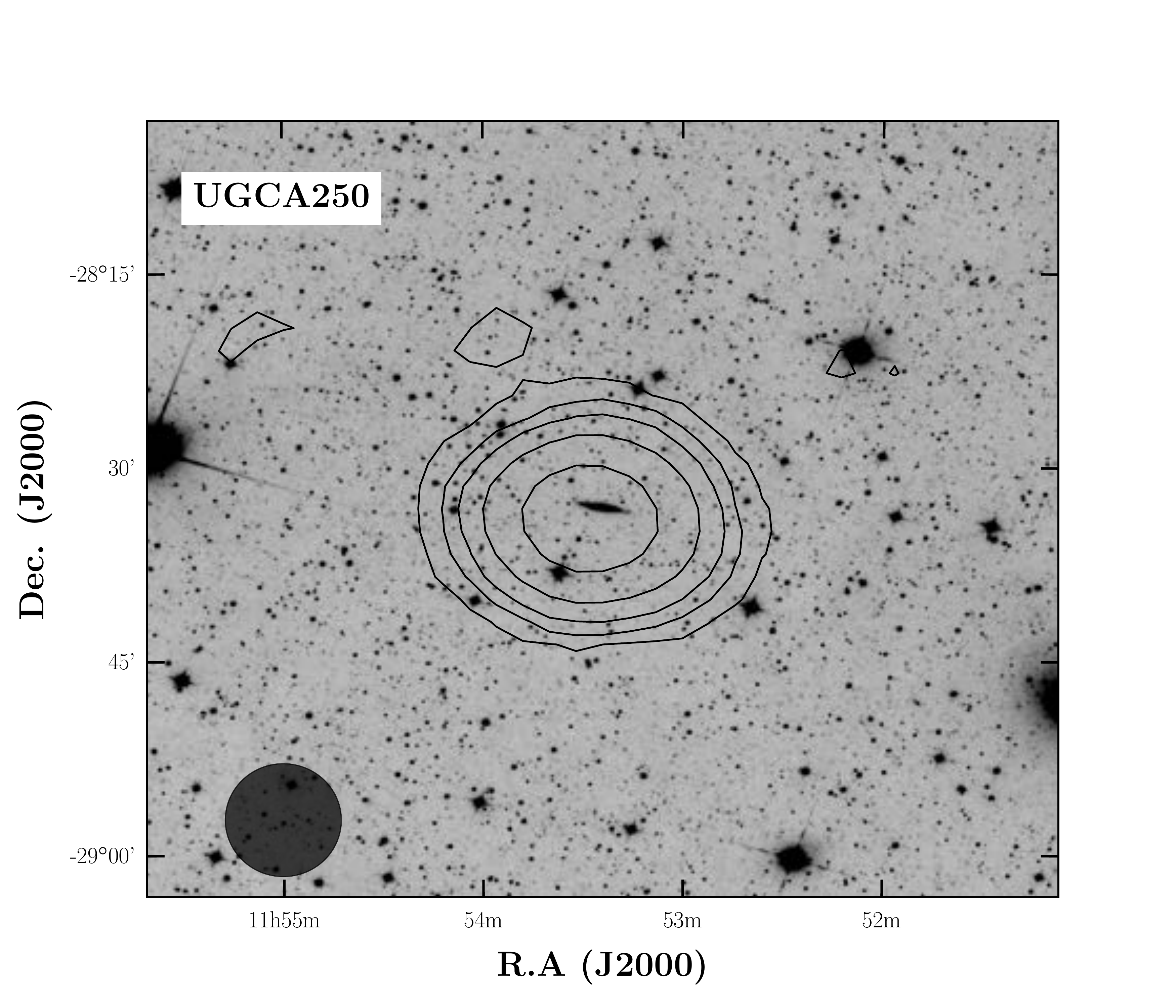}
	\includegraphics[width=\columnwidth]{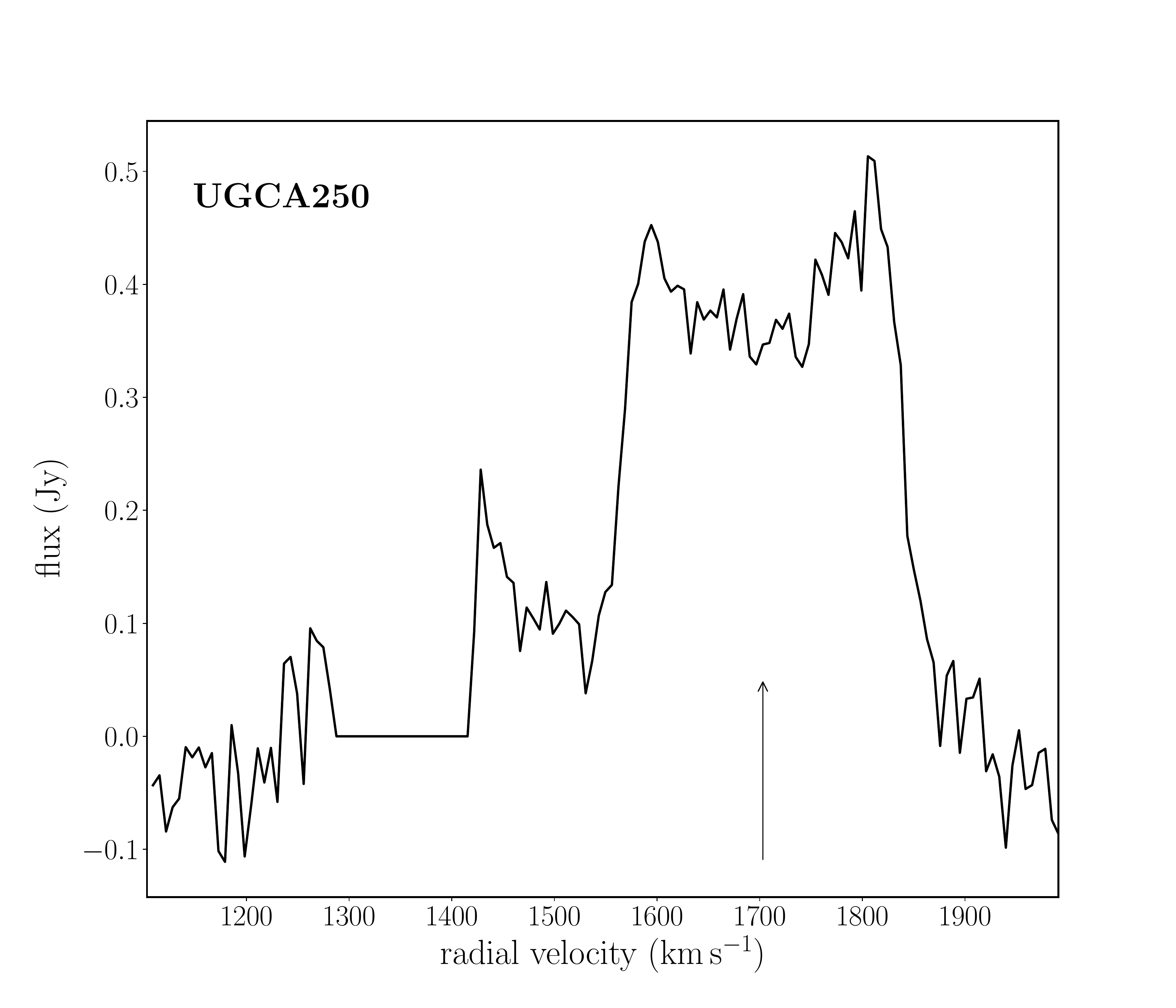}
    \caption{UGCA 250}
    \label{app:ugca250}
\end{figure*}
\begin{figure*}
	\includegraphics[width=\columnwidth]{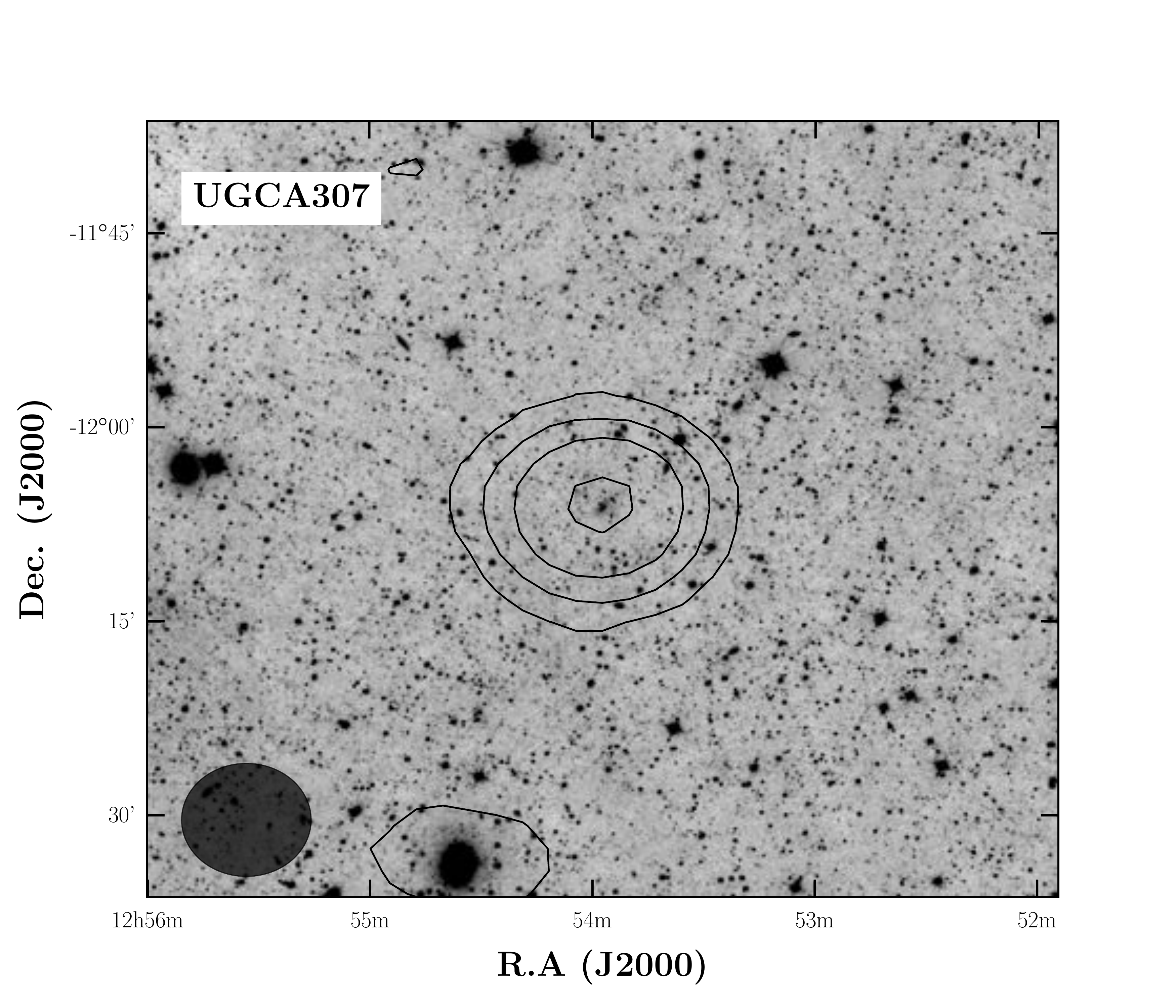}
	\includegraphics[width=\columnwidth]{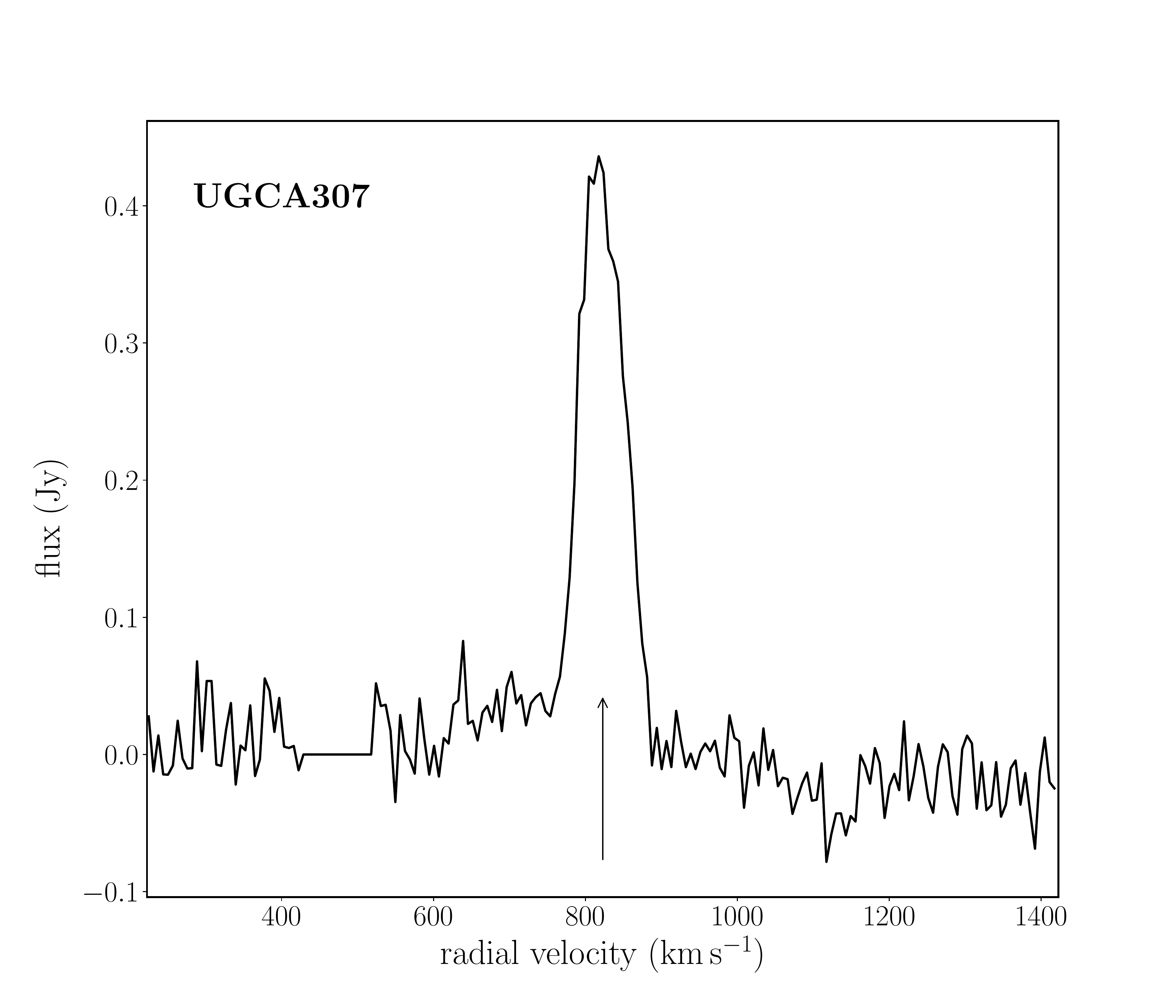}
    \caption{UGCA 307}
    \label{app:ugca307}
\end{figure*}
\begin{figure*}
	\includegraphics[width=\columnwidth]{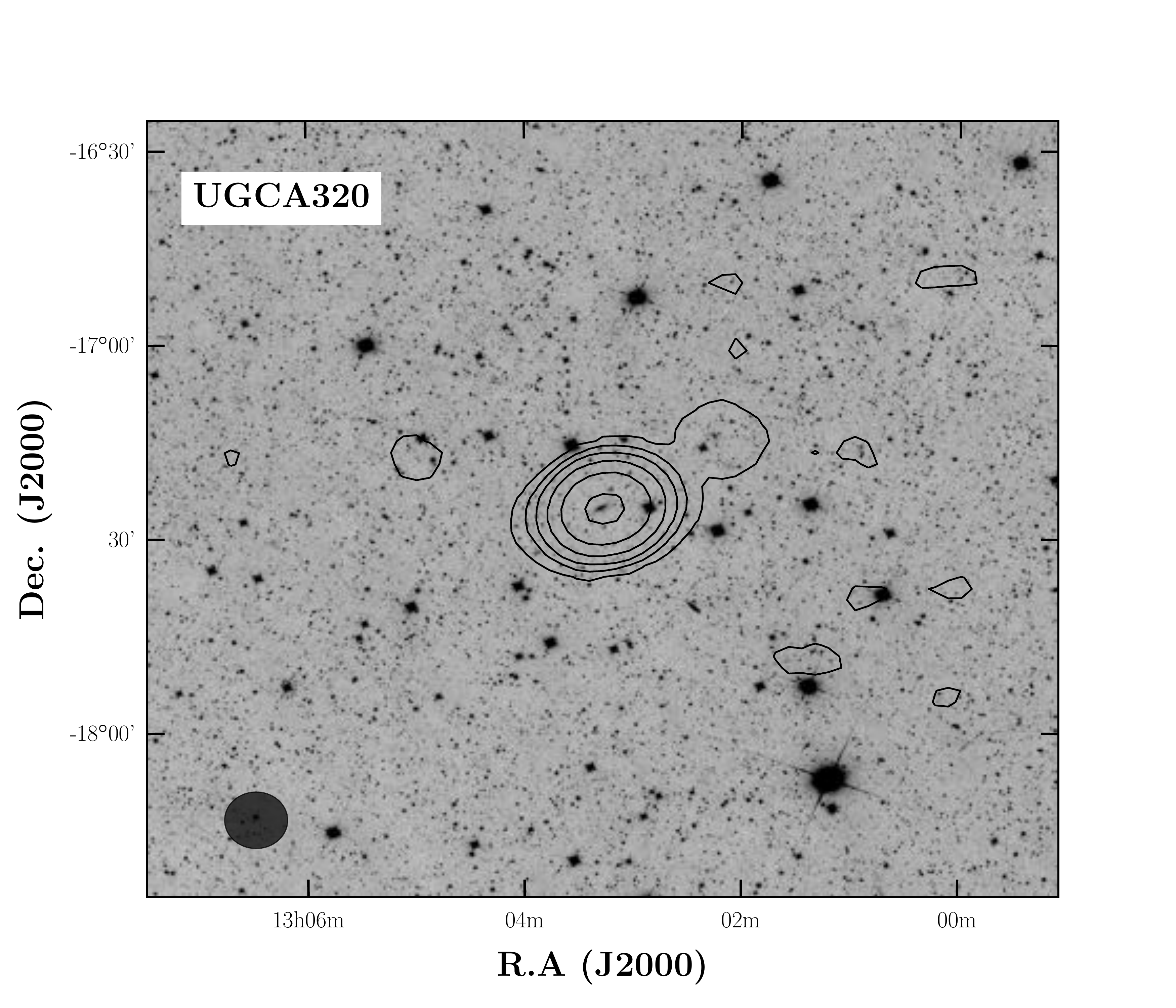}
	\includegraphics[width=\columnwidth]{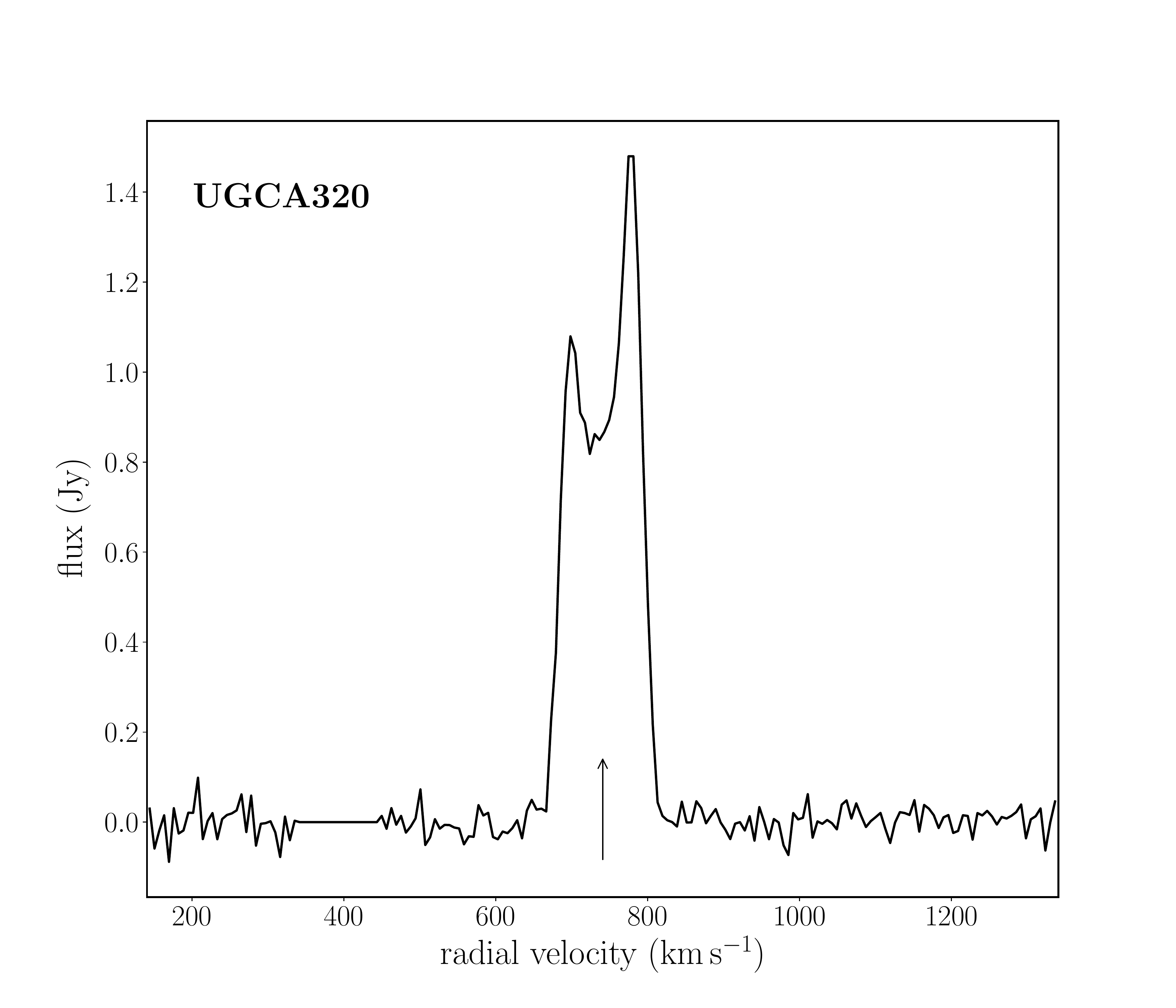}
    \caption{The central contours represent the position of UGCA 320, while the adjacent contour on the upper right side shows the galaxy UGCA 319.}
    \label{app:ugca320}
\end{figure*}

\begin{figure*}
	\includegraphics[width=\columnwidth]{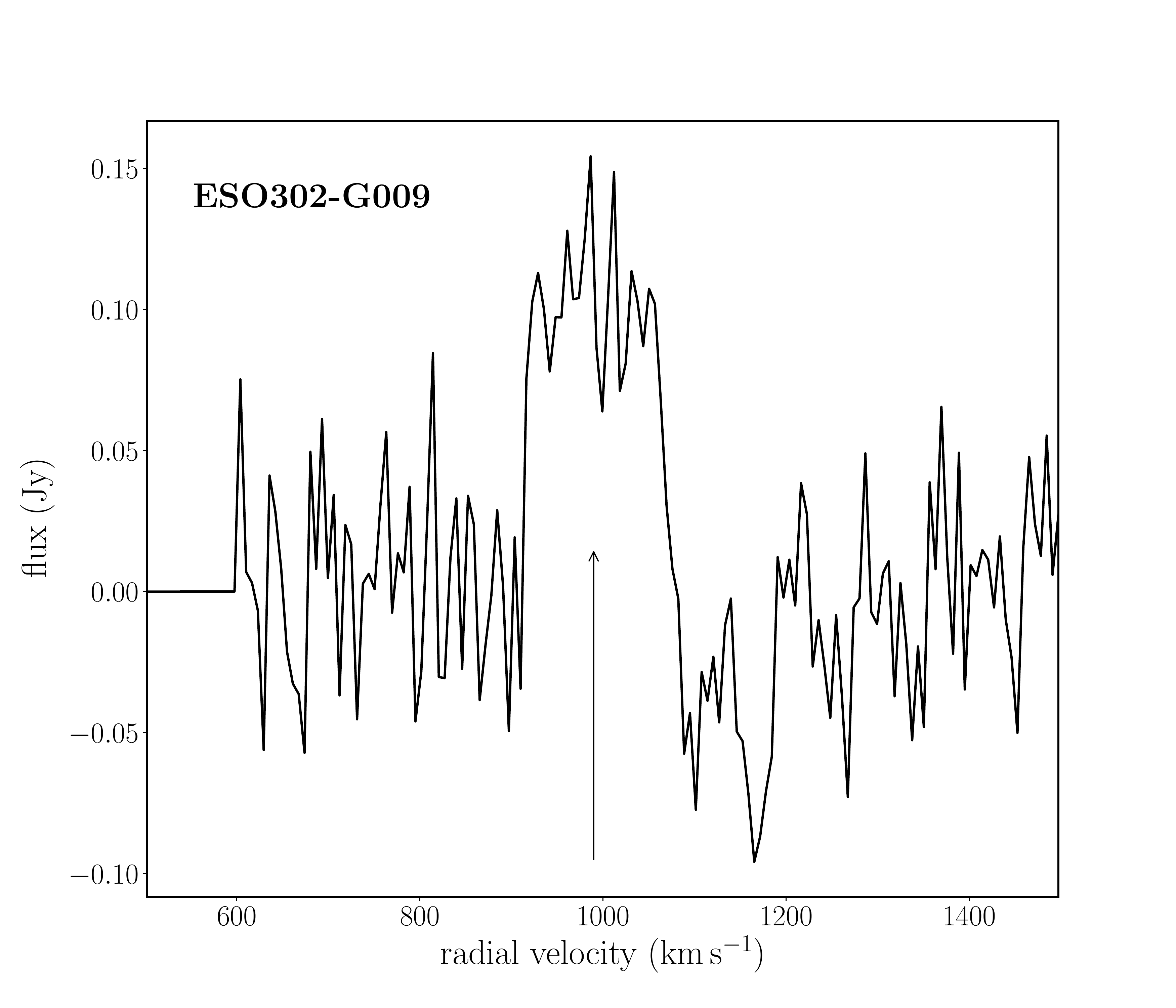}
	\includegraphics[width=\columnwidth]{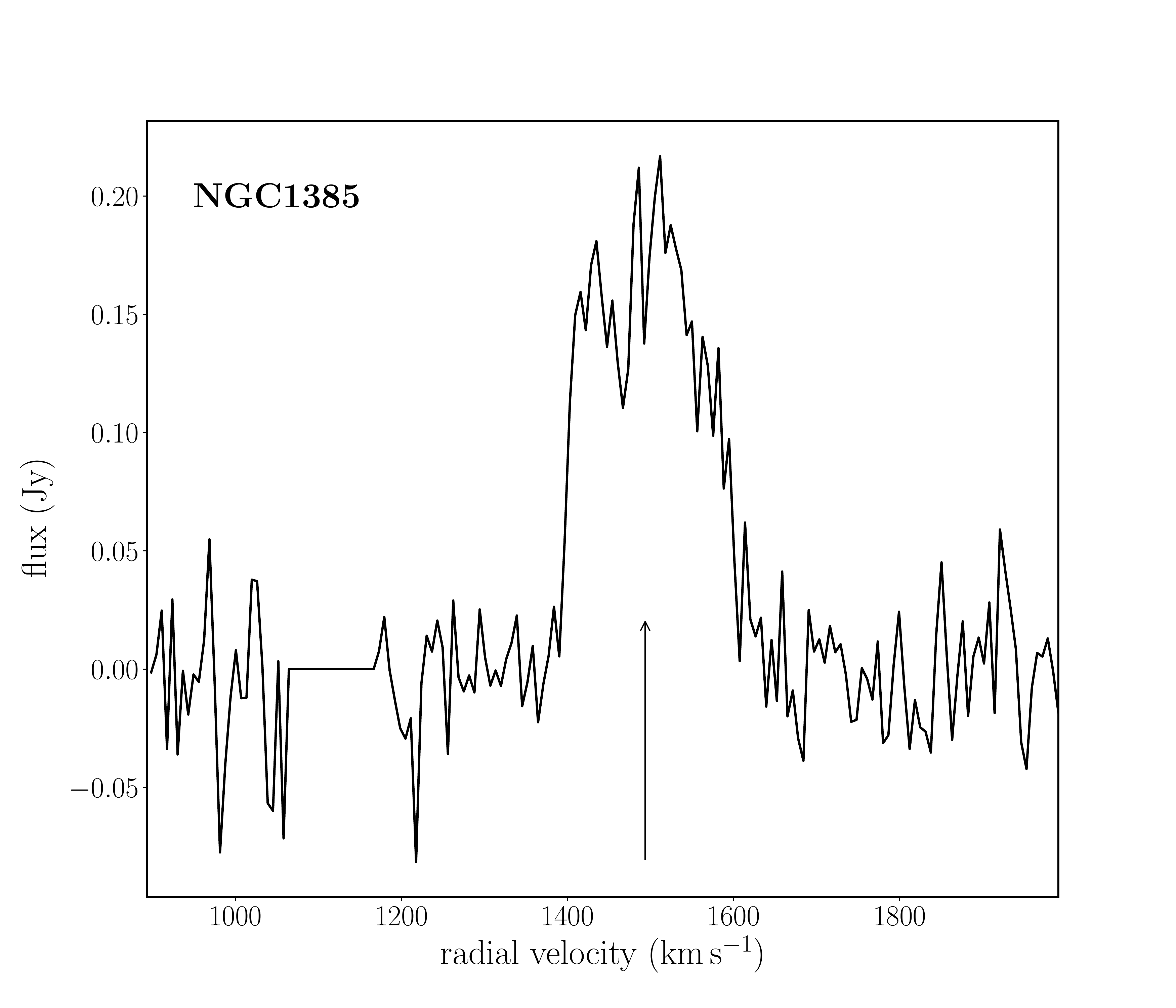}
	\includegraphics[width=\columnwidth]{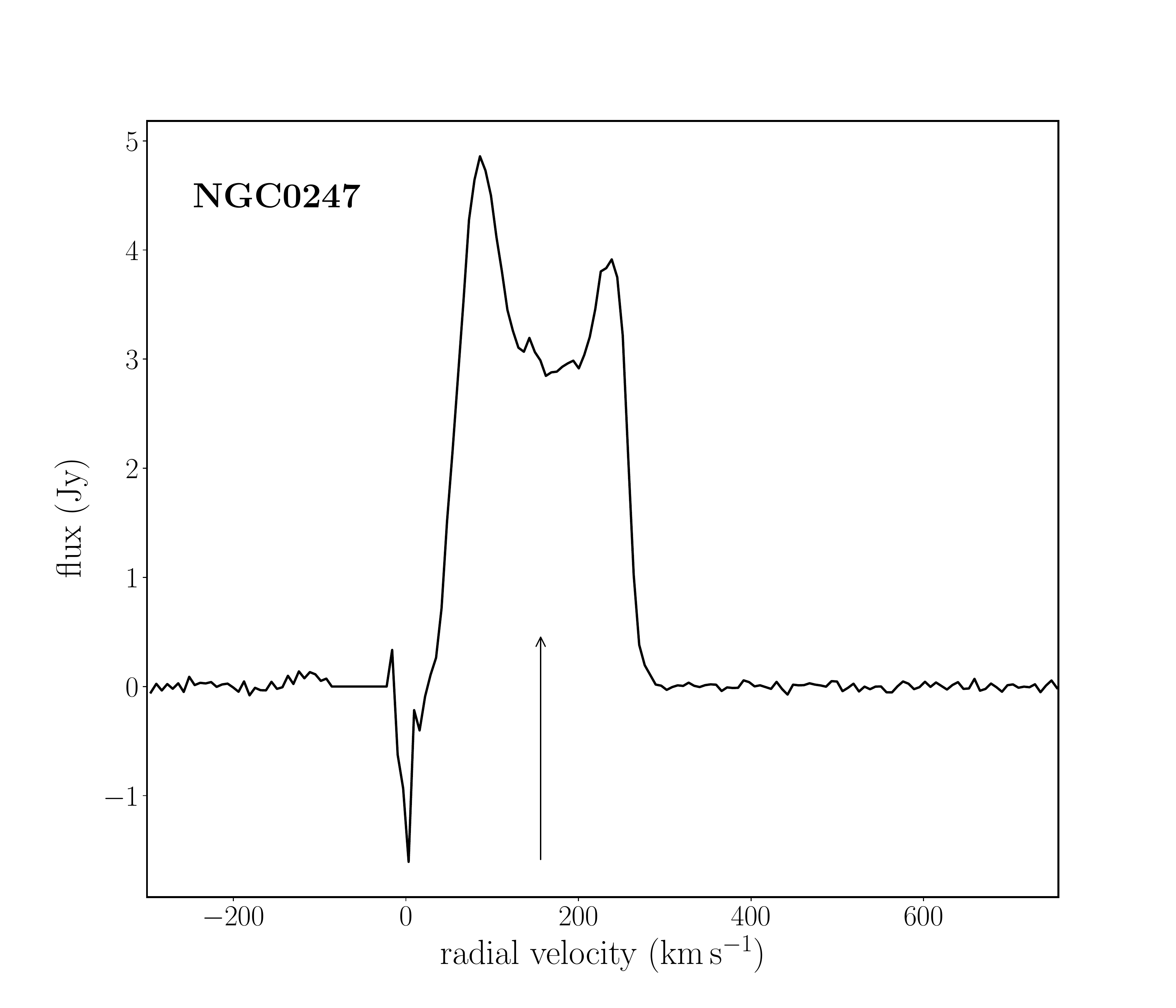}
	\includegraphics[width=\columnwidth]{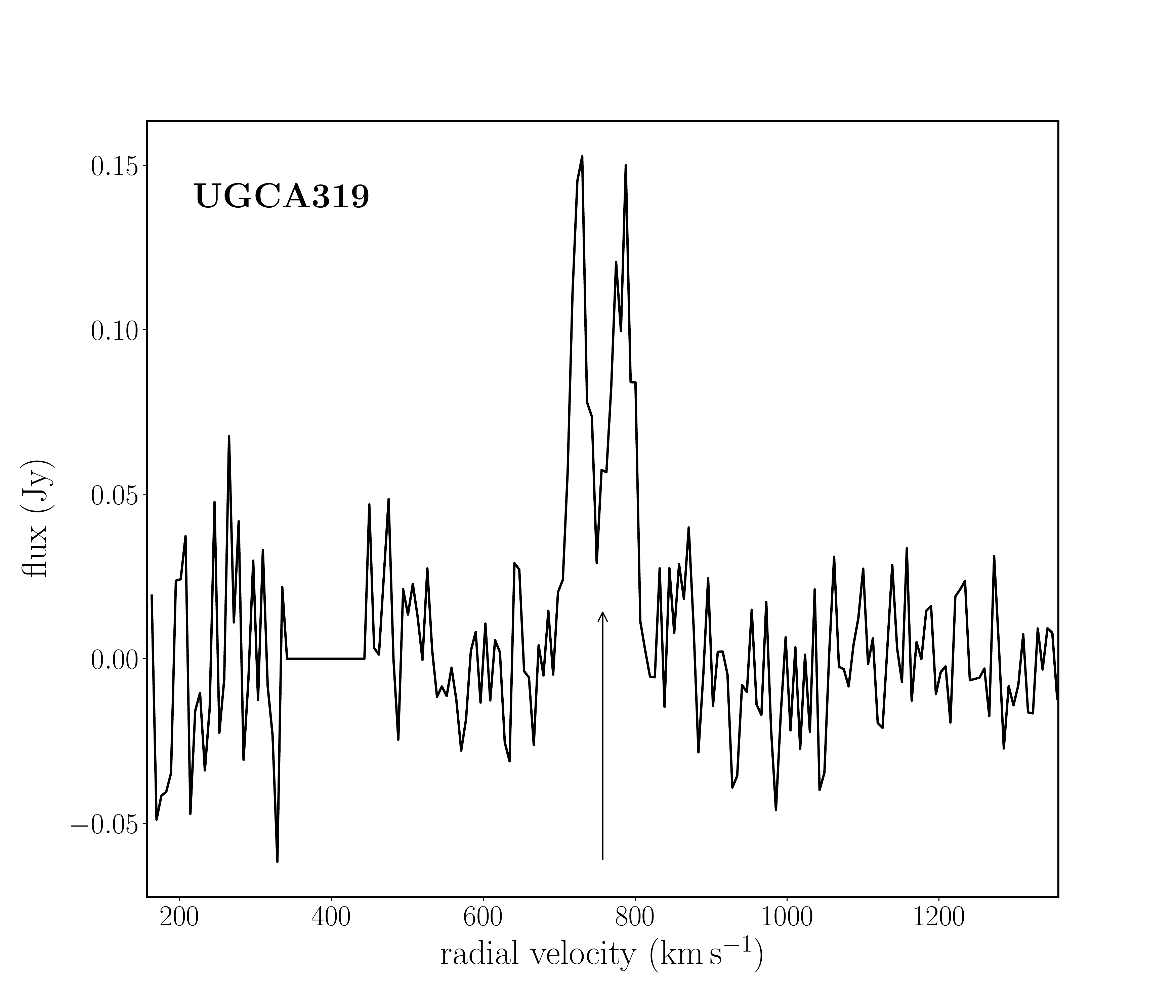}
    \caption{The global \hi\, profiles of the secondary sources ESO 302-G009 ({\it upper left}), NGC 1385 ({\it upper right}), NGC 247 ({\it lower left}) and UGCA 319 ({\it lower right}).}
    \label{app:profiles}
\end{figure*}